\documentclass{aastex62}
\usepackage{amssymb}            
\usepackage{graphicx,times}
\usepackage{epstopdf}
\usepackage{natbib}
\usepackage{float}
\usepackage{subfigure}
\usepackage{bbding}

\shorttitle{Galactic $^{14}$NH$_{3}$/$^{15}$NH$_{3}$}
\shortauthors{Chen et al.}

\begin{document}

\title{\textbf{Interstellar nitrogen isotope ratios: New NH$_{3}$ data from the Galactic center out to the Perseus arm}}

\correspondingauthor{JiangShui Zhang}
\email{jszhang@gzhu.edu.cn}

\author[0000-0001-8980-9663]{J. L. Chen}
\affil{Center for Astrophysics, Guangzhou University, Guangzhou, 510006, PR China; jszhang@gzhu.edu.cn} 

\author[0000-0002-5161-8180]{J. S. Zhang}
\affil{Center for Astrophysics, Guangzhou University, Guangzhou, 510006, PR China; jszhang@gzhu.edu.cn}

\author[0000-0002-7495-4005]{C. Henkel}
\affil{Max-Planck-Institut f{\"u}r Radioastronomie, Auf dem H{\"u}gel 69, D-53121 Bonn, Germany}
\affil{Astronomy Department, King Abdulaziz University, P.O. Box 80203, 21589 Jeddah, Saudi Arabia}

\author[0000-0001-5574-0549]{Y. T. Yan}
\affil{Max-Planck-Institut f{\"u}r Radioastronomie, Auf dem H{\"u}gel 69, D-53121 Bonn, Germany}
\affil{Center for Astrophysics, Guangzhou University, Guangzhou, 510006, PR China; jszhang@gzhu.edu.cn} 

\author{H.Z. Yu}
\affil{Center for Astrophysics, Guangzhou University, Guangzhou, 510006, PR China; jszhang@gzhu.edu.cn} 

\author[0000-0002-9829-8655]{J. J. Qiu}
\affil{School of Physics and Astronomy, Sun Yat-sen University, Guangzhou 510275, PR China}

\author[0000-0002-4154-4309]{X. D. Tang}
\affil{Xinjiang Astronomical Observatory, Chinese Academy of Sciences, 830011 Urumqi, PR China}
\affil{Key Laboratory of Radio Astronomy, Chinese Academy of Sciences, PR China}

\author{J. Wang}
\affil{Center for Astrophysics, Guangzhou University, Guangzhou, 510006, PR China; jszhang@gzhu.edu.cn} 

\author{W. Liu}
\affil{Center for Astrophysics, Guangzhou University, Guangzhou, 510006, PR China; jszhang@gzhu.edu.cn} 

\author{Y. X. Wang}
\affil{Center for Astrophysics, Guangzhou University, Guangzhou, 510006, PR China; jszhang@gzhu.edu.cn} 

\author{Y. H. Zheng}
\affil{University of Chinese Academy of Sciences, Beijing 100049, PR China}
\affil{National Astronomical Observatories, CAS, Beijing 100012, PR China}
\affil{Center for Astrophysics, Guangzhou University, Guangzhou, 510006, PR China; jszhang@gzhu.edu.cn} 

\author{J. Y. Zhao}
\affil{Center for Astrophysics, Guangzhou University, Guangzhou, 510006, PR China; jszhang@gzhu.edu.cn} 

\author{Y. P. Zou}
\affil{Center for Astrophysics, Guangzhou University, Guangzhou, 510006, PR China; jszhang@gzhu.edu.cn}

\begin{abstract}
Our aim is to measure the interstellar $^{14}$N/$^{15}$N ratio across the Galaxy, to establish a standard data set on interstellar ammonia isotope ratios and to provide new constraints on the Galactic chemical evolution. 
The ($J$, $K$) = (1, 1), (2, 2) and (3, 3) lines of $^{14}$NH$_{3}$ and $^{15}$NH$_{3}$ were observed with the Shanghai Tianma 65 m radio telescope (TMRT) and the Effelsberg-100 m telescope toward a large sample of 210 sources. 
141 of these sources were detected by the TMRT in $^{14}$NH$_{3}$. 
8 out of them were also detected in $^{15}$NH$_{3}$. For 10 of the 36 sources with strong NH$_{3}$ emission, the Effelsberg-100 m telescope detected successfully their $^{15}$NH$_{3}$(1, 1) lines, including 3 sources (G081.7522, W51D and Orion-KL) with detections by the TMRT telescope. 
Thus, a total of 15 sources are detected in both the $^{14}$NH$_{3}$ and $^{15}$NH$_{3}$ lines. 
Line and physical parameters for these 15 sources are derived, including optical depths, rotation and kinetic temperatures, and total column densities. 
$^{14}$N/$^{15}$N isotope ratios were determined from the $^{14}$NH$_{3}$/$^{15}$NH$_{3}$ abundance ratios. 
Isotope ratios obtained from both telescopes agree for a given source within the uncertainties and no dependence on heliocentric distance and kinetic temperature is seen. 
$^{14}$N/$^{15}$N ratios tend to increase with galactocentric distance, confirming a radial nitrogen isotope gradient. 
This is consistent with results from recent Galactic chemical model calculations, including the impact of super-AGB stars and novae.
\end{abstract}





\section{Introduction}\label{sec:intro}

Galactic chemical evolution (GCE) is a powerful tool to study the stellar evolution history in the Milky Way \citep{Milam05}. 
Determining isotopic ratios as a function of distance to the Galactic center ($D_{\rm GC}$), it is possible to trace back the star formation history and/or initial mass function (IMF) along the Galactic plane with different $D_{\rm GC}$ \citep{Wilson94, Zhang18}. 
Nitrogen is the fifth most common element in the universe \citep{Colzi18a}. 
The abundance ratio of its two stable isotopes, $^{14}$N and $^{15}$N, is critical to our understanding of the GCE and the origin of the solar system. 
$^{14}$N/$^{15}$N ratios are believed to be a good indicator of stellar nucleosynthesis and the mixing that subsequently occurs since the two isotopes are not synthesized in the same way \citep[e.g.,][]{Audouze75,Wilson99}.

Both $^{14}$N and $^{15}$N can be produced in the carbon-nitrogen-oxygen (CNO) cycle, which is one of the major reaction sequences of stellar hydrogen burning \citep{Wiescher10}.  $^{15}$N is believed to enrich the interstellar medium (ISM) mainly during nova outbursts, being synthesized by the hot CNO cycle \citep{Clayton03,Romano17,Colzi18a}. $^{14}$N can be created from $^{13}$C or $^{17}$O in the cold CNO cycle and should mostly be a secondary product. However, a primary component of $^{14}$N can also be formed in the so-called Hot Bottom Burning (HBB) of asymptotic giant branch (AGB) stars \citep[e.g.][]{Izzard04}. Thus $^{14}$N may be a more primary product with respect to $^{15}$N. 
Different origins of the two nitrogen isotopes should lead to an increase of $^{14}$N/$^{15}$N ratios with galactocentric distance \citep{Dahmen95,Adande12}, as predicted by models of Galactic chemical evolution \citep[][]{Romano03,Romano17}.

Previous measurements of $^{14}$N/$^{15}$N ratios in the interstellar medium are based on spectral radio lines of different molecular tracers.
A study of HCN from the Galactic disk by \citet{Dahmen95} revealed ratios that slightly increase with increasing galactocentric distance, with values of $\sim$400 in the local ISM. 
Subsequently, observations of CN, C$^{15}$N, HN$^{13}$C and H$^{15}$NC lines reported by \citet{Adande12} led to $^{14}$N/$^{15}$N ratios with increasing $D_{\rm GC}$ (290 $\pm$ 40 near the solar circle).
For the most recent study, \citet{Colzi18a} derived $^{14}$N/$^{15}$N ratios from HN$^{13}$C, H$^{15}$NC, H$^{13}$CN and HC$^{15}$N, which also show a trend with increasing values at larger $D_{\rm GC}$ (375 $\pm$ 50 in the local ISM). 
However, the slope of the increasing $^{14}$N/$^{15}$N ratio with $D_{\rm GC}$ is still a matter of debate.
For the Galactic center region, there are only few direct observations.
Based on H$^{13}$CN and HC$^{15}$N observations, \citet{Wannier81} obtained a $^{14}$N/$^{15}$N ratio \textgreater 510 toward Sgr A. 
An even larger value of $\sim$1000, this time from $^{14}$NH$_{3}$/$^{15}$NH$_{3}$ data, was reported by \citet{Gusten85} toward Galactic center clouds, leading to a ratio surpassing that from the solar system by a factor of four.
Towards yet another source in the Galactic center region, Sgr B(N), \citet{Mills18} measured a much lower $^{14}$N/$^{15}$N ratio ($\sim$200) in the N2 hot core, through VLA observations of $^{14}$NH$_{3}$ and $^{15}$NH$_{3}$. 
This is consistent with extrapolated values in the Galactic center region of 123 $\pm$ 37 \citep{Adande12} and 250 $\pm$ 67 \citep{Colzi18a}, from their proposed radial Galactic trends, respectively.
However, systematic errors in the results from \citet{Mills18} (e.g. optically thick transitions) could not be excluded.
Moreover, galactocentric distances lower than 4 kpc are excluded in Galactic chemical evolution models due to the peculiarity and complexity of this region \citep[e.g., ][]{Romano17,Romano19}.

From a theoretical point of view, chemical isotopic fractionation might be significant in star forming regions. 
This is still not well understood.
Current models predict different degrees of fractionation depending on the N-bearing molecular species. 
However, these models have faced difficulties explaining discrepant $^{14}$N/$^{15}$N ratios with large variations toward different astrophysically relevant interstellar sources \citep[e.g., ][]{Charnley02,Roueff15,Furuya18,Colzi19}. 
Thus more observations from as many molecular species as possible and more modeling work, including the Galactic center region, are required. 

Many observations used a double isotope ratio also including $^{12}$C/$^{13}$C, which may enhance uncertainties related to the abundance of $^{14}$N/$^{15}$N \citep[e.g. CN, HCN or HNC, ][]{Colzi18a, Loison20}.
With nitrogen isotope ratios usually surpassing 100, any analysis using HCN, HNC or N$_{2}$H$^{+}$ may be hampered by the fact that either the main species is optically thick, that the rare species remains undetectable or that double isotope ratios have to be implemented. 
Hyperfine splitting, even if present, is in most cases not wide enough to allow for the determination of opacities in the potentially optically thick lines of the main species. 
In this context, CN and NH$_{3}$ are the notable exceptions, providing a direct evaluation of line opacities in the critical $^{14}$N bearing main species. In the following we focus on ammonia (NH$_{3}$). 
Allowing for the observation of many lines in a limited frequency interval, permitting the determination of optical depths, rotational temperatures and total column densities, it is one of the best tools to directly determine $^{14}$N/$^{15}$N isotope ratios. 
Moreover, so far proposed abundance gradients established across the body of the Milky Way may have been greatly affected by the uncertain distances of the targets \citep[e. g.,][]{Adande12,Colzi18a}. 
Now, however, NH$_{3}$ can be measured in a large number of sources with well determined distances (see Sect. \ref{sec:distance}).

Therefore, with the Tianma-65 m (TMRT) and Effelsberg-100 m telescopes, we performed observations of the ($J, K$) = (1, 1), (2, 2) and (3, 3) inversion lines of NH$_{3}$ and $^{15}$NH$_{3}$ toward a large sample of star formation regions, covering $D_{\rm GC}$ distance bins from the Galactic center out to $\sim$10 kpc. 
Based on comparisons of nitrogen isotope ratios obtained from different molecular species, i.e. NH$_{3}$, HCN and HNC \citep[for the latter, see ][]{Colzi18a}, systematic discrepancies could be related to the choice of molecular species to quantify for the first time also chemical aspects being caused by potential fractionation \citep{Roueff15,Viti19} on a Galaxy-wide scale. 
In Sect. \ref{sec:obser}, sample and observations are introduced. 
An analysis of observational data and main results are presented in Sect. \ref{sec:result}. 
In Sect. \ref{sec:discussion}, these results are discussed in the light of possible physical and chemical effects and are compared with previous studies. 
A brief summary is provided in Sect. \ref{sec:summary}.

\section{TARGETS, OBSERVATIONS, AND DATA REDUCTION}\label{sec:obser}

\subsection{Sample selection and distance}\label{sec:distance}

A total of 210 sources was chosen from previously studied strong NH$_{3}$ sources \citep[e.g.,][]{Wyrowski96,Longmore07,Rosolowsky09,Lis10,Cyganowski13,Reid14}. 
Sources have accurate distance values, including 113 sources from trigonometric parallax measurements and 97 from the Parallax-Based Distance Calculator \citep{Reid14,Reid19}. 
Using a Bayesian approach, sources are assigned to arms based on their ($l,b,v$) coordinates with respect to arm signatures seen in CO and HI surveys. 
The most reasonable distance (near or far) can be derived through a full distance probability density function from the Parallax-Based Distance Calculator, considering a source's kinematic distance, displacement from the plane, and proximity to individual parallax sources.
We believe that it is an important improvement to reveal radial variations of $^{14}$N/$^{15}$N in an unbiased way. The heliocentric distance was used to calculate the galacocentric distance of targets \citep{Roman09},
\begin{equation}
D_{\rm GC} = \sqrt{[R_{0}\cos(l)-d]^{2}+R_{0}^{2}\sin^{2}(l)}.
\end{equation}
$l$ is the Galactic longitude, $R_{0}$ and $d$ are the distance of the Sun from the Galactic center \citep[8.122 $\pm$ 0.031 kpc, from the][]{Gravity18} and of the targeted source from the Sun \citep{Reid14}, respectively.The error in the distance to the Galactic center is so small that it will be neglected in the following. 

The sample includes star forming regions at different evolutionary stages, including sources associated with InfraRed Dark Clouds (IRDCs), massive young stellar objects (YSOs) and H II regions \footnote{\url{http://simbad.u-strasbg.fr/simbad/}}, which are used to better constrain radial trends of the Galactic $^{14}$N/$^{15}$N isotope ratio.
The source list is presented in the Appendix.

\subsection{Observations}

\subsubsection{Tianma 65 m observations}
For our sample of 210 sources, we made observations of the ($J, K$) = (1, 1), (2, 2) and (3, 3) lines of $^{14}$NH$_{3}$ and $^{15}$NH$_{3}$ (see Table \ref{tab:para}), firstly by the Shanghai Tianma 65 m radio telescope (TMRT) in 2019 April, May, November and December, with a beam size of $\sim$50$^{''}$.
A cryogenically cooled K-band (17.9 - 26.2 GHz) receiver was employed, and the Digital Backend System, DIBAS, was used for recording \citep[see][]{Li16}. 
The DIBAS mode 22 was adopted for observations, with eight spectral windows, to cover the $^{14}$NH$_{3}$ and $^{15}$NH$_{3}$ lines simultaneously, each with a bandwidth of 23.4 MHz (16384 channels), supplying a spectral resolution of 1.43 kHz ($\sim$0.02 km s$^{-1}$). 
The active surface system of the primary dish and a subreflector were used to improve the aperture efficiency. 
Observations were performed in position switching mode.
The system temperature was 100 – 200 K on an antenna temperature scale ($T^{*}_{\rm A}$).
The main beam brightness temperature ($T_{\rm mb}$) can be obtained from the antenna temperature scale by $T_{\rm mb}$ = $T_{\rm A}^{*}$/$\eta _{b}$, where $\eta _{b}$ is the main beam efficiency correction factor, with a mean value of $\sim$0.6 \citep{Mei20}.   
The on-source integration time was about 0.1 – 3.0 hr for each of our sources.

\subsubsection{Effelsberg 100 m observations}
We used the Effelsberg 100 m telescope\footnote{The 100-m telescope at Effelsberg is operated by the Max-Planck-Institut für Radioastronomie (MPIFR) on behalf of the Max-Planck Gesellschaft (MPG).} to observe the ($J$, $K$) = (1, 1), (2, 2) and (3, 3) lines of $^{14}$NH$_{3}$ and $^{15}$NH$_{3}$ (see Table \ref{tab:para}) toward 36 selected sources with strong $^{14}$NH$_{3}$ signals from previous TMRT observations in 2019 December and 2020 January.
The newly installed Fast Fourier Transform Spectrometer (FFTS) was used as backend. 
Initially, observations were carried out in the Hi-res (olution) mode, with 2 spectral windows covering the $^{14}$NH$_{3}$ and $^{15}$NH$_{3}$ lines simultaneously, each with a bandwidth of 300 MHz (65536 channels), resulting in a spectral resolution of 4.6 kHz ($\sim$0.066 km s$^{-1}$). The Low-res mode was adopted for later observations, with 4 spectral windows, each with a bandwidth of 2 GHz (65536 channels), supplying a spectral resolution of 38.1 kHz ($\sim$0.6 km s$^{-1}$).
The system temperature was 90 - 250 K on an antenna temperature scale. 
The beam size is close to 40" near 23 GHz.
Strong continuum sources (e.g., NGC 7027 and 3C 286) were used to calibrate spectral line flux densities. Standard 23 GHz flux densities of 5.6 Jy and 2.5 Jy were adopted for NGC 7027 and 3C 286, respectively \citep{Ott94}.
The main beam brightness temperature $T_{\rm mb}$ (K) scale can be determined from the observed flux density (Jy) by a conversion factor, which is 1.7 K Jy$^{-1}$ at 18.5 GHz, 1.5 K Jy$^{-1}$ at 22 GHz, and 1.4 K Jy$^{-1}$ at 23.7 GHz \citep{Gong15}.
The spectra were obtained in a position switching mode. The on-source integration time was, depending on line strength, 0.3 -- 2 hr for each source. 
The focus was checked every few hours. Pointing was obtained every two hours toward nearby pointing sources (e.g., 3C 123, or NGC 7027). 

\startlongtable

\begin{deluxetable}{lcccc}
	
	\tablecaption{The parameters of $^{14}$NH$_{3}$ and $^{15}$NH$_{3}$ transition lines.\label{tab:para}}
	
	\tablehead{\colhead{Line} & \colhead{Frequency} & \colhead{log$_{10}(A_{ij})$\tablenotemark{a}} & \colhead{$E_{u}/k$\tablenotemark{b}} & \colhead{$g_{u}$\tablenotemark{c}} \\
		\colhead{} & \colhead{(MHz)} & \colhead{} & \colhead{(K)} & \colhead{}
	}
	\startdata
	$^{14}$NH$_{3}$(1, 1)& 23694.5   &      -6.76650           &   24.35   &       6      \\
	$^{14}$NH$_{3}$(2, 2)& 23722.6   &      -6.31125           &   65.34   &       10      \\
	$^{14}$NH$_{3}$(3, 3)& 23870.1   &      -6.25203           &   124.73   &       28      \\
	$^{15}$NH$_{3}$(1, 1)& 22624.9   &      -6.83680           &   23.82   &       6        \\
	$^{15}$NH$_{3}$(2, 2)& 22649.8   &      -6.71062           &   64.93   &       10       \\
	$^{15}$NH$_{3}$(3, 3)& 22789.4   &      -6.65097           &   123.91   &       28       \\
	\enddata
	\tablenotetext{a}{Einstein coefficient for spontaneous emission.}
	\tablenotetext{b}{Energy of the upper level above the ground state.}	
	\tablenotetext{c}{Upper state degeneracy.}
	\tablecomments{The parameters are from the JPL Molecular Spectroscopy Catalog \citep{Pickett98}.}
\end{deluxetable}

\subsection{Data reduction}

The Continuum and Line Analysis Single-dish Software (CLASS) of the Grenoble Image and Line Data Analysis Software packages \footnote{\url{http://http://www.iram.fr/IRAMFR/GILDAS/}} (GILDAS, e. g., Guilloteau \& Lucas 2000) was used to reduce the spectral line data. 
After subtracting baselines and applying Hanning smoothing, the line parameters are obtained from Gaussian fits for detected lines (signal-to-noise S/N \textgreater 3 sigma), with a spectral resolution of $\sim$0.78 km\,s$^{-1}$ for TMRT and 0.70 km\,s$^{-1}$ for Effelsberg observations, respectively.

\clearpage

\startlongtable

\begin{deluxetable}{lccccccccc}
	
	\tablecaption{Observational parameters of the ($J, K$) = (1, 1), (2, 2) and (3, 3) inversion lines of $^{14}$NH$_{3}$ and $^{15}$NH$_{3}$ obtained from Gaussian fits \label{tab:15NH3para_TMRT-Eff}}.
	
	\tablehead{\colhead{Object}	&	\colhead{Telescope}	&	\colhead{$\alpha$(2000) $\delta$(2000)}	 		&	\colhead{Total time}	&	\colhead{Molecule}	&	\colhead{r.m.s.}	&	\colhead{$ \int{T_{\rm mb}{\rm{d}} v}$} 			&	\colhead{$V_{\rm LSR}$}			&	\colhead{$\Delta$ V}			&	\colhead{$T_{\rm mb}$}	\\
		&		&	($^h \; ^m \; ^s$)	 	($^{\circ} \; ^{\prime} \; ^{\prime\prime}$)	&	(min)	&		&	(mK)	&	\colhead{(${\rm K} \, {\rm km} \, {\rm s}^{-1}$)}			&	\colhead{($\rm km \, \rm s^{-1}$)}			&	\colhead{($\rm km \, \rm s^{-1}$)}			&	(K)	
		}
	\colnumbers
	\startdata
	G032.04	&	T	&	18:49:36.3	 	-00:45:37.1	&	11	&	$^{14}$NH$_{3}$(1,1)	&	46.9	&	34.9	$\pm$	0.3	&	94.7	$\pm$	0.8	&	3.2	$\pm$	0.8	&	3.93	\\
	&		&		 		&		&	$^{14}$NH$_{3}$(2,2)	&	65.6	&	6.10	$\pm$	0.18	&	94.19	$\pm$	0.06	&	3.16	$\pm$	0.14	&	1.80	\\
	&		&		 		&		&	$^{14}$NH$_{3}$(3,3)	&	62.0	&	3.90	$\pm$	0.19	&	94.17	$\pm$	0.12	&	4.7	$\pm$	0.3	&	0.77	\\
	&		&		 		&		&	$^{15}$NH$_{3}$(1,1)	&	77.7	&	0.8	$\pm$	0.2	&	95.6	$\pm$	0.3	&	2.5	$\pm$	0.6	&	0.37	\\
	&		&		 		&		&	$^{15}$NH$_{3}$(2,2)	&	77.4	&	…			&	…			&	…			&	…	\\
	&		&		 		&		&	$^{15}$NH$_{3}$(3,3)	&	78.4	&	…			&	…			&	…			&	…	\\
	G053.23	&	T	&	19:29:33.2	 	+18:01:00.6	&	65	&	$^{14}$NH$_{3}$(1,1)	&	25.6	&	14.33	$\pm$	0.13	&	24.3	$\pm$	0.8	&	1.2	$\pm$	0.8	&	4.02	\\
	&		&		 		&		&	$^{14}$NH$_{3}$(2,2)	&	28.7	&	1.19	$\pm$	0.04	&	23.7	$\pm$	0.2	&	1.09	$\pm$	0.04	&	1.04	\\
	&		&		 		&		&	$^{14}$NH$_{3}$(3,3)	&	29.4	&	…			&	…			&	…			&	…	\\
	&		&		 		&		&	$^{15}$NH$_{3}$(1,1)	&	72.8	&	0.09	$\pm$	0.03	&	23.8	$\pm$	0.2	&	1.0	$\pm$	0.5	&	0.08	\\
	&		&		 		&		&	$^{15}$NH$_{3}$(2,2)	&	38.1	&	…			&	…			&	…			&	…	\\
	&		&		 		&		&	$^{15}$NH$_{3}$(3,3)	&	135	&	…			&	…			&	…			&	…	\\
	G081.75	&	T	&	20:39:02.0	 	+42:24:58.6	&	360	&	$^{14}$NH$_{3}$(1,1)	&	19.5	&	35.8	$\pm$	1.0	&	-3.8	$\pm$	0.8	&	2.2	$\pm$	0.8	&	6.04	\\
	&		&		 		&		&	$^{14}$NH$_{3}$(2,2)	&	18.9	&	9.70	$\pm$	0.03	&	-4.29	$\pm$	0.06	&	2.23	$\pm$	0.01	&	3.09	\\
	&		&		 		&		&	$^{14}$NH$_{3}$(3,3)	&	14.4	&	2.98	$\pm$	0.03	&	-4.07	$\pm$	0.01	&	2.75	$\pm$	0.04	&	1.02	\\
	&		&		 		&		&	$^{15}$NH$_{3}$(1,1)	&	26.3	&	0.14	$\pm$	0.02	&	-4.39	$\pm$	0.16	&	2.3	$\pm$	0.4	&	0.06	\\
	&		&		 		&		&	$^{15}$NH$_{3}$(2,2)	&	19.0	&	…			&	…			&	…			&	…	\\
	&		&		 		&		&	$^{15}$NH$_{3}$(3,3)	&	16.2	&	…			&	…			&	…			&	…	\\
	&	E	&	20:39:02.0	 	+42:24:58.6	&	235	&	$^{14}$NH$_{3}$(1,1)	&	27.3	&	27.4	$\pm$	0.5	&	-3.9	$\pm$	0.7	&	2.0	$\pm$	0.7	&	4.81	\\
	&		&		 		&		&	$^{14}$NH$_{3}$(2,2)	&	83.3	&	7.7	$\pm$	0.2	&	-4.38	$\pm$	0.01	&	1.94	$\pm$	0.01	&	2.75	\\
	&		&		 		&		&	$^{14}$NH$_{3}$(3,3)	&	13.4	&	2.14	$\pm$	0.03	&	-4.17	$\pm$	0.02	&	2.40	$\pm$	0.04	&	0.84	\\
	&		&		 		&		&	$^{15}$NH$_{3}$(1,1)	&	49.3	&	0.09	$\pm$	0.02	&	-4.4	$\pm$	0.3	&	2.2	$\pm$	0.8	&	0.06	\\
	&		&		 		&		&	$^{15}$NH$_{3}$(2,2)	&	12.1	&	…			&		…		&		…		&	…	\\
	&		&		 		&		&	$^{15}$NH$_{3}$(3,3)	&	11.7	&	…			&		…		&		…		&	…	\\
	G121.29	&	T	&	00:36:47.3	 	+63:29:02.2	&	331	&	$^{14}$NH$_{3}$(1,1)	&	14.7	&	27.02	$\pm$	0.13	&	-17.6	$\pm$	0.4	&	2.2	$\pm$	0.4	&	4.74	\\
	&		&		 		&		&	$^{14}$NH$_{3}$(2,2)	&	16.2	&	6.45	$\pm$	0.05	&	-18.1	$\pm$	0.4	&	2.4	$\pm$	0.4	&	1.85	\\
	&		&		 		&		&	$^{14}$NH$_{3}$(3,3)	&	12.6	&	1.74	$\pm$	0.03	&	-18.01	$\pm$	0.03	&	3.05	$\pm$	0.07	&	0.53	\\
	&		&		 		&		&	$^{15}$NH$_{3}$(1,1)	&	36.7	&	0.3	$\pm$	0.1	&	-19.5	$\pm$	0.3	&	1.7	$\pm$	0.7	&	0.30	\\
	&		&		 		&		&	$^{15}$NH$_{3}$(2,2)	&	29.3	&	…			&	…			&	…			&	…	\\
	&		&		 		&		&	$^{15}$NH$_{3}$(3,3)	&	34.7	&	…			&	…			&	…			&	…	\\
	G30.70	&	T	&	18:47:36.1	 	-02:00:58.2	&	15	&	$^{14}$NH$_{3}$(1,1)	&	69.9	&	90.9	$\pm$	1.4	&	91.2	$\pm$	0.4	&	4.9	$\pm$	0.4	&	6.12	\\
	&		&		 		&		&	$^{14}$NH$_{3}$(2,2)	&	73.6	&	34.8	$\pm$	0.7	&	90.6	$\pm$	0.4	&	5.2	$\pm$	0.4	&	4.06	\\
	&		&		 		&		&	$^{14}$NH$_{3}$(3,3)	&	67.6	&	21.6	$\pm$	0.3	&	90.7	$\pm$	0.4	&	5.0	$\pm$	0.4	&	3.99	\\
	&		&		 		&		&	$^{15}$NH$_{3}$(1,1)	&	56.7	&	0.89	$\pm$	0.08	&	89.5	$\pm$	0.2	&	1.0	$\pm$	0.4	&	0.18	\\
	&		&		 		&		&	$^{15}$NH$_{3}$(2,2)	&	54.2	&	…			&	…			&	…			&	…	\\
	&		&		 		&		&	$^{15}$NH$_{3}$(3,3)	&	58.6	&	…			&	…			&	…			&	…	\\
	NGC\,6334\,I	&	T	&	17:20:53.3	 	-35:47:01.2	&	33	&	$^{14}$NH$_{3}$(1,1)	&	114.0	&	245	$\pm$	4	&	-6.9	$\pm$	0.5	&	5.2	$\pm$	0.5	&	15.34	\\
	&		&		 		&		&	$^{14}$NH$_{3}$(2,2)	&	49.4	&	68.7	$\pm$	1.1	&	-7.5	$\pm$	0.6	&	5.9	$\pm$	0.6	&	5.68	\\
	&		&		 		&		&	$^{14}$NH$_{3}$(3,3)	&	35.9	&	80.7	$\pm$	1.4	&	-7.2	$\pm$	0.6	&	5.8	$\pm$	0.6	&	7.72	\\
	&		&		 		&		&	$^{15}$NH$_{3}$(1,1)	&	46.6	&	1.70	$\pm$	0.19	&	-6.7	$\pm$	0.3	&	4.5	$\pm$	1.1	&	0.42	\\
	&		&		 		&		&	$^{15}$NH$_{3}$(2,2)	&	46.9	&	0.82	$\pm$	0.11	&	-6.9	$\pm$	0.3	&	4.5	$\pm$	0.6	&	0.31	\\
	&		&		 		&		&	$^{15}$NH$_{3}$(3,3)	&	42.3	&	2.41	$\pm$	0.15	&	-6.81	$\pm$	0.13	&	4.4	$\pm$	0.4	&	0.48	\\
	Orion-KL	&	T	&	05:35:14.4	 	-05:22:29.6	&	94	&	$^{14}$NH$_{3}$(1,1)	&	36.8	&	152.2	$\pm$	1.1	&	7.7	$\pm$	0.4	&	5.5	$\pm$	0.4	&	8.06	\\
	&		&		 		&		&	$^{14}$NH$_{3}$(2,2)	&	42.3	&	146.4	$\pm$	1.2	&	7.1	$\pm$	0.4	&	10.2	$\pm$	0.4	&	8.26	\\
	&		&		 		&		&	$^{14}$NH$_{3}$(3,3)	&	33.1	&	173.2	$\pm$	1.4	&	7.0	$\pm$	0.4	&	13.0	$\pm$	0.4	&	9.78	\\
	&		&		 		&		&	$^{15}$NH$_{3}$(1,1)	&	54.5	&	2.02	$\pm$	0.10	&	7.3	$\pm$	0.4	&	10.7	$\pm$	1.2	&	0.30	\\
	&		&		 		&		&	$^{15}$NH$_{3}$(2,2)	&	29.1	&	1.58	$\pm$	0.08	&	7.4	$\pm$	0.4	&	6.8	$\pm$	0.9	&	0.29	\\
	&		&		 		&		&	$^{15}$NH$_{3}$(3,3)	&	26.5	&	3.11	$\pm$	0.11	&	7.24	$\pm$	0.16	&	8.3	$\pm$	0.4	&	0.38	\\
	&	E	&	05:35:14.4	 	-05:22:29.6	&	40	&	$^{14}$NH$_{3}$(1,1)	&	66.5	&	168.2	$\pm$	1.3	&	7.2	$\pm$	0.6	&	7.7	$\pm$	0.6	&	7.46	\\
	&		&		 		&		&	$^{14}$NH$_{3}$(2,2)	&	51.1	&	160.2	$\pm$	1.4	&	7.0	$\pm$	0.6	&	11.8	$\pm$	0.6	&	8.12	\\
	&		&		 		&		&	$^{14}$NH$_{3}$(3,3)	&	49.3	&	170.5	$\pm$	1.7	&	6.8	$\pm$	0.6	&	12.4	$\pm$	0.6	&	8.45	\\
	&		&		 		&		&	$^{15}$NH$_{3}$(1,1)	&	16.1	&	2.06	$\pm$	0.10	&	6.7	$\pm$	0.2	&	9.6	$\pm$	0.6	&	0.29	\\
	&		&		 		&		&	$^{15}$NH$_{3}$(2,2)	&	31.8	&	1.84	$\pm$	0.13	&	6.8	$\pm$	3.9	&	6.6	$\pm$	9.3	&	0.26	\\
	&		&		 		&		&	$^{15}$NH$_{3}$(3,3)	&	23.6	&	2.49	$\pm$	0.09	&	6.87	$\pm$	0.12	&	7.8	$\pm$	0.3	&	0.34	\\
	W51D	&	T	&	19:23:40.1	 	14:31:07.1	&	72	&	$^{14}$NH$_{3}$(1,1)	&	93.0	&	84.9	$\pm$	1.4	&	60.1	$\pm$	0.6	&	5.8	$\pm$	0.6	&	4.68	\\
	&		&		 		&		&	$^{14}$NH$_{3}$(2,2)	&	24.2	&	37.1	$\pm$	0.3	&	60.3	$\pm$	0.6	&	6.5	$\pm$	0.6	&	2.26	\\
	&		&		 		&		&	$^{14}$NH$_{3}$(3,3)	&	29.7	&	52.3	$\pm$	0.4	&	60.5	$\pm$	0.6	&	7.6	$\pm$	0.6	&	3.05	\\
	&		&		 		&		&	$^{15}$NH$_{3}$(1,1)	&	15.6	&	0.70	$\pm$	0.13	&	59.1	$\pm$	0.6	&	5.5	$\pm$	1.5	&	0.17	\\
	&		&		 		&		&	$^{15}$NH$_{3}$(2,2)	&	16.9	&	0.42	$\pm$	0.02	&	58.6	$\pm$	0.2	&	6.2	$\pm$	0.5	&	0.06	\\
	&		&		 		&		&	$^{15}$NH$_{3}$(3,3)	&	10.7	&	0.76	$\pm$	0.07	&	59.1	$\pm$	0.4	&	6.8	$\pm$	1.2	&	0.09	\\
	&	E	&	19:23:39.8	 	14:31:10.1	&	153	&	$^{14}$NH$_{3}$(1,1)	&	13.0	&	34.0	$\pm$	0.3	&	59.4	$\pm$	0.6	&	7.7	$\pm$	0.6	&	1.74	\\
	&		&		 		&		&	$^{14}$NH$_{3}$(2,2)	&	8.54	&	24.1	$\pm$	0.1	&	59.91	$\pm$	0.10	&	7.11	$\pm$	0.04	&	1.57	\\
	&		&		 		&		&	$^{14}$NH$_{3}$(3,3)	&	12.5	&	25.4	$\pm$	0.1	&	59.8	$\pm$	0.6	&	10.4	$\pm$	0.6	&	1.44	\\
	&		&		 		&		&	$^{15}$NH$_{3}$(1,1)	&	18.4	&	0.49	$\pm$	0.20	&	60.3	$\pm$	0.9	&	7.5	$\pm$	1.4	&	0.16	\\
	&		&		 		&		&	$^{15}$NH$_{3}$(2,2)	&	18.1	&	0.34	$\pm$	0.04	&	59.0	$\pm$	0.8	&	8.8	$\pm$	1.7	&	0.05	\\
	&		&		 		&		&	$^{15}$NH$_{3}$(3,3)	&	15.8	&	1.51	$\pm$	0.13	&	59.8	$\pm$	0.4	&	7.4	$\pm$	1.0	&	0.10	\\
	G016.92	&	E	&	18:18:08.5	 	-13:45:05.7	&	60	&	$^{14}$NH$_{3}$(1,1)	&	19.6	&	10.2	$\pm$	0.1	&	21.0	$\pm$	0.6	&	2.7	$\pm$	0.6	&	1.43	\\
	&		&		 		&		&	$^{14}$NH$_{3}$(2,2)	&	15.6	&	2.17	$\pm$	0.07	&	20.6	$\pm$	0.1	&	2.93	$\pm$	0.11	&	0.70	\\
	&		&		 		&		&	$^{14}$NH$_{3}$(3,3)	&	40.4	&	0.88	$\pm$	0.05	&	20.34	$\pm$	0.12	&	4.4	$\pm$	0.3	&	0.19	\\
	&		&		 		&		&	$^{15}$NH$_{3}$(1,1)	&	17.0	&	0.51	$\pm$	0.13	&	20.6	$\pm$	0.2	&	2.4	$\pm$	1.0	&	0.22	\\
	&		&		 		&		&	$^{15}$NH$_{3}$(2,2)	&	20.2	&	…			&	…			&	…			&	…	\\
	&		&		 		&		&	$^{15}$NH$_{3}$(3,3)	&	18.8	&	…			&	…			&	…			&	…	\\
	G10.47	&	E	&	18:08:38.2	 	-19:51:49.6	&	60	&	$^{14}$NH$_{3}$(1,1)	&	29.9	&	54.32	$\pm$	0.13	&	67.55	$\pm$	0.13	&	12.93	$\pm$	0.17	&	2.32	\\
	&		&		 		&		&	$^{14}$NH$_{3}$(2,2)	&	34.4	&	38.2	$\pm$	0.3	&	66.9	$\pm$	0.5	&	8.8	$\pm$	0.4	&	1.96	\\
	&		&		 		&		&	$^{14}$NH$_{3}$(3,3)	&	39.2	&	41.0	$\pm$	0.3	&	67.1	$\pm$	0.5	&	10.0	$\pm$	0.5	&	2.20	\\
	&		&		 		&		&	$^{15}$NH$_{3}$(1,1)	&	35.8	&	0.68	$\pm$	0.08	&	64.9	$\pm$	0.2	&	2.8	$\pm$	0.4	&	0.23	\\
	&		&		 		&		&	$^{15}$NH$_{3}$(2,2)	&	61.2	&	2.3	$\pm$	0.2	&	66.1	$\pm$	2.2	&	9.1	$\pm$	0.9	&	0.28	\\
	&		&		 		&		&	$^{15}$NH$_{3}$(3,3)	&	34.1	&	2.04	$\pm$	0.07	&	65.2	$\pm$	0.2	&	11.4	$\pm$	0.8	&	0.27	\\
	G188.79	&	E	&	06:09:06.9	 	+21:50:41.4	&	180	&	$^{14}$NH$_{3}$(1,1)	&	6.38	&	4.40	$\pm$	0.02	&	-0.51	$\pm$	0.58	&	3.2	$\pm$	0.6	&	0.61	\\
	&		&		 		&		&	$^{14}$NH$_{3}$(2,2)	&	8.46	&	1.06	$\pm$	0.03	&	-0.85	$\pm$	0.04	&	3.07	$\pm$	0.09	&	0.33	\\
	&		&		 		&		&	$^{14}$NH$_{3}$(3,3)	&	7.09	&	0.60	$\pm$	0.02	&	-0.59	$\pm$	0.06	&	3.28	$\pm$	0.13	&	0.17	\\
	&		&		 		&		&	$^{15}$NH$_{3}$(1,1)	&	7.85	&	0.15	$\pm$	0.04	&	-0.53	$\pm$	0.23	&	3.5	$\pm$	1.1	&	0.09	\\
	&		&		 		&		&	$^{15}$NH$_{3}$(2,2)	&	7.58	&	…			&	…			&	…			&	…	\\
	&		&		 		&		&	$^{15}$NH$_{3}$(3,3)	&	6.71	&	…			&	…			&	…			&	…	\\
	G35.14	&	E	&	18:58:07.0	 	01:37:11.9	&	24	&	$^{14}$NH$_{3}$(1,1)	&	62.9	&	45.8	$\pm$	0.5	&	33.9	$\pm$	0.5	&	4.4	$\pm$	0.5	&	3.98	\\
	&		&		 		&		&	$^{14}$NH$_{3}$(2,2)	&	52.2	&	9.35	$\pm$	0.03	&	33.5	$\pm$	0.1	&	4.45	$\pm$	0.06	&	1.55	\\
	&		&		 		&		&	$^{14}$NH$_{3}$(3,3)	&	66.8	&	4.69	$\pm$	0.10	&	33.83	$\pm$	0.14	&	4.65	$\pm$	0.13	&	0.91	\\
	&		&		 		&		&	$^{15}$NH$_{3}$(1,1)	&	59.7	&	0.54	$\pm$	0.14	&	34.7	$\pm$	0.2	&	1.9	$\pm$	0.8	&	0.33	\\
	&		&		 		&		&	$^{15}$NH$_{3}$(2,2)	&	52.7	&	…			&	…			&	…			&	…	\\
	&		&		 		&		&	$^{15}$NH$_{3}$(3,3)	&	52.7	&	…			&	…			&	…			&	…	\\
	NGC\,1333	&	E	&	03:29:11.6	 	31:13:26.0	&	176	&	$^{14}$NH$_{3}$(1,1)	&	10.5	&	11.0	$\pm$	0.2	&	7.49	$\pm$	0.04	&	1.57	$\pm$	0.01	&	2.76	\\
	&		&		 		&		&	$^{14}$NH$_{3}$(2,2)	&	11.9	&	1.37	$\pm$	0.02	&	6.94	$\pm$	0.01	&	1.69	$\pm$	0.03	&	0.57	\\
	&		&		 		&		&	$^{14}$NH$_{3}$(3,3)	&	12.0	&	0.67	$\pm$	0.06	&	6.69	$\pm$	0.18	&	3.8	$\pm$	0.6	&	0.10	\\
	&		&		 		&		&	$^{15}$NH$_{3}$(1,1)	&	11.5	&	0.09	$\pm$	0.03	&	7.0	$\pm$	0.4	&	2.4	$\pm$	0.8	&	0.03	\\
	&		&		 		&		&	$^{15}$NH$_{3}$(2,2)	&	10.3	&	…			&	…			&	…			&	…	\\
	&		&		 		&		&	$^{15}$NH$_{3}$(3,3)	&	11.8	&	…			&	…			&	…			&	…	\\
	SGR A	&	E	&	17:47:52.7	 	-28:59:59.9	&	60	&	$^{14}$NH$_{3}$(1,1)	&	30.9	&	8.71	$\pm$	0.09	&	17.7	$\pm$	0.1	&	2.54	$\pm$	0.08	&	1.39	\\
	&		&		 		&		&	$^{14}$NH$_{3}$(2,2)	&	30.0	&	1.21	$\pm$	0.11	&	17.34	$\pm$	0.13	&	2.4	$\pm$	0.3	&	0.41	\\
	&		&		 		&		&	$^{14}$NH$_{3}$(3,3)	&	24.5	&	0.66	$\pm$	0.09	&	17.8	$\pm$	0.3	&	3.2	$\pm$	0.5	&	0.20	\\
	&		&		 		&		&	$^{15}$NH$_{3}$(1,1)	&	41.6	&	0.37	$\pm$	0.14	&	17.9	$\pm$	0.2	&	2.5	$\pm$	0.5	&	0.15	\\
	&		&		 		&		&	$^{15}$NH$_{3}$(2,2)	&	28.7	&	…			&	…			&	…			&	…	\\
	&		&		 		&		&	$^{15}$NH$_{3}$(3,3)	&	48.9	&	…			&	…			&	…			&	…	\\
	Barnard-1b	&	E	&	03:33:20.8	 	+31:07:34.0	&	184	&	$^{14}$NH$_{3}$(1,1)	&	17.6	&	18.4	$\pm$	0.1	&	6.68	$\pm$	0.06	&	1.23	$\pm$	0.06	&	4.78	\\
	&		&		 		&		&	$^{14}$NH$_{3}$(2,2)	&	16.8	&	1.46	$\pm$	0.01	&	6.16	$\pm$	0.01	&	1.04	$\pm$	0.01	&	1.00	\\
	&		&		 		&		&	$^{14}$NH$_{3}$(3,3)	&	10.0	&	0.11	$\pm$	0.02	&	6.1	$\pm$	0.3	&	3.1	$\pm$	0.7	&	0.03	\\
	&		&		 		&		&	$^{15}$NH$_{3}$(1,1)	&	11.1	&	0.25	$\pm$	0.06	&	6.15	$\pm$	0.10	&	1.2	$\pm$	0.2	&	0.20	\\
	&		&		 		&		&	$^{15}$NH$_{3}$(2,2)	&	9.58	&	…			&	…			&	…			&	…	\\
	&		&		 		&		&	$^{15}$NH$_{3}$(3,3)	&	11.3	&	…			&	…			&	…			&	…	\\
	\enddata
	\tablecomments{Column(1): source name; Column(2): T: TMRT - 65 m; E: Effelsberg-100 m; Column(3): Right ascension (J2000) and Declination (J2000); Column(4): total integration time; Column(5): molecular line; Column(6): the Root-Mean-Square ({\it rms}) noise value for channel widths of 0.78 km s$^{-1}$ (TMRT) or 0.70 km s$^{-1}$ (Effelsberg); Column(7): the integrated line intensity covering all groups of hyperfine components from Gaussian fitting. For three sources with blended components in the $^{14}$NH$_{3}$ spectra, their spectrum can not be fitted well by Gaussian fitting. Thus we determined their integrated intensities by summing line intensities over the entire velocity interval needed to cover the main and the satellite features (Orion-KL: -45 to 51 km s$^{-1}$; W51D: 13 to 105 km s$^{-1}$ and G10.47: 20 to 113 km s$^{-1}$) using the first moment by the "Print" command in CLASS; Column(8): LSR velocity; Column(9): line width (FWHM); Column(10): Main beam brightness peak temperature. }
\end{deluxetable}

\begin{figure*}
	\centering
	{\includegraphics[width=4.5cm]{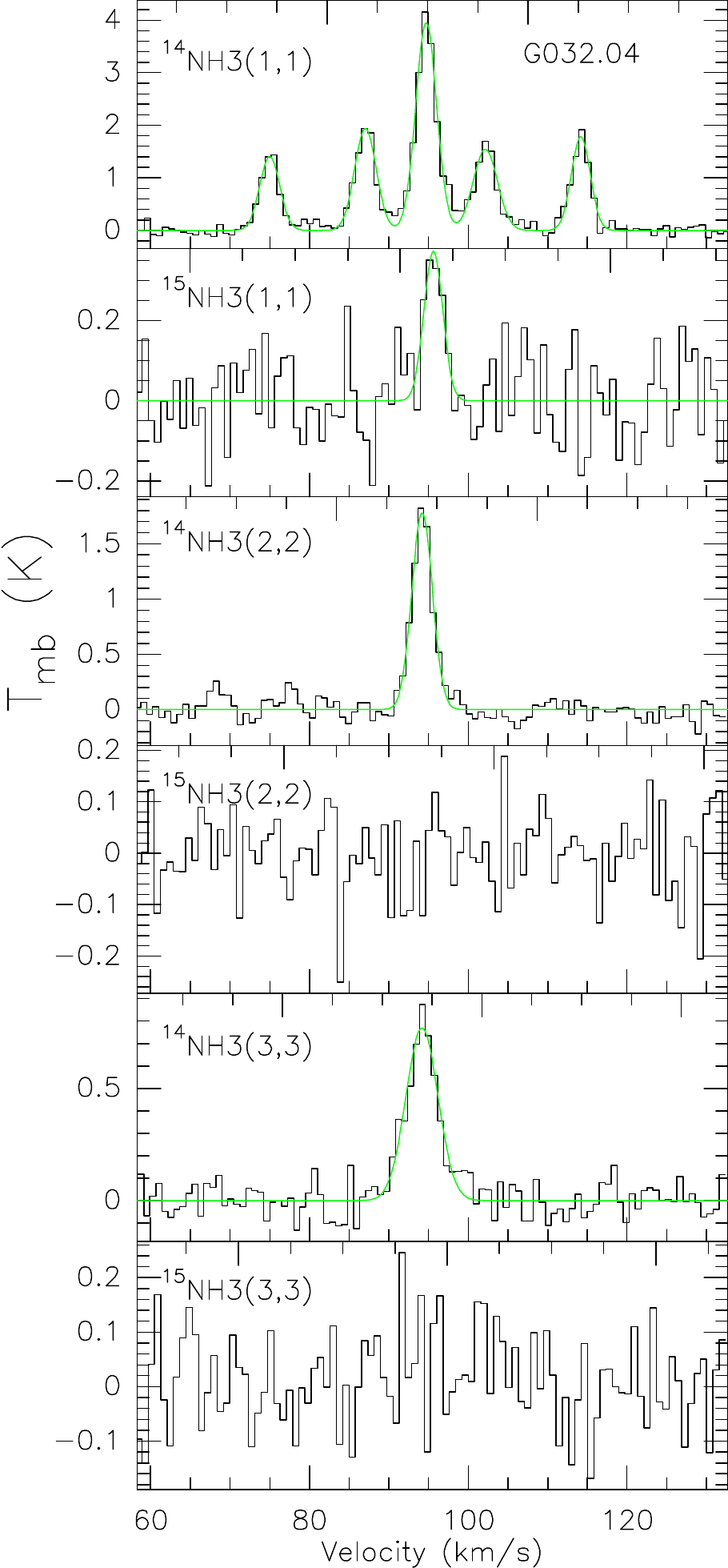}}
	{\includegraphics[width=4.3cm]{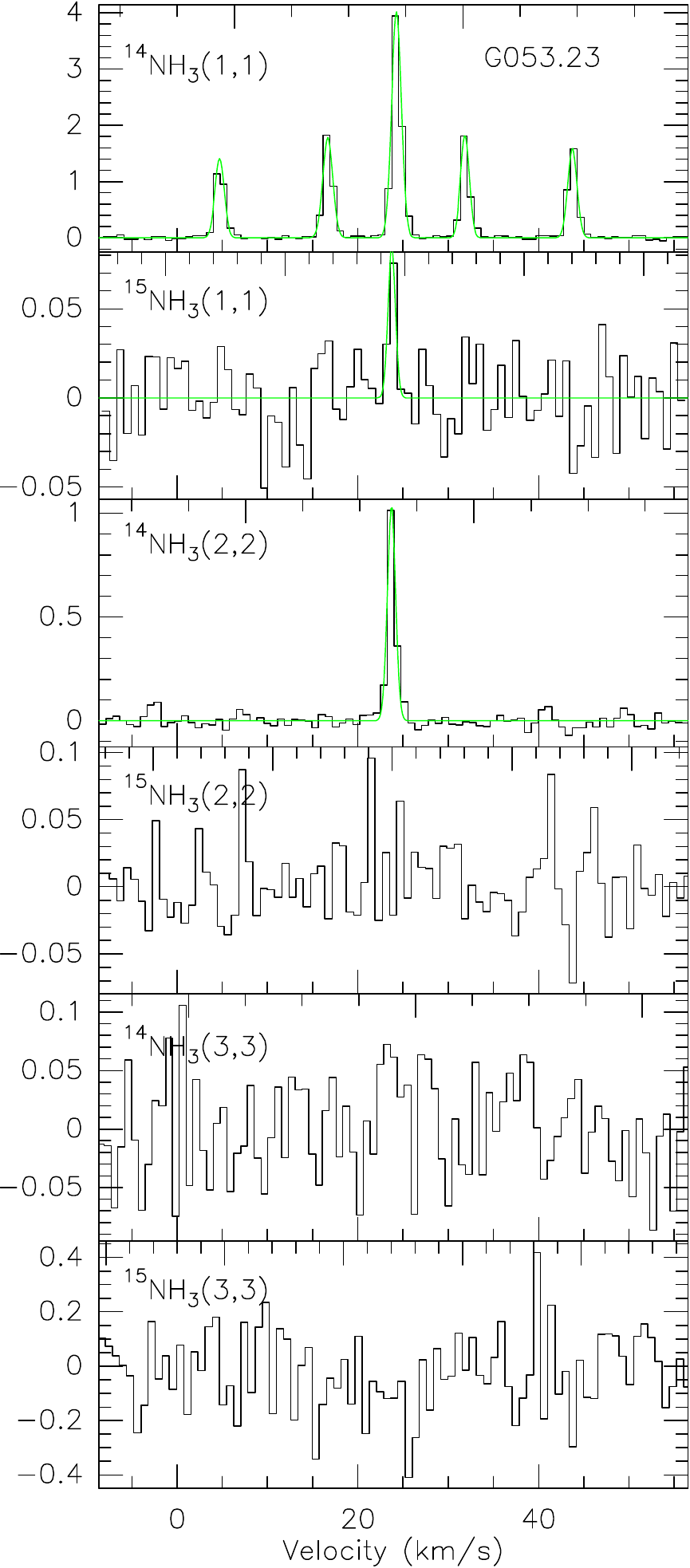}}
	{\includegraphics[width=4.3cm]{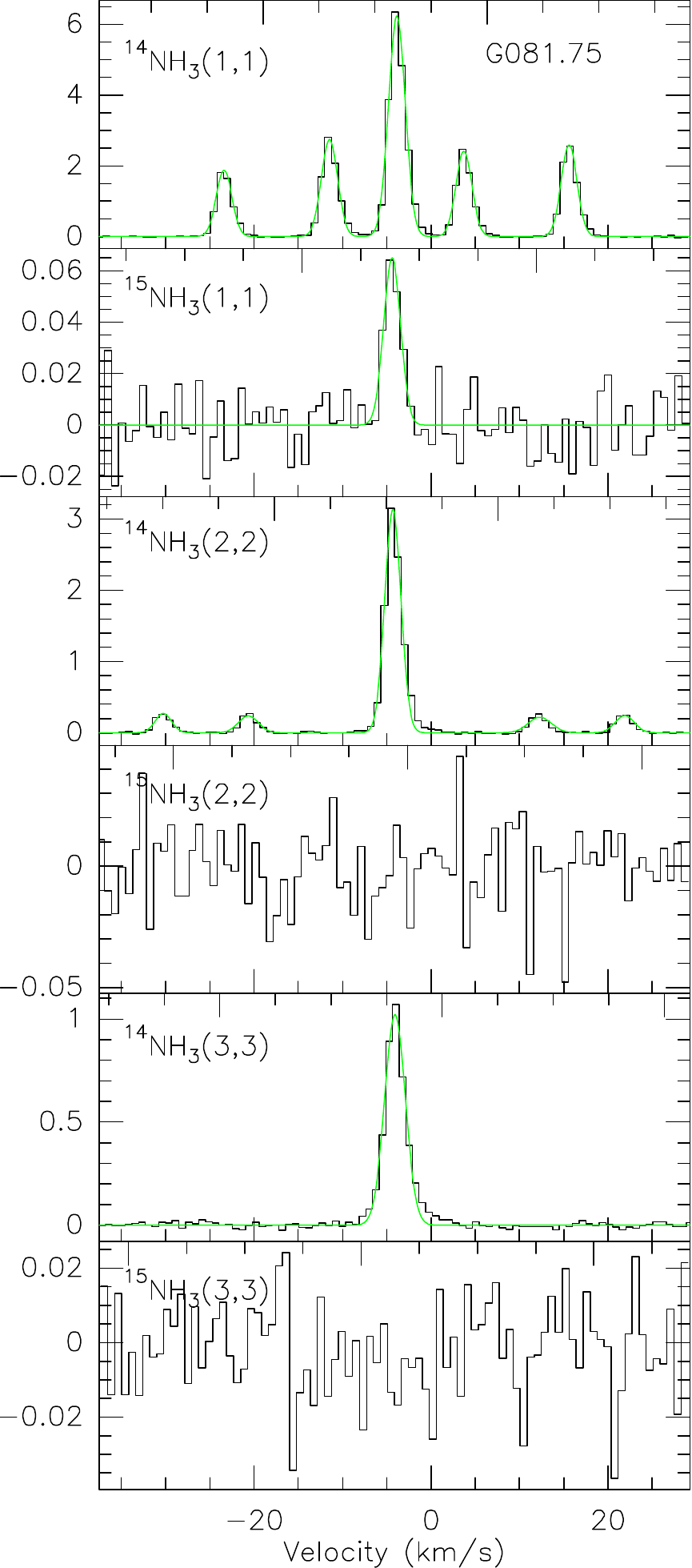}}
	{\includegraphics[width=4.3cm]{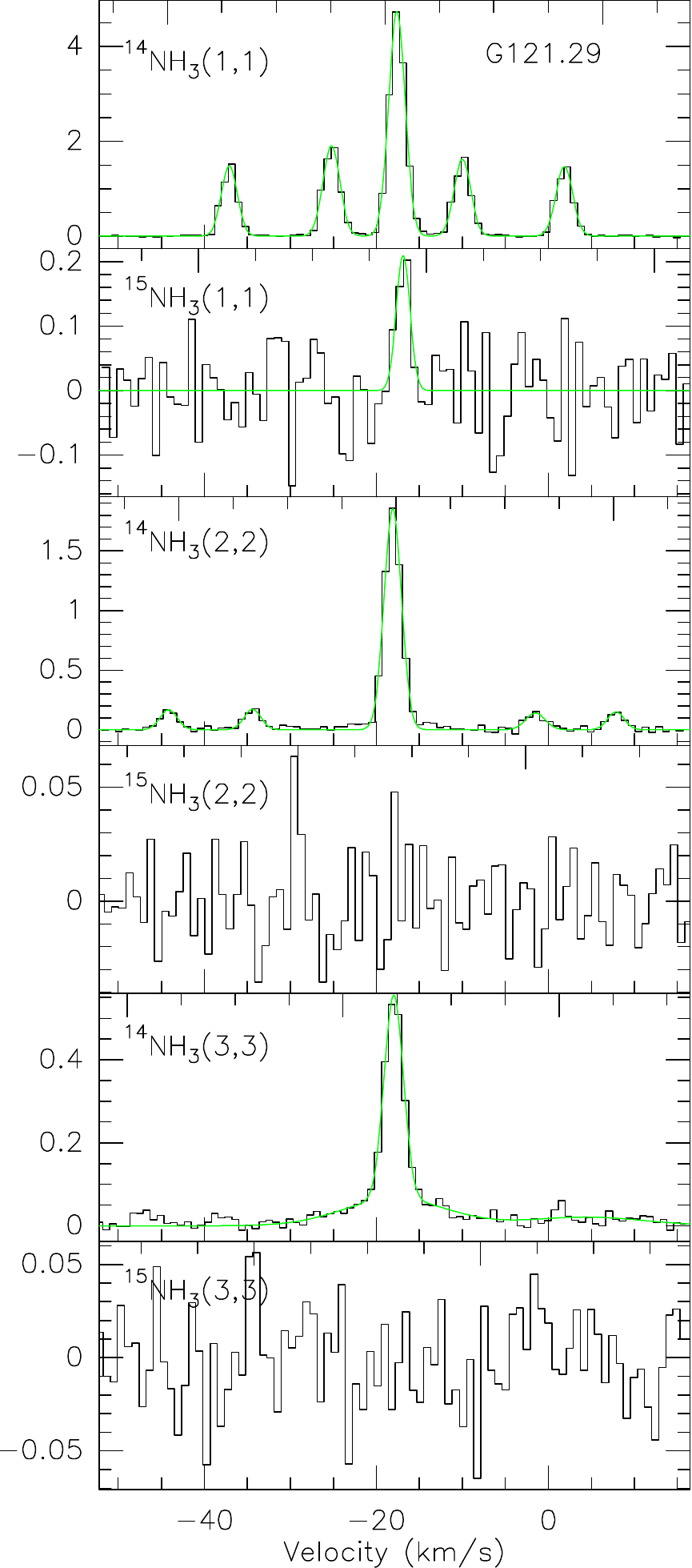}}
	{\includegraphics[width=4.55cm]{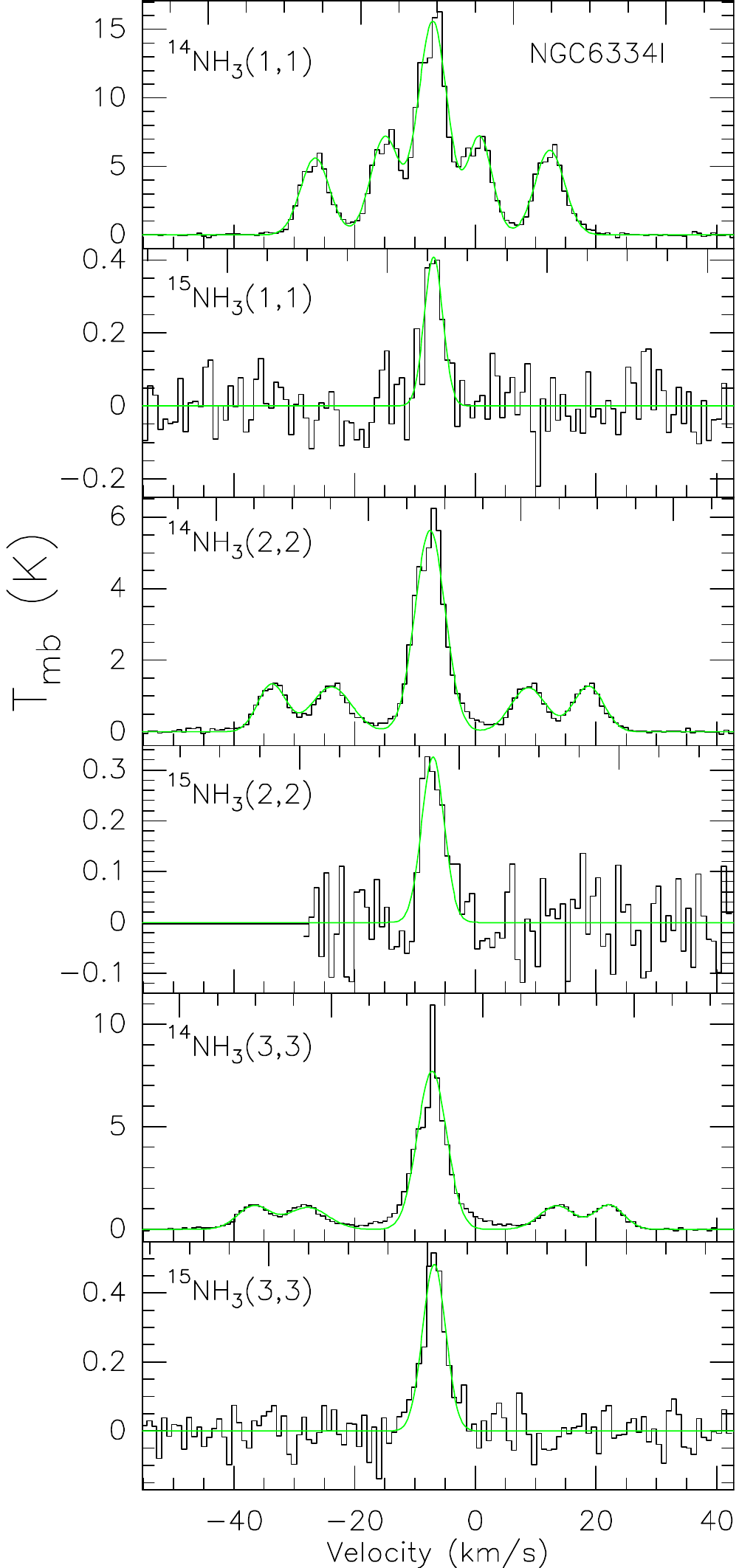}}
	{\includegraphics[width=4.35cm]{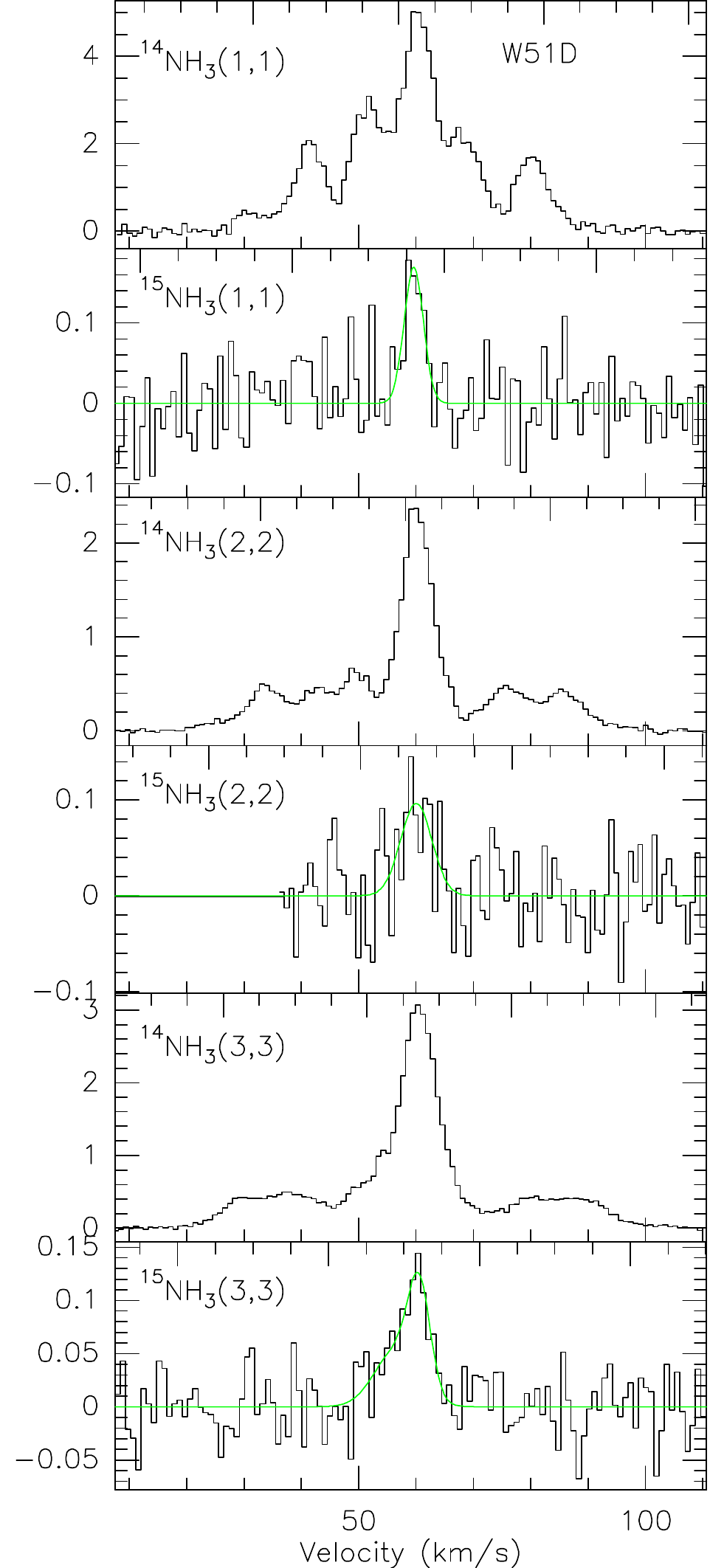}}
	{\includegraphics[width=4.35cm]{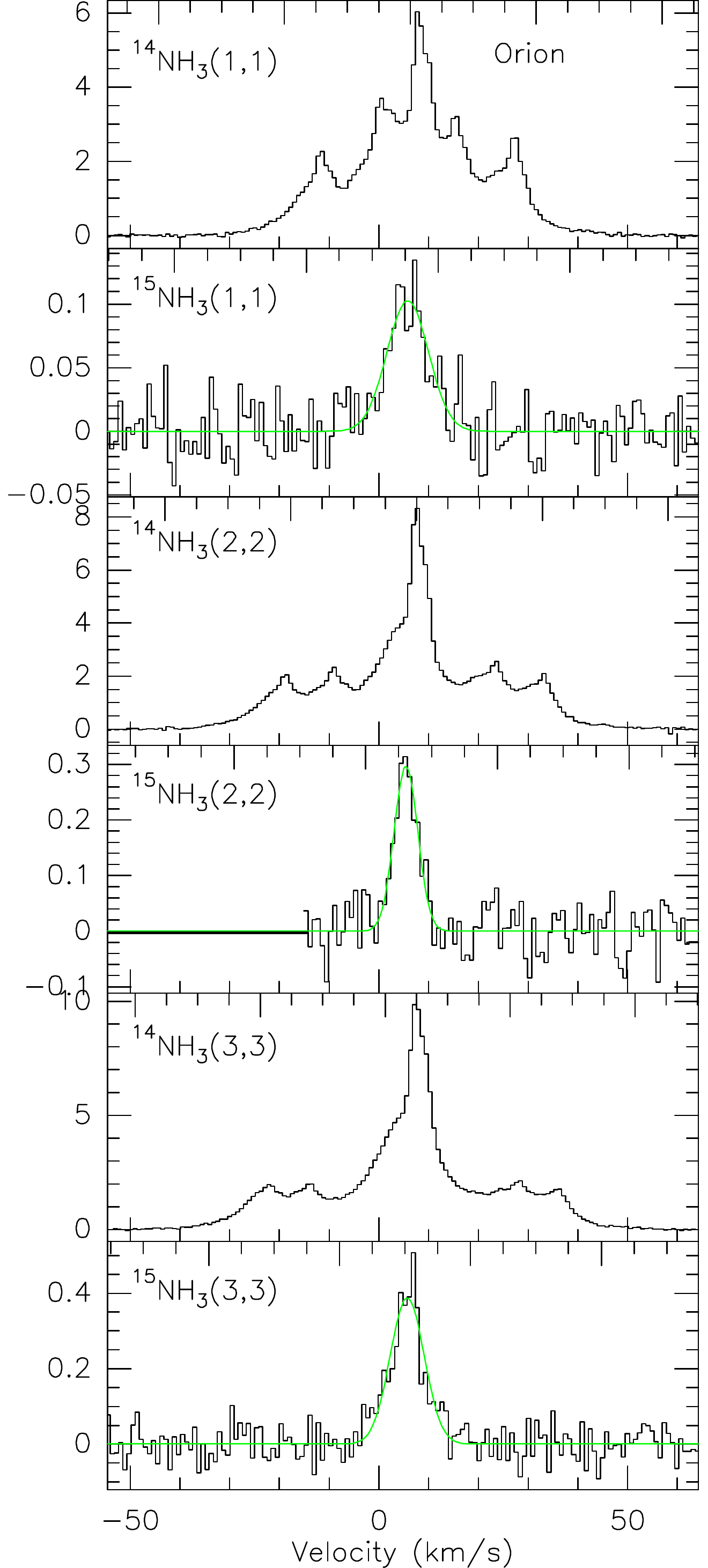}}
	{\includegraphics[width=4.15cm]{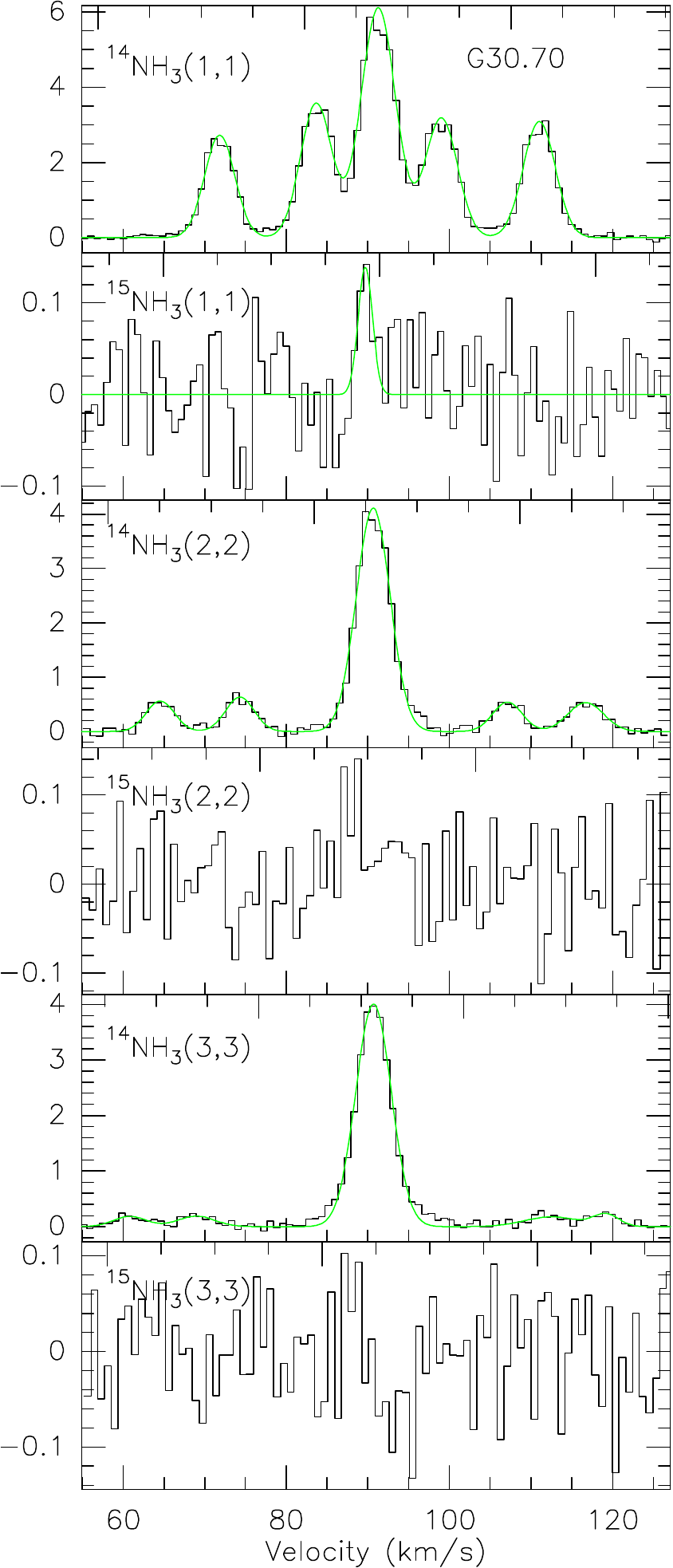}}
	\caption{TMRT spectra of those 8 sources with detected $^{15}$NH$_{3}$ lines, after subtracting baselines and applying Hanning smoothing leading to 0.78 km s$^{-1}$ wide channels. Green lines show Gaussian fits.}
	\label{15NH3fig_TMRT}
\end{figure*}


\clearpage
\begin{figure*}
	\centering
	{\includegraphics[width=3.75cm]{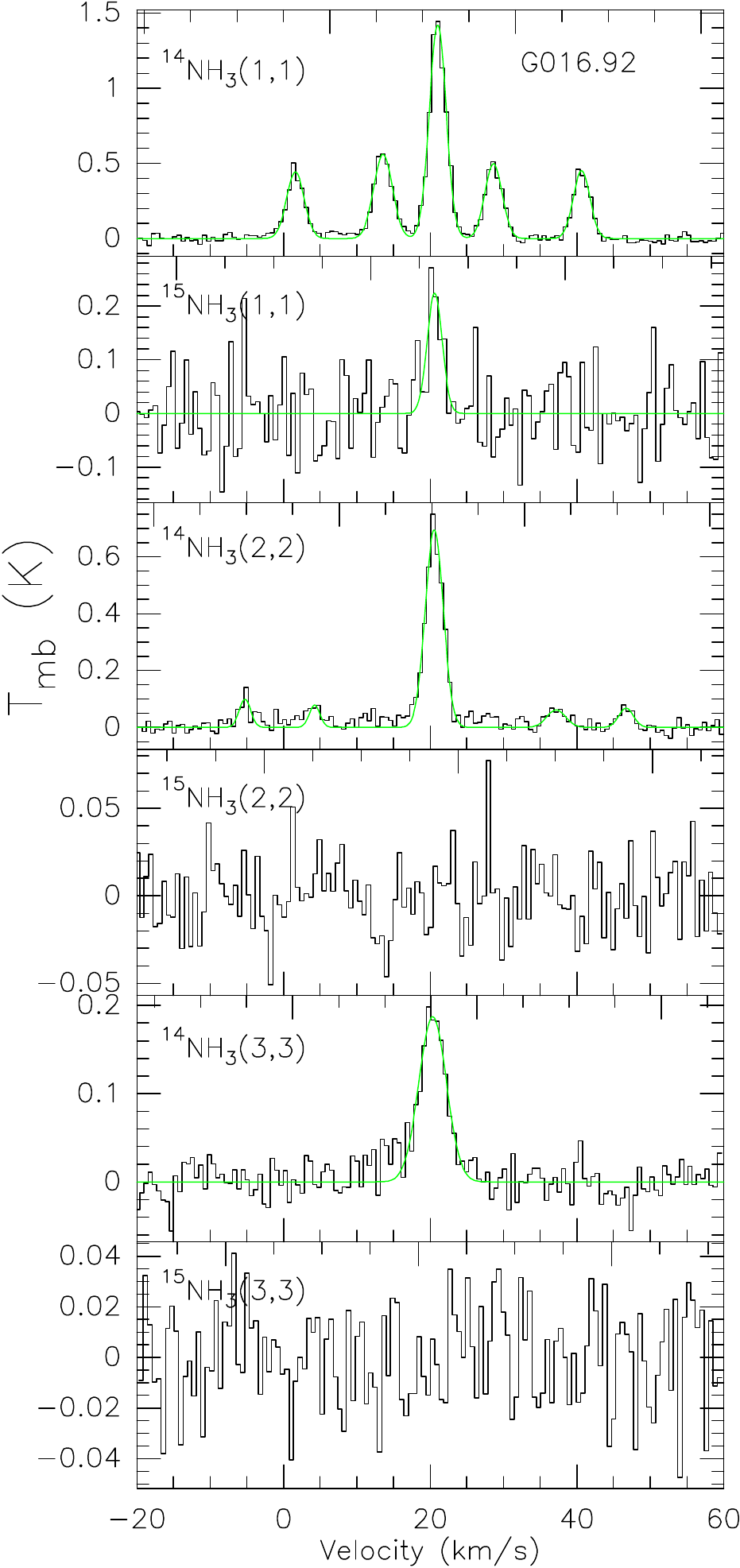}}
	{\includegraphics[width=3.45cm]{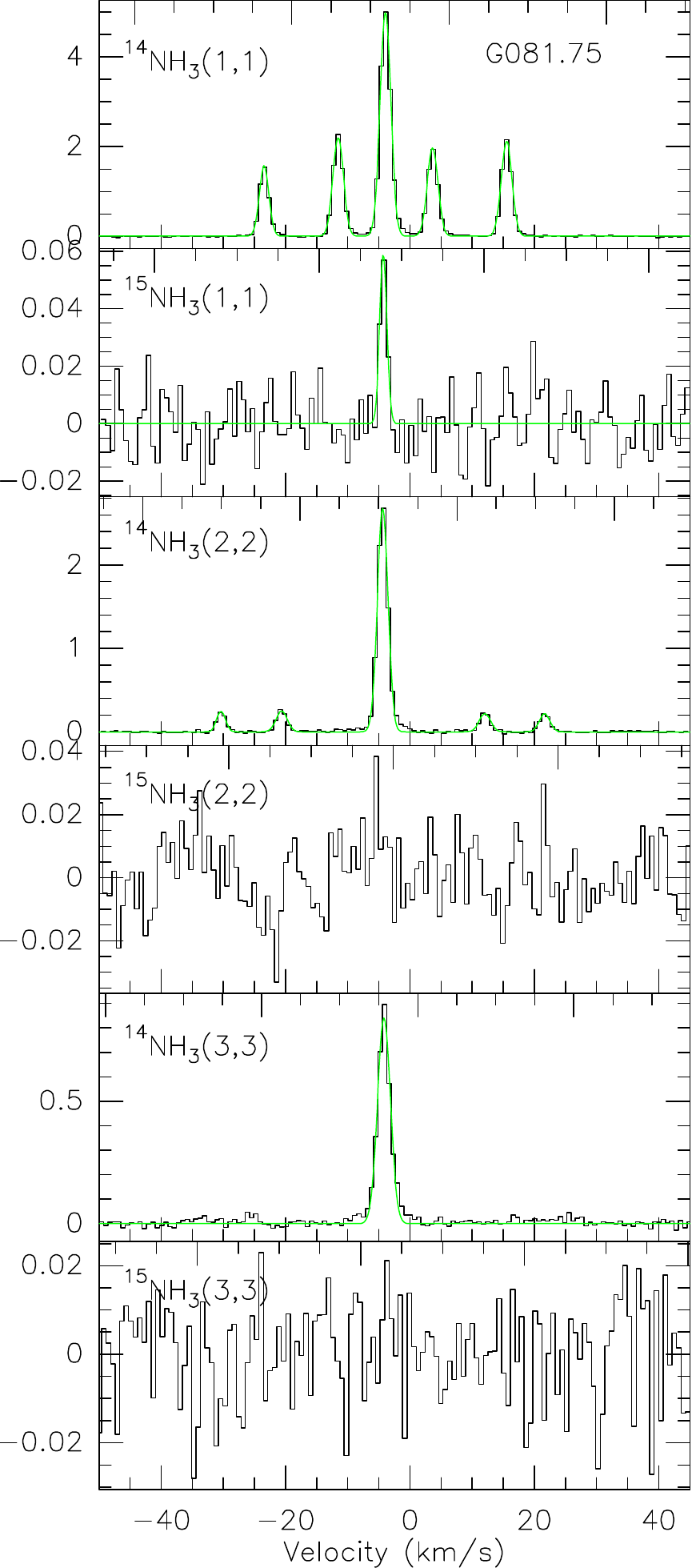}}
	{\includegraphics[width=3.4cm]{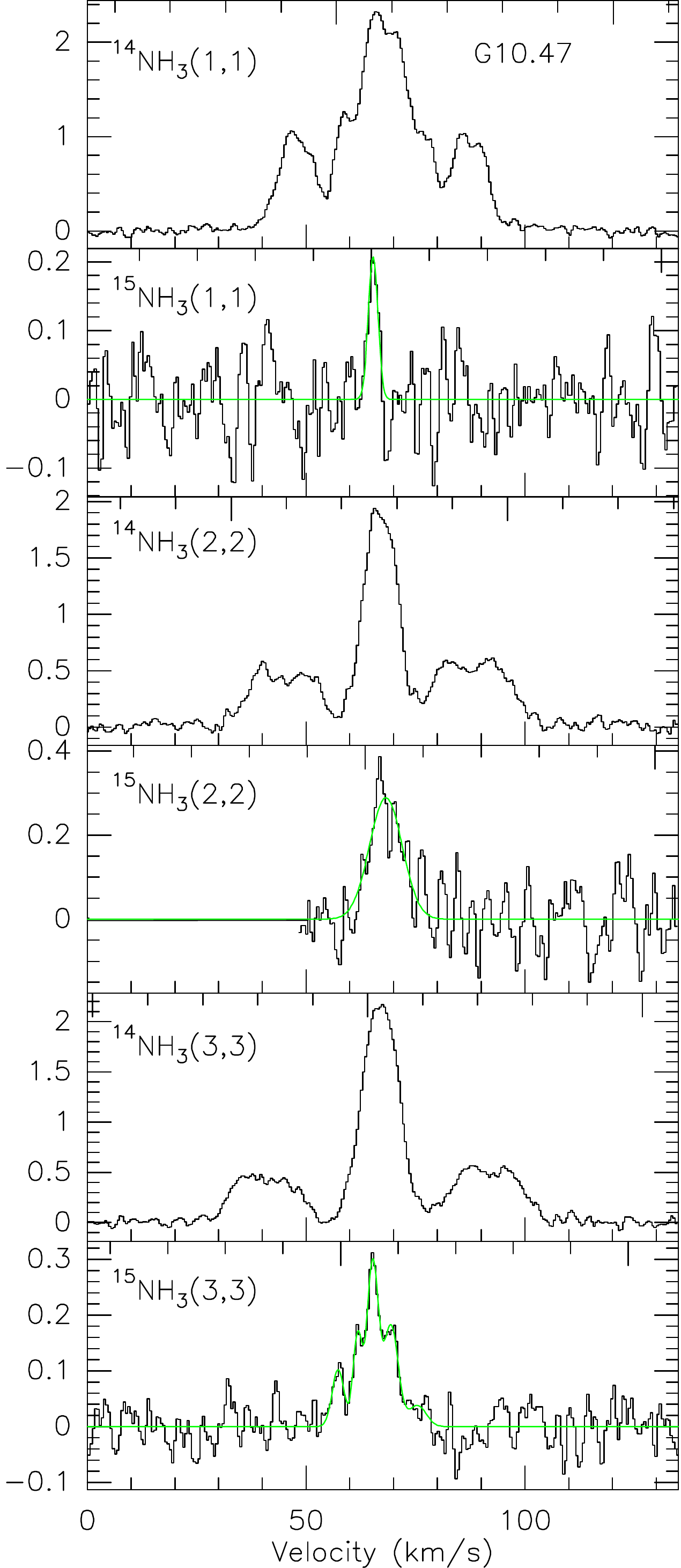}}
	{\includegraphics[width=3.5cm]{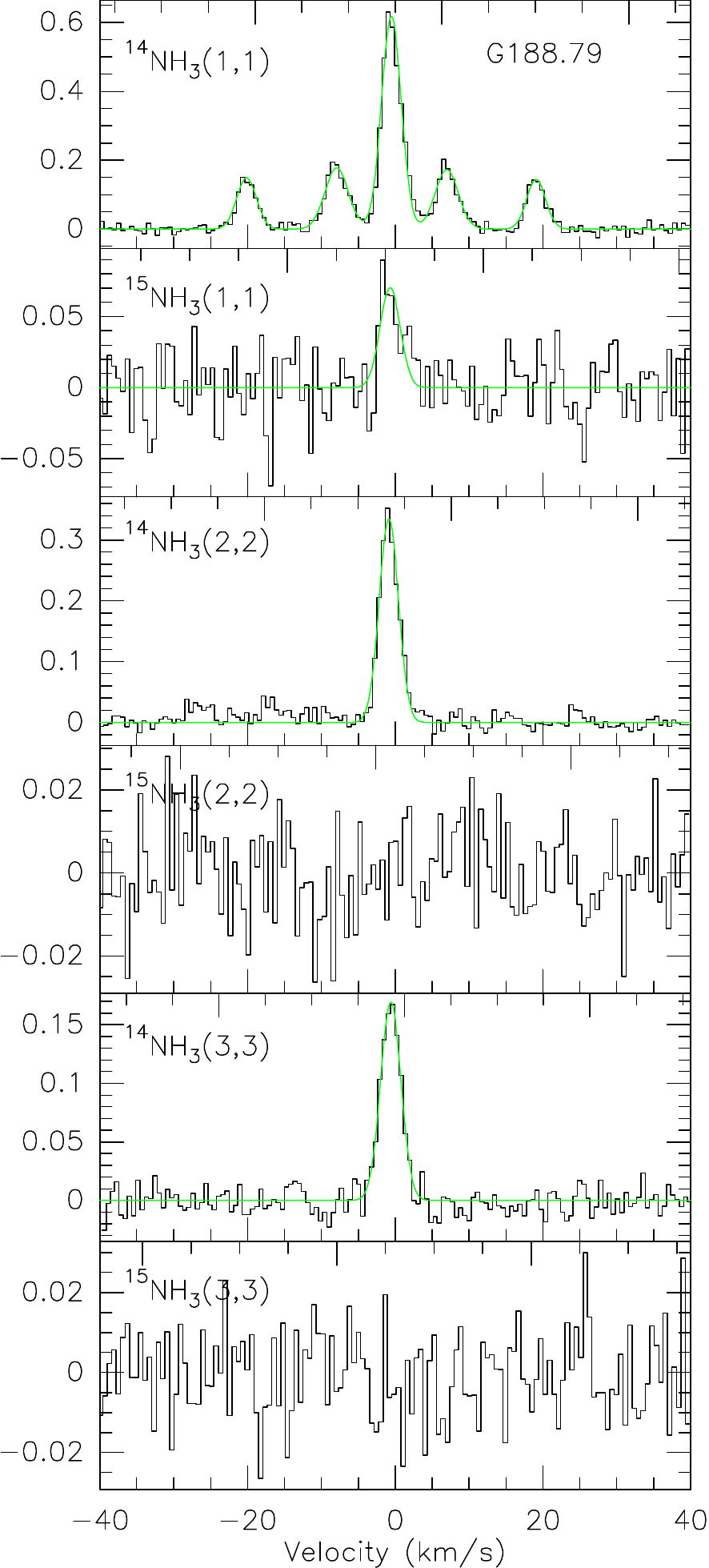}}
	{\includegraphics[width=3.4cm]{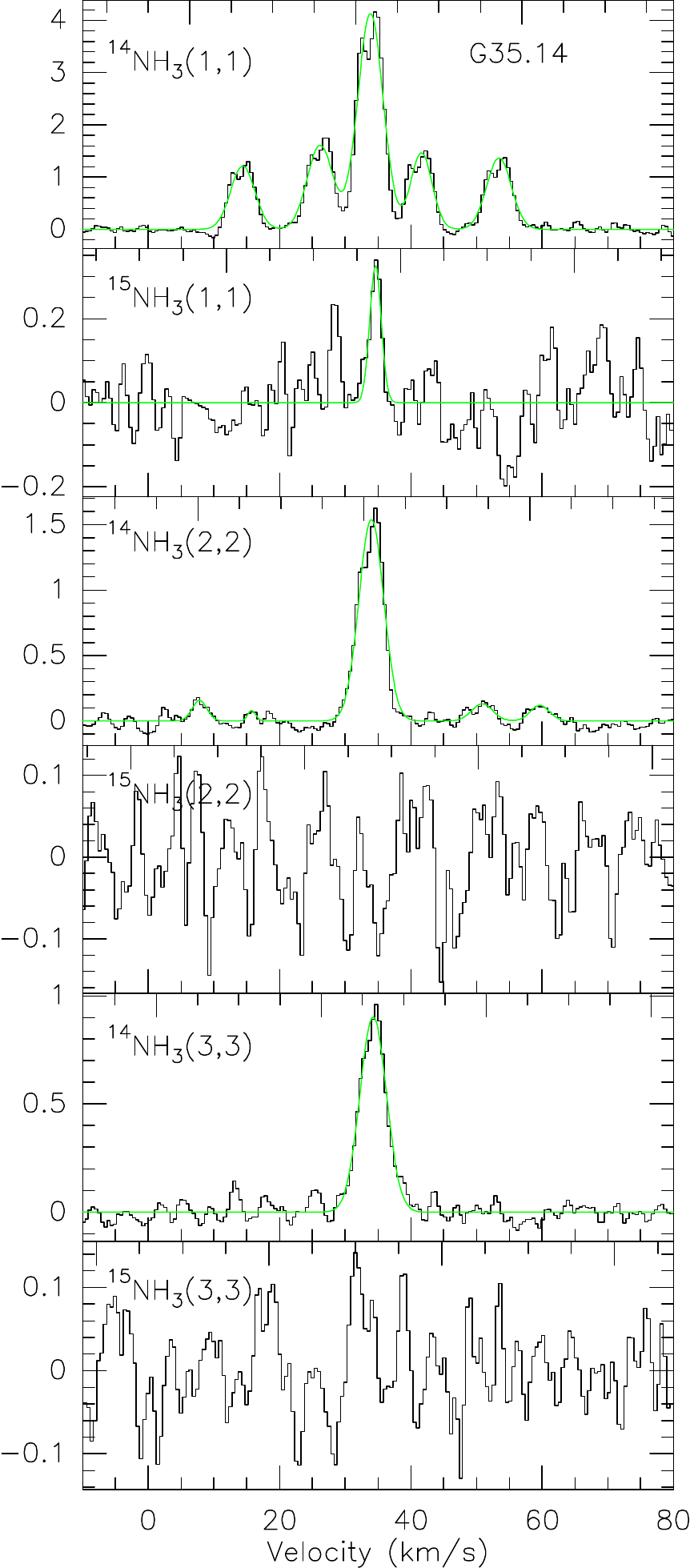}}
	{\includegraphics[width=3.55cm]{ORI_3-eps-converted-to.pdf}}
	{\includegraphics[width=3.4cm]{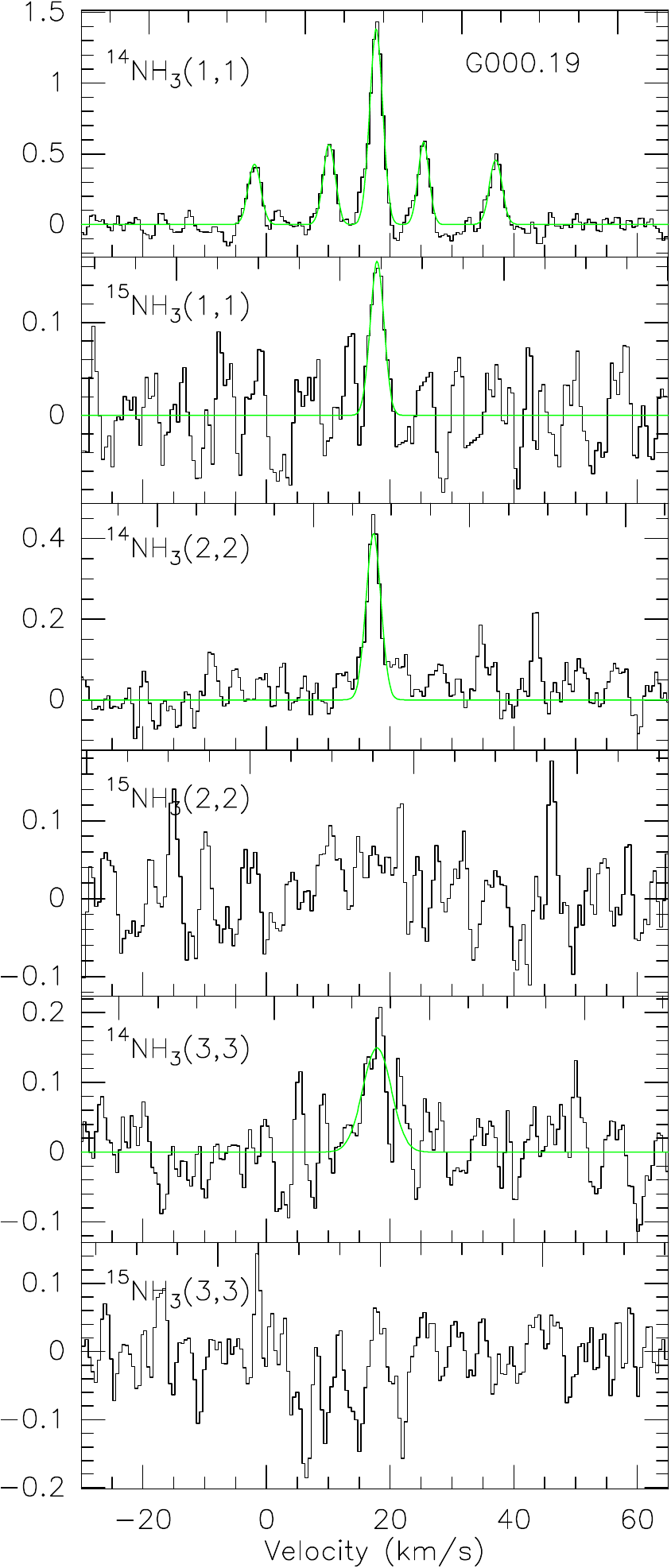}}
	{\includegraphics[width=3.55cm]{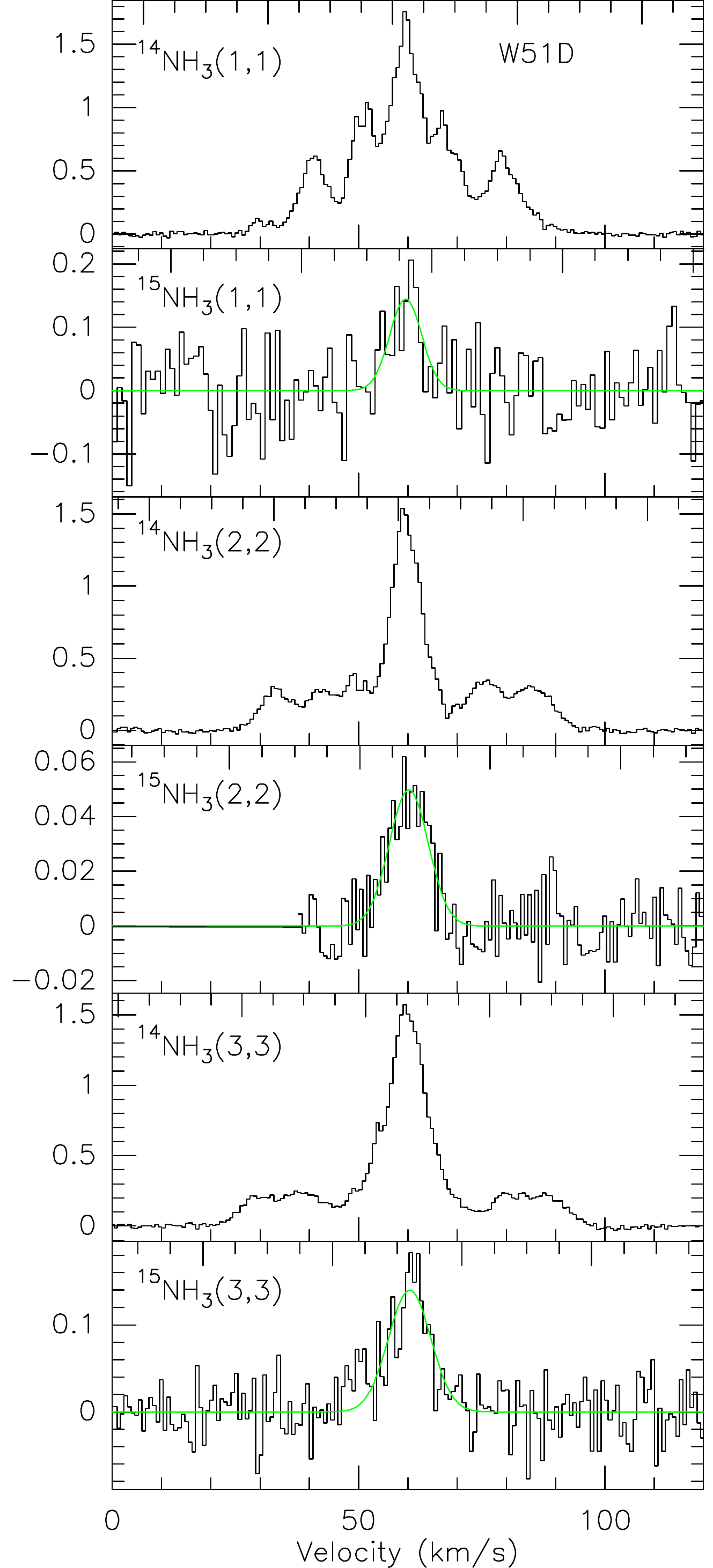}}
	{\includegraphics[width=3.45cm]{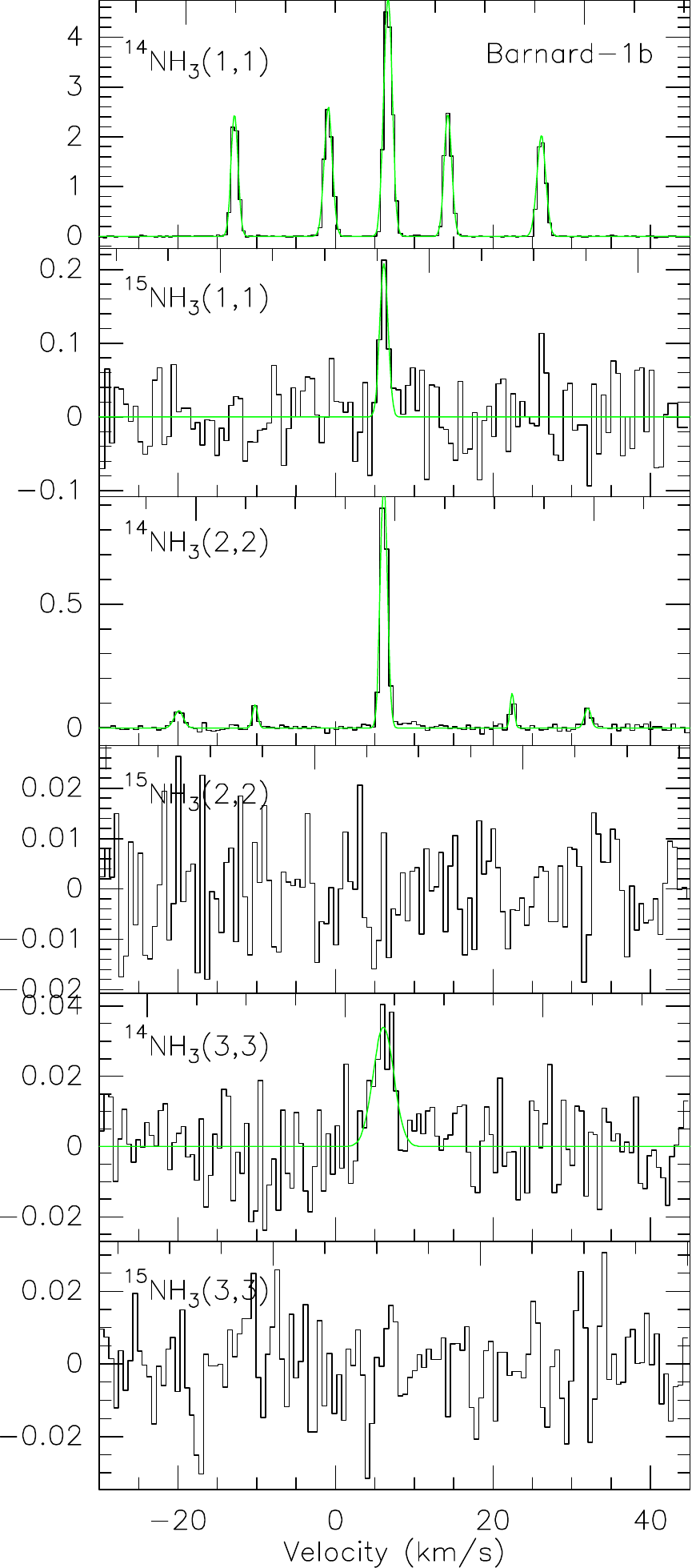}}
	{\includegraphics[width=3.45cm]{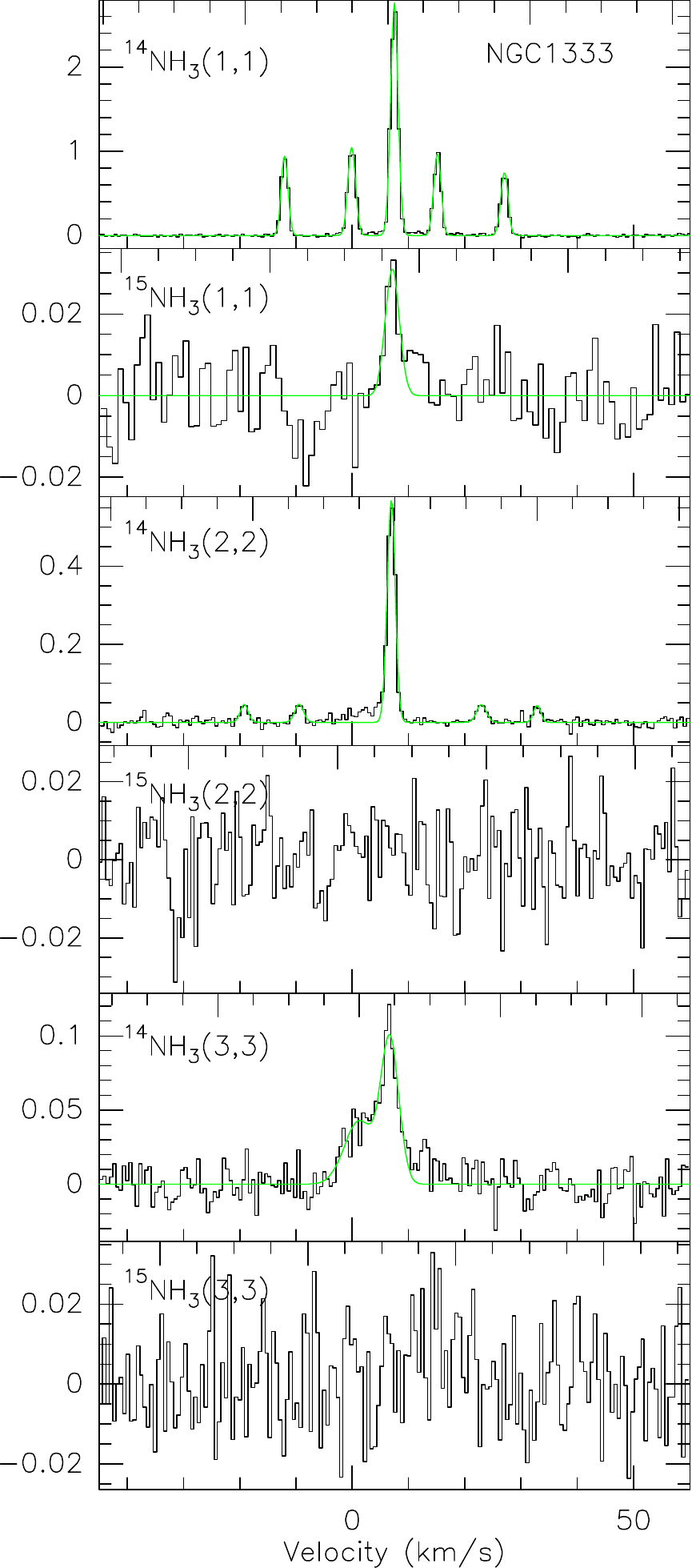}}
	\caption{Effelsberg spectra of those 10 sources with detected $^{15}$NH$_{3}$ lines, after subtracting baselines and applying Hanning smoothing leading to 0.7 km s$^{-1}$ wide channels. Green lines show Gaussian fits.}
	\label{15NH3fig_Eff}
\end{figure*}

\section{Results}\label{sec:result}

\subsection{Detections and non-detections}\label{sec:detection}

Toward the TMRT 65 m 210 targets, 141 sources were detected in at least one of the $^{14}$NH$_{3}$ lines. Among them, 8 sources were also successfully detected in the $^{15}$NH$_{3}$(1, 1) line. The $^{15}$NH$_{3}$(2, 2) and (3, 3) lines are also detected in three sources (NGC\,6334\,I, W51D and Orion-KL, see Fig. \ref{15NH3fig_TMRT}). 

Among the 36 targets with strong $^{14}$NH$_{3}$ emission (Tianma flux density \textgreater 1.5 Jy), the Effelsberg-100 m telescope detected successfully the $^{15}$NH$_{3}$(1, 1) lines toward 10 sources. The $^{15}$NH$_{3}$(2, 2) and (3, 3) lines are also detected in three of them (W51D Orion-KL and G10.47, see Fig. \ref{15NH3fig_Eff}).

Combinations of both TMRT and Effelsberg observations lead to 15 sources with detections of $^{15}$NH$_{3}$ lines, including 3 sources (G081.7522, W51D and Orion-KL) with detections by both telescopes. 
The spectral line parameters of these 15 sources are listed in Table \ref{tab:15NH3para_TMRT-Eff}.
For G30.70 with the FWHM of $^{15}$NH$_{3}$(1, 1) being much smaller than that of $^{14}$NH$_{3}$(1, 1), the integrated line intensity of $^{15}$NH$_{3}$(1, 1) was taken assuming the ratio of integrated line intensities equals the ratio of the peak values of the main beam brightness temperature.
For those sources without detection of $^{15}$NH$_{3}$, the upper limit of the line was estimated from the {\it rms} value, which will be used for later analysis (see Sect. \ref{sec:ratio}). 

\subsection{Line and physical parameters of sources with detections of $^{14}$NH$_{3}$ and $^{15}$NH$_{3}$}
For those 15 sources with both $^{14}$NH$_{3}$ and $^{15}$NH$_{3}$ line detections, we determine the line and physical parameters here, including the optical depth, the total column density, and the rotation and kinetic temperatures.

\subsubsection{Optical depth}\label{sec:optical}

In view of the expected large nitrogen isotope ratios, there is reason to expect that in clouds with detected $^{15}$NH$_{3}$ emission, the $^{14}$NH$_{3}$ lines may be optically thick, leading to a non-linear correlation between integrated intensity and molecular column density. We have therefore determined the optical depth of the $^{14}$NH$_{3}$ inversion lines using two methods, the so-called intensity ratio method and the hyperfine (HF) structure line fits. Spectroscpic information on the $^{14}$NH$_{3}$(1, 1) hyperfine components is listed in Table \ref{tab:14HF}.
\startlongtable

\begin{deluxetable}{lcccccc}
	
	\tablecaption{Spectroscopic information related to the $^{14}$NH$_{3}$(1, 1) hyperfine components.\label{tab:14HF}}
	\tablehead{\colhead{Hyperfine}	&	\colhead{HFC}	&	\colhead{$F^{'} \rightarrow F$}	&	\colhead{$F_{1}^{'} \rightarrow F_{1}$}	&	\colhead{Frequency$^{b}$}	&	\colhead{Relative }	&	\colhead{Velocity}	\\
		\colhead{ Group $^{a}$}	&	\colhead{number}	&	\colhead{}	&	\colhead{}	&	\colhead{(kHz)}	&	\colhead{Intensities $^{c}$}	&	\colhead{(km s$^{-1}$)}	\\	
	}
	\startdata
	osg.1	&	1	&	1/2, 1/2	&	(0,1)	&	-1568.49	&	1/27	&	-19.84	\\
	&	2	&	1/2, 3/2	&	(0,1)	&	-1526.96	&	2/37	&	-19.32	\\
	\hline													
	isg.1	&	3	&	3/2, 1/2	&	(2,1)	&	-623.31	&	5/108	&	-7.89	\\
	&	4	&	5/2, 3/2	&	(2,1)	&	-590.92	&	1/12	&	-7.47	\\
	&	5	&	3/2, 3/2	&	(2,1)	&	-580.92	&	1/108	&	-7.35	\\
	\hline													
	mg	&	6	&	1/2, 1/2	&	(1,1)	&	-36.54	&	1/54	&	-0.46	\\
	&	7	&	3/2, 1/2	&	(1,1)	&	-25.54	&	1/108	&	-0.32	\\
	&	8	&	5/2, 3/2	&	(2,2)	&	-24.39	&	1/60	&	-0.31	\\
	&	9	&	3/2, 3/2	&	(2,2)	&	-14.98	&	3/20	&	-0.19	\\
	&	10	&	1/2, 3/2	&	(1,1)	&	5.85	&	1/108	&	0.07	\\
	&	11	&	5/2, 5/2	&	(2,2)	&	10.52	&	7/30	&	0.13	\\
	&	12	&	3/2, 3/2	&	(1,1)	&	16.85	&	5/108	&	0.21	\\
	&	13	&	3/2, 5/2	&	(2,2)	&	19.93	&	1/60	&	0.25	\\
	\hline													
	isg.2	&	14	&	1/2, 3/2	&	(1,2)	&	571.79	&	5/108	&	7.23	\\
	&	15	&	3/2, 3/2	&	(1,2)	&	582.79	&	1/108	&	7.37	\\
	&	16	&	3/2, 5/2	&	(1,2)	&	617.70	&	1/12	&	7.81	\\
	\hline													
	osg.2	&	17	&	1/2, 1/2	&	(1,0)	&	1534.05	&	1/27	&	19.41	\\
	&	18	&	3/2, 1/2	&	(1,0)	&	1545.05	&	2/27	&	19.55	\\
	\enddata
	\tablecomments{$^{(a)}$ mg = main group of hyperfine components, isg = group of inner satellite hyperfine components, osg = group of outer satellite hyperfine components; $^{(b)}$ The frequencies in column (5) are given relative to 23694.5 MHz. $^{(c)}$ The hyperfine intensities are taken from \citet{Mangum15} and \citet{Wang20}. The sum of these intensities is 1.0. }
\end{deluxetable}

\textbf{Intensity ratio method}: Assuming the same excitation temperature and beam filling factor for all transitions under conditions of local thermodynamic equilbrium (LTE), the optical depth of the $^{14}$NH$_{3}$(1, 1) line can be estimated from the measured intensity ratio between its main and satellite components \citep[e. g.,][]{Ho83, Mangum92}, respectively: 
\begin{equation}                                             
\frac{T_{\rm mb}(1, 1, m)}{T_{\rm mb}(1, 1, s)} = \frac{1 - e^{-\tau(1, 1, m)}}{1 - e^{-a\tau(1, 1, m)}}.
\end{equation} 
where $T_{\rm mb}$ is the peak value of the main beam brightness temperature; $m$ and $s$ refer to the main and satellite group components, respectively; $\tau$($J, K, m$) is the optical depth of the main group of hyperfine components, $a$ represents the expected intensity ratios of the satellite to the main group of hyperfine components, 0.278 and 0.222 for the inner and outer groups of HF features under conditions of LTE and optically thin line emission \citep[see][]{Ho83,Mangum15,Zhou20}.
For our sample, we used formula (2) to calculate the $^{14}$NH$_{3}$(1, 1) optical depth, according to the $T_{\rm mb}$ peak values of the main component and inner and outer satellite components. The mean value of the four inner and outer calculated NH$_{3}$(1, 1) optical depths was taken for each source (Table \ref{tab:tau and T}). 

For the $^{14}$NH$_{3}$(2, 2) and (3, 3) transitions, there are only two sources with clearly separated satellite components (NGC\,6334\,I and G30.70). Here we also used the measured intensity ratio of the main and satellite components to determine their optical depth. For other sources, we could not use this method, since the hyperfine satellite components were too weak to be detected.
However, the optical depth can be estimated from the intensity ratio of the main group of (2, 2) hyperfine components to that of the (1, 1) transition, assuming equal excitation temperatures and beam filling factors \citep{Mangum92}:
\begin{equation}                                             
\frac{T_{\rm mb}(1, 1, m)}{T_{\rm mb}(2, 2, m)} = \frac{1 - e^{-\tau(1, 1, m)}}{1 - e^{-\tau(2, 2, m)}}
\end{equation} 

\textbf{HF fit method}: The optical depth of the main hyperfine component can also be determined by the HF fit method in CLASS ("method" command). Here, the excitation temperature and Gaussian opacity profiles are assumed to be the same for all HF components. $\tau_{tot}$ (the opacity summed over all the hyperfine components) can also be retrieved from CLASS. The relation between $\tau_{tot}$ and $\tau$($J, K, m$) can be found in Eq. A8 by \citet{Mangum92}.
Through adjusting parameters to fit the observed spectra, we derived the optical depths of the $^{14}$NH$_{3}$(1, 1), (2, 2) and (3, 3) transitions for 12 sources with $^{15}$NH$_{3}$ detections (see Table \ref{tab:tau and T}). 
For the other 3 sources among the sample (G10.47, Orion-KL, and W51D), the (1, 1), (2, 2) and (3, 3) optical depths could not be determined by the "method" fits in CLASS, due to overlap of the satellite HF components with their main group of HF components (see Table \ref{tab:tau and T}). 

Comparisons of the opacities derived from the two different methods, the intensity ratio and HF fit methods, reveal consistency with those from the HF fits in CLASS. Therefore we chose to take the optical depth from the intensity ratio method, where results could be obtained from all sources with detected $^{15}$NH$_3$ emission for our analysis.

\clearpage

\startlongtable

 \setlength{\tabcolsep}{1mm}{
\begin{deluxetable}{lccccccccccc}
	\tablecaption{Observational parameters of NH$_{3}$ measured with the TMRT and the Efelsberg-100 m telescope.\label{tab:tau and T}}
	
	\tablehead{\colhead{Object}	&\colhead{Telescope}		&	\multicolumn{3}{c}{Intensity ratio method}			&			\multicolumn{3}{c}{HF fitting}			&\colhead{$T_{rot}^{ir}$}		&\colhead{$T_{rot}^{rd}$}		&\colhead{$T_{rot}^{hf}$}		&\colhead{$T_{k}$}		\\
		&		&\colhead{$\tau$(1, 1)}		&\colhead{$\tau$(2, 2)}		&\colhead{$\tau$(3, 3)}		&\colhead{$\tau$(1,  1)}		&\colhead{$\tau$(2, 2)}		&\colhead{$\tau$(3, 3)}		&\colhead{K}		&\colhead{K}		&\colhead{K}		&\colhead{K}
	}
	\colnumbers
	\startdata
	G032.04	&	TMRT	&	1.73(0.06)	&	0.54(0.02)	&	0.21(0.01)	&	1.83(0.16)	&	0.2(0.4)	&	0.1(0.3)	&	22(10)	&	11	(6)	&	19(3)	&	21(4)	\\
	G053.23	&	TMRT	&	1.44(0.04)	&	0.21(0.01)	&	...	&	1.05(0.06)	&	0.11 (0.04)	&	…	&	16(8)	&	9	(3)	&	11(5)	&	12(4)	\\
	G081.75	&	TMRT	&	1.35(0.01)	&	0.54(0.01)	&	0.12(0.01)	&	1.56(0.15)	&	0.47(0.06)	&	0.13(0.04)	&	27(9)	&	13	(5)	&	19(2)	&	22(3)	\\
	&	Effelsberg	&	1.47(0.01)	&	0.53(0.01)	&	0.12(0.01)	&	1.65(0.12)	&	0.42(0.08)	&	0.15(0.04)	&	29(10)	&	13	(7)	&	19(3)	&	22(3)	\\	
	G121.29	&	TMRT	&	1.02(0.01)	&	0.33(0.01)	&	0.16(0.01)	&	1.1(30.01)	&	0.48(0.03)	&	0.12(0.12)	&	20(9)	&	13	(6)	&	17(3)	&	20(4)	\\
	G30.70	&	TMRT	&	2.23(0.03)	&	1.38(0.02)	&	0.43(0.01)	&	3.27(0.06)	&	2.23(0.11)	&	0.12(0.04)	&	26(8)	&	14	(5)	&	20(2)	&	25(2)	\\
	NGC\,6334\,I	&	TMRT	&	1.75(0.02)	&	0.42(0.01)	&	0.42(0.01)	&	2.06(0.03)	&	4.72(0.19)	&	4.3(0.4)	&25(11)	&	12	(6)	&	14(1)	&	17(1)	\\
	Orion-KL	&	TMRT	&	2.4(0.4)	&	4.8(0.8)	&	7.4(1.0)	&	…	&	…	&	…	&36(17)	&	42	(11)	&	27(3)	&	40(4)	\\
	&	Effelsberg	&	2.56(0.01)	&	5.52(0.02)	&	8.56(0.02)	&	…	&	…	&	…	&35(14)	&	46	(10)	&	26(3)	&	37(4)	\\
	W51D	&	TMRT	&	2.15(0.05)	&	0.51(0.01)	&	0.86(0.01)	&	…	&	…	&	…	&21(12)	&	14	(10)	&	18(4)	&	22(4)	\\
	&	Effelsberg	&	2.1(0.2)	&	1.56(0.02)	&	1.61(0.02)	&	…	&	…	&	…	&30(22)	&	20	(16)	&	24(4)	&	34(5)	\\
	G016.92	&	Effelsberg	&	1.14(0.03)	&	0.41(0.01)	&	0.12(0.01)	&	1.27(0.07)	&	1.0(0.3)	&	0.1(0.3)	&27(13)	&	12	(6)	&	19(8)	&	23(9)	\\
	G10.47	&	Effelsberg	&	2.21(0.05)	&	1.35(0.03)	&	1.80(0.03)	&	…	&	…	&	…	&29(8)	&	18	(6)	&	24(5)	&	32(6)	\\
	G188.79	&	Effelsberg	&	0.22(0.01)	&	0.16(0.01)	&	0.13(0.01)	&	0.23(0.05)	&	0.1(0.5)	&	0.10(0.03)	&21(10)	&	14	(6)	&	21(17)	&	28(22)	\\
	G35.14	&	Effelsberg	&	1.16(0.02)	&	0.51(0.01)	&	0.26(0.01)	&	1.42(0.02)	&	1.12(0.19)	&	0.2(0.8)	&29(19)	&	12	(4)	&	20(2)	&	24(2)	\\
	NGC\,1333	&	Effelsberg	&	1.01(0.02)	&	0.12(0.01)	&	0.15(0.01)	&	1.1(0.2)	&	0.7(0.2)	&	0.13(0.11)	&19(7)	&	10	(4)	&	12(5)	&	14(5)	\\
	G000.19	&	Effelsberg	&	1.12(0.10)	&	0.25(0.02)	&	0.11(0.02)	&	1.22(0.16)	&	2.6(1.1)	&	0.1(0.5)	&17(5)	&	11	(3)	&	16(8)	&	18(9)	\\
	Barnard-1b	&	Effelsberg	&	2.25(0.02)	&	0.22(0.01)	&	0.11(0.02)	&	2.82(0.02)	&	0.33(0.10)	&	1.16(0.10)	&	14(4)	&	8	(3)	&	10(3)	&	11(3)	\\
	\enddata
	\tablecomments{Column (1): source name; Column (2): telescope; Column (3) - (5): the peak optical depths of the ($J$, $K$) = (1, 1), (2, 2) and (3, 3) main group of hyperfine components of $^{14}$NH$_{3}$, from the intensity ratio method; Columns (6) -- (8): peak optical depths from the HF fitting procedure provided by CLASS; Column (9): the rotational temperature $T_{rot}^{ir}$, from the intensity ratio method; Column (10): $T_{rot}^{rd}$ from the rotation diagram method; Column (11): $T_{rot}^{hf}$ from the improved HF fitting method (see Sect. \ref{sec:Temp}) for 11 sources, that of NGC\,6334\,I from RADEX calculation and that of the remaining three sources with blended spectral features ( G10.47, Orion-KL and W51D) from the HyperFine Group Ratio (HFGR) method (see details in Sect. \ref{sec:Temp}); Column (12): the kinetic temperature calculated from the empirical formula displayed in Appendix B of \citet{Tafalla04}.}
\end{deluxetable}
}

\subsubsection{Temperature}\label{sec:Temp}
\textbf{Rotational temperature}: NH$_{3}$ inversion lines have been widely used as tracers of the temperature in molecular clouds \citep{Li03,Tafalla04,Mangum15,Wang20}. Three main methods, either starting from observed or modeled spectra, are used to estimate the rotational temperature, including the intensity ratio method, the rotation diagram method and the improved HF fitting method. These are outlined below:

\textbf{a) Intensity ratio method}: This method was described in Sect. \ref{sec:optical} to determine the optical depth, according to the intensity ratio of main and satellite components from the observed spectra. 
According to the determined opacities and the measured brightness temperatures of the (1, 1) and (2, 2) main groups of hyperfine components, the rotational temperature T$^{21}_{rot}$ can be derived \citep{Ho83, Mangum92, Ragan11} by: 
\begin{equation}    
T_{rot} = -41.5
\bigg[ln\big(-\frac{0.283}{\tau (1, 1, m) }ln\Big[1-\frac{T_{\rm mb}(2, 2, m)}{T_{\rm mb}(1, 1, m)}(1-e^{-\tau(1, 1, m)} ) \Big]\big)\bigg]^{-1}.
\end{equation}
The $T_{rot}$ results for those 15 sources with $^{15}$NH$_{3}$ detections are listed in column 9 of Table \ref{tab:tau and T}.

\textbf{b) Rotation diagram method}: For optically thin lines in LTE, the relation of the column density and energy above the ground state in the upper inversion doublet (the two states of a given inversion doublet are only about 1 K apart) with the corresponding values for the lower inversion doublet can be determined on the basis of measured line temperatures. The rotation diagram,  i.e., a plot of the upper level column density per statistical weight of a number of molecular energy levels, as a function of their energy above the ground state, is frequently used to estimate the temperature and the total column density \citep[e. g.,][]{Mangum92, Goldsmith99}. 
For the ($J, K$) = (1, 1) and (2, 2) transitions of $^{14}$NH$_{3}$,  the column density in the upper state $N_{u}$ for both transitions can be written, assuming optically thin emission:
\begin{equation} 
N_{u}(1, 1) = \frac{8\pi kv^{2}_{1}\int T_{\rm mb}^{11}dv}{hc^{3}A_{1}}=\frac{6.37\times 10^{7}}{[K^{-1}cm^{-3}s]} \times \frac{\int T_{\rm mb}^{11}dv}{[Kcms^{-1}]} 
\end{equation}
\begin{equation} 
N_{u}(2, 2) = \frac{8\pi kv^{2}_{2}\int T_{\rm mb}^{22}dv}{hc^{3}A_{2}}=\frac{2.24\times 10^{7}}{[K^{-1}cm^{-3}s]} \times \frac{\int T_{\rm mb}^{22}dv}{[Kcms^{-1}]},
\end{equation}
where $N_{u}$(1, 1) and $N_{u}$(2, 2) are the column density in the upper state for the ($J, K$) = (1, 1) and (2, 2), respectively; $k$ is the Boltzmann constant, $c$ is the speed of light and $h$ is the Planck constant. $A_{ul}$ is the Einstein coefficiant for spontaneous emission, which was obtained from the JPL Molecular Spectroscopy Catalog \citep{Pickett98} and is listed in Table \ref{tab:para}. 

Optical depths of the $^{14}$NH$_{3}$ transition lines of our sources are mostly large ($\geq$ 1 for 14 out of 15 sources in NH$_{3}$(1, 1), see Table \ref{tab:tau and T}), so that the assumption of optically thin emission only provides lower limits to the $^{14}$NH$_{3}$ abundances and may underestimate the real abundance ratio $^{14}$NH$_{3}$/$^{15}$NH$_{3}$. Thus an optical depth correction should be considered for the upper state column density \citep{Goldsmith99, Mei20} yielding: $N_{u}^{'}$ = $N_{u}\tau(J, K, m) /(1-exp(-\tau(J, K, m)))$.
For the (1, 1) and (2, 2) lines, the relation between the opacity-corrected total column density $N_{t}$ and $N_{u}^{'}$ in a Boltzmann distribution should be: 
\begin{equation}
ln\frac{N_{t}}{Q(T_{rot})} = ln\frac{N_{u}^{'}(1, 1)}{g_{1}}+ \frac{E_{1}}{kT_{rot}}
\end{equation}
and
\begin{equation}
ln\frac{N_{t}}{Q(T_{rot})} = ln\frac{N_{u}^{'}(2, 2)}{g_{2}}+ \frac{E_{2}}{kT_{rot}}.
\end{equation}
where $g_{u}$ and $E_{u}$ are the degeneracy and the energy of the upper state, respectively (see Table \ref{tab:para}).
$Q$($T_{rot}$) is the partition function from the JPL Molecular Spectroscopy Catalog \citep{Pickett98}.

Thus 
\begin{equation}
ln\frac{N_{u}^{'}(1, 1)}{g_{1}} - ln\frac{N_{u}^{'}(2, 2)}{g_{2}}  = 
ln\frac{\tau_{11} \int T_{\rm mb}^{11}dv}{1-exp(-\tau_{11})} - ln\frac{\tau_{22} \int T_{\rm mb}^{22}dv}{1-exp(-\tau_{22})} + 1.56 = \frac{E_{2}}{kT_{rot}} - \frac{E_{1}}{kT_{rot}}. 
\end{equation}

\begin{figure*}
	\centering
	{\includegraphics[width=4.4cm]{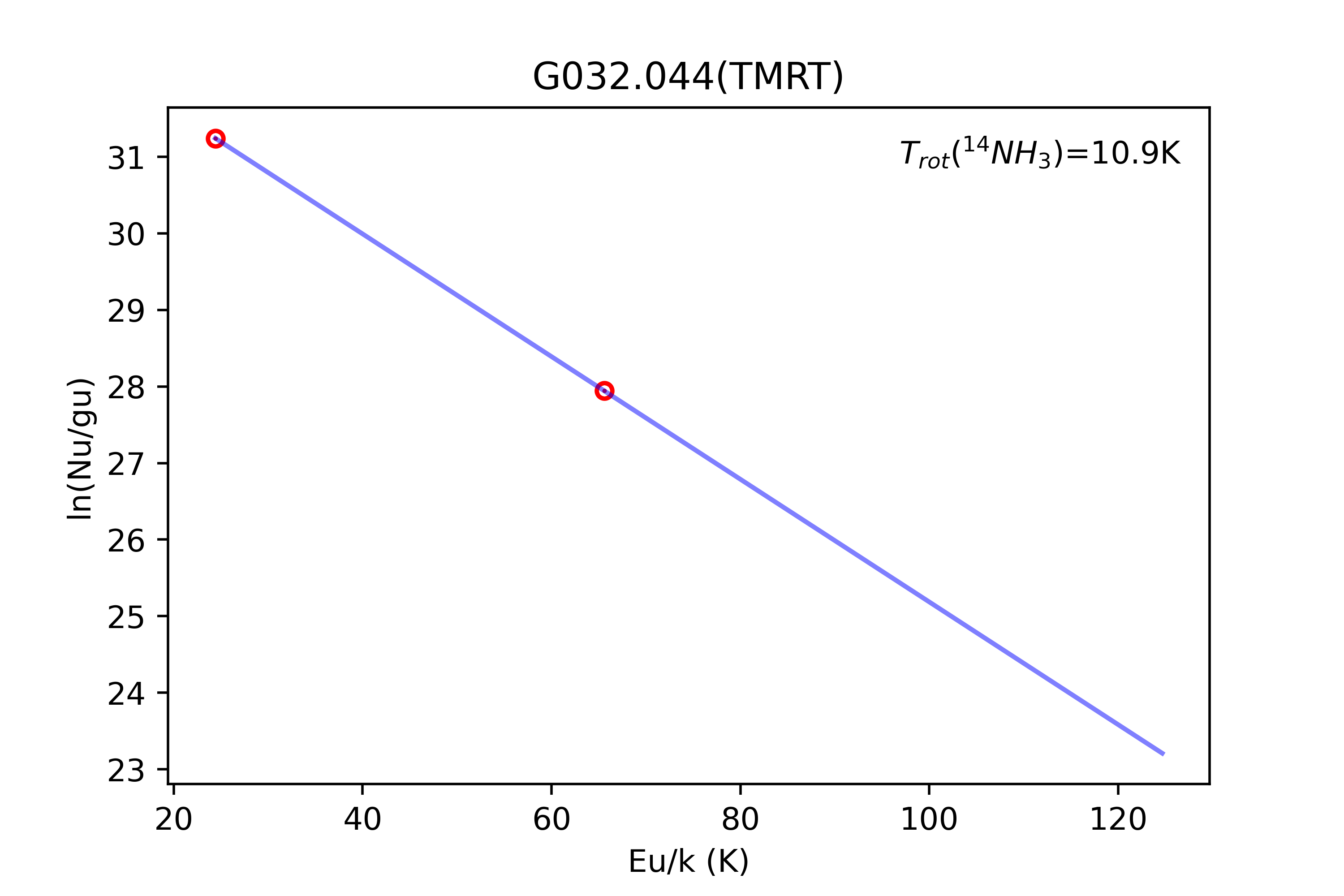}}
	{\includegraphics[width=4.4cm]{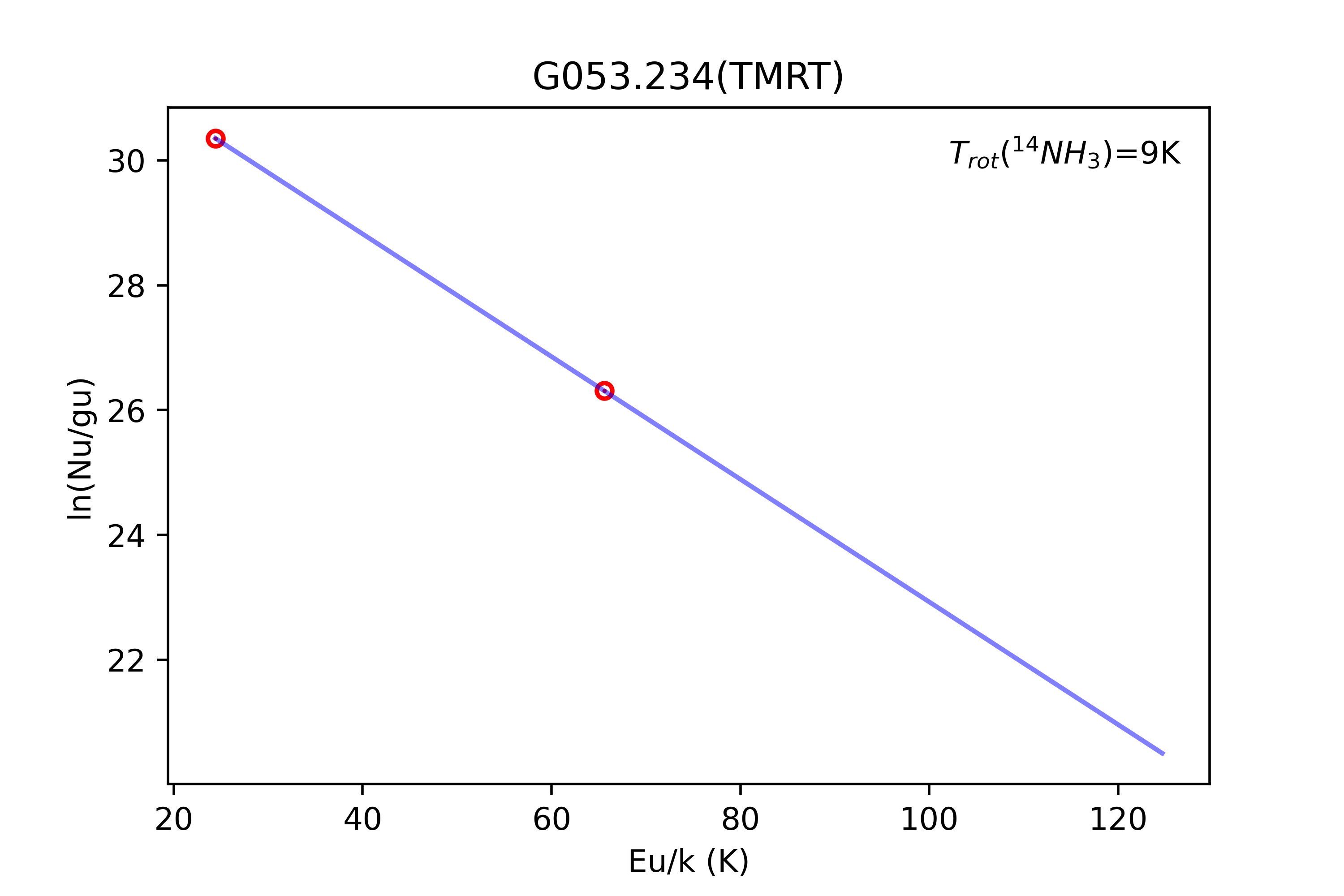}}
	{\includegraphics[width=4.4cm]{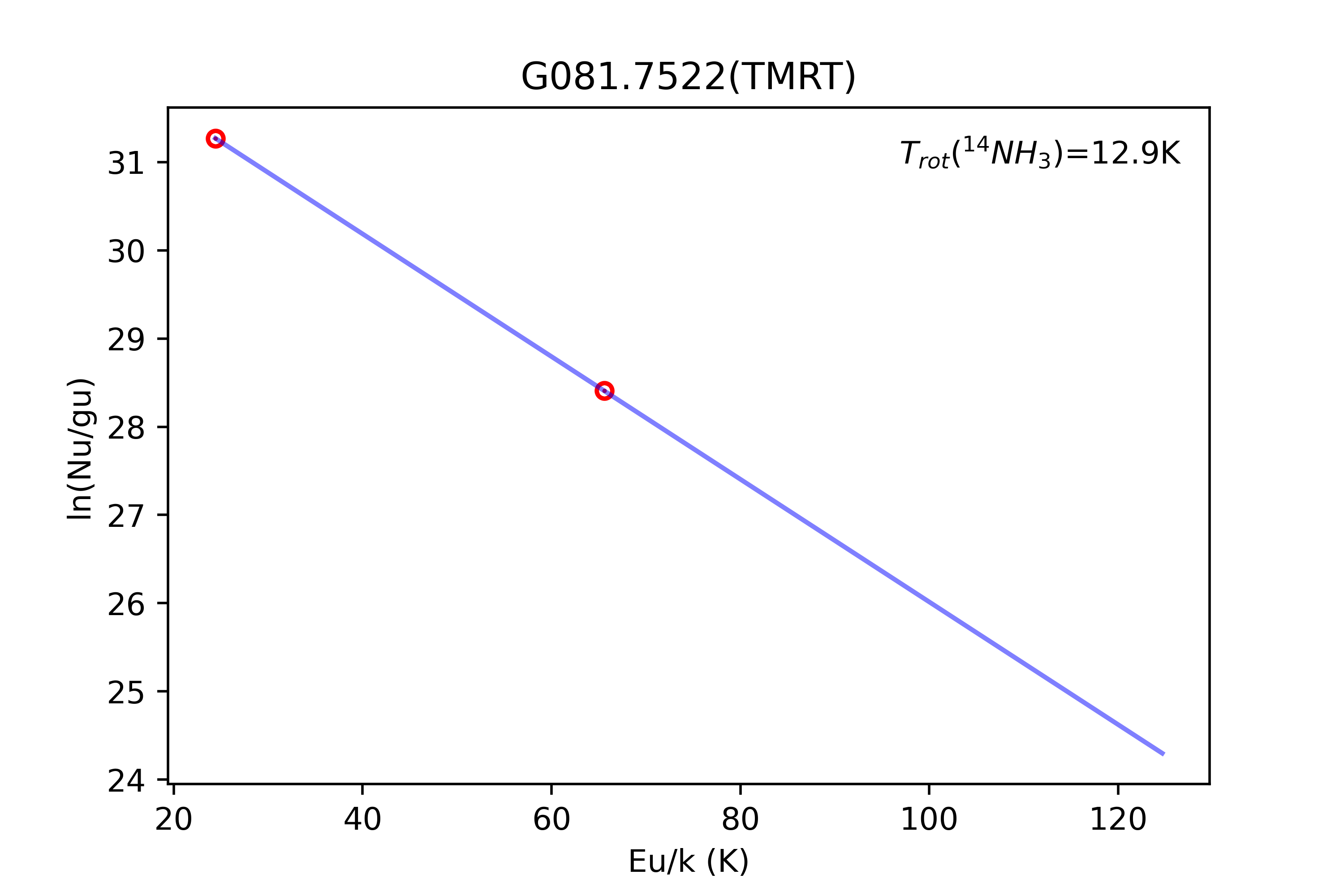}}
	{\includegraphics[width=4.4cm]{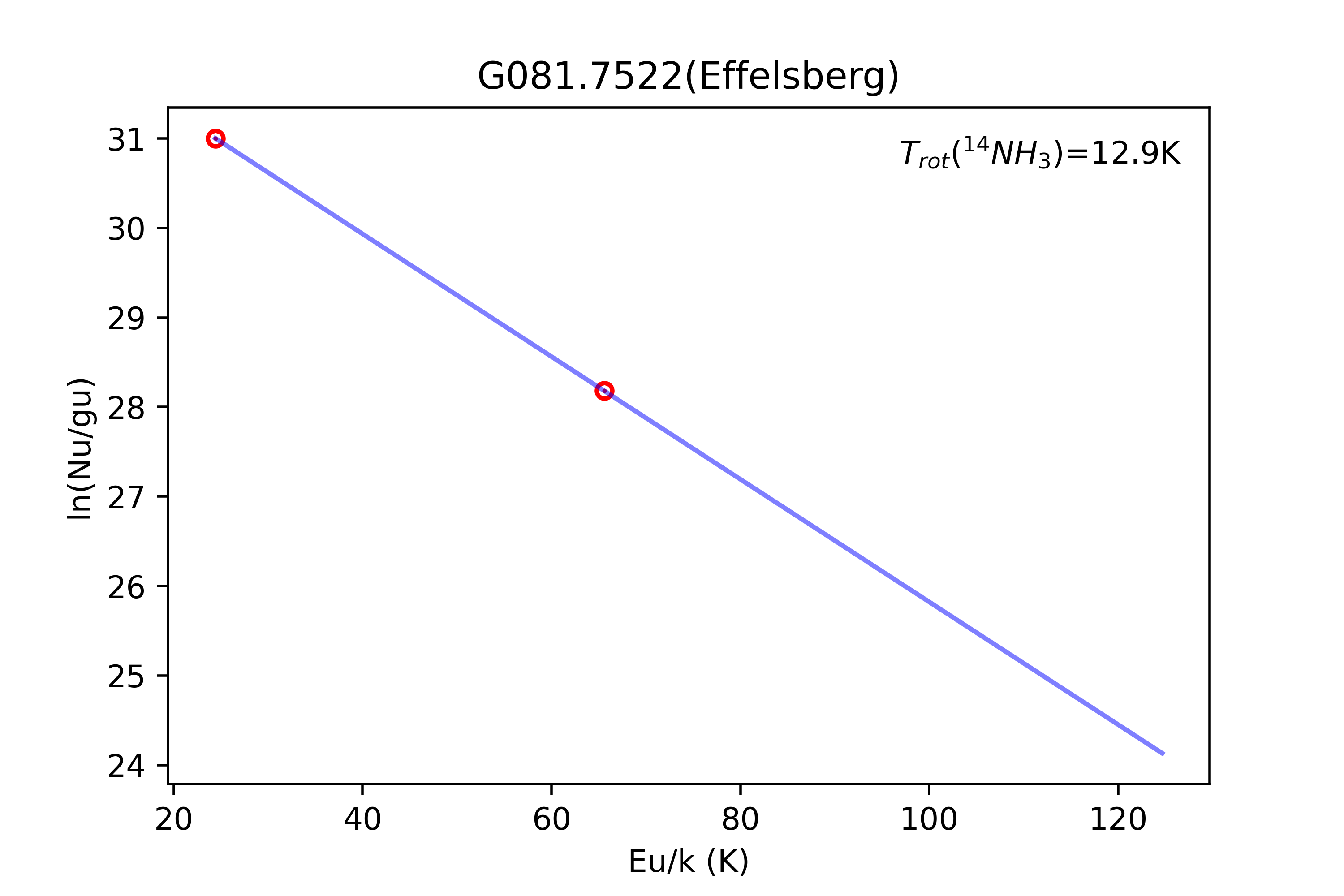}}
	{\includegraphics[width=4.4cm]{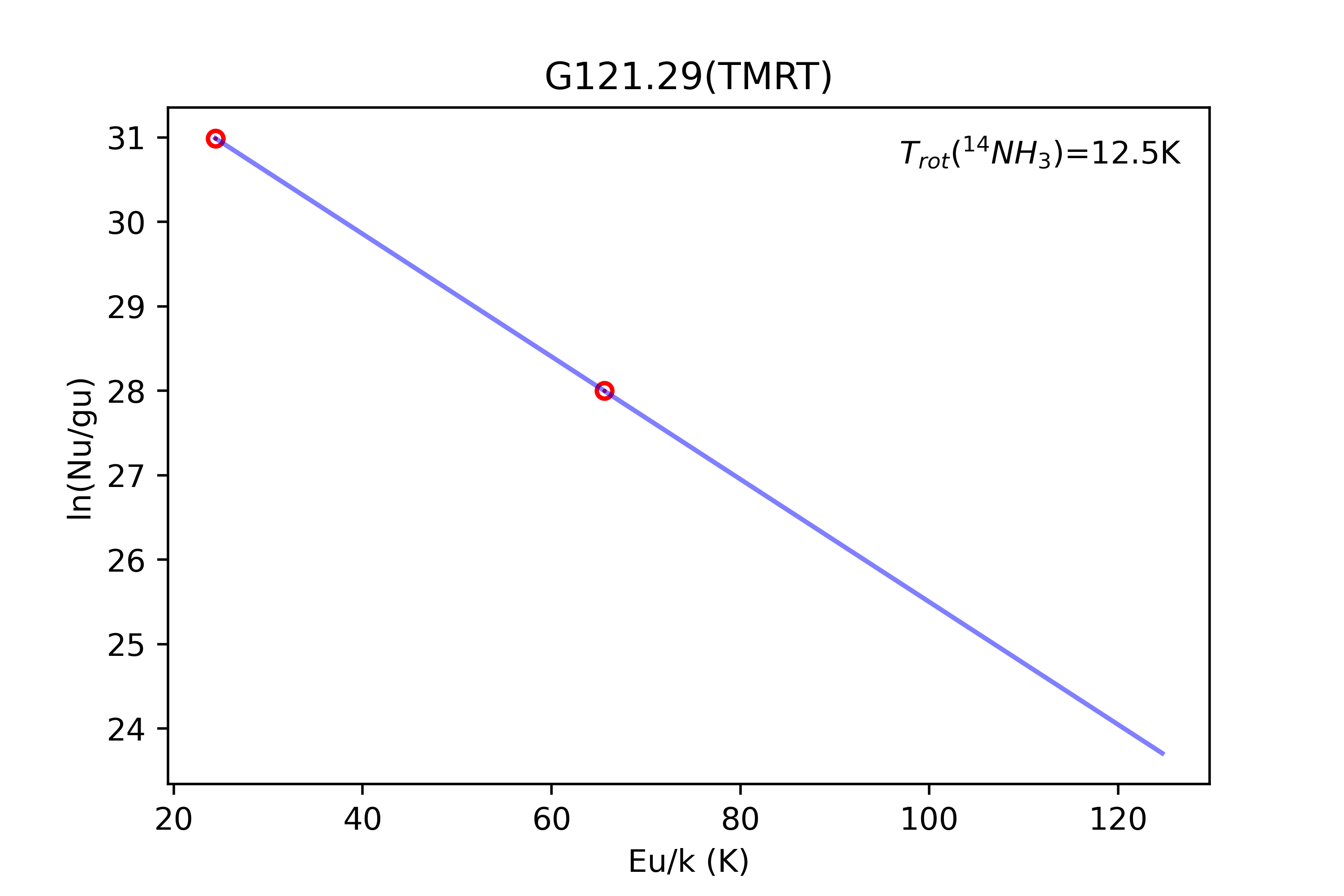}}
	{\includegraphics[width=4.4cm]{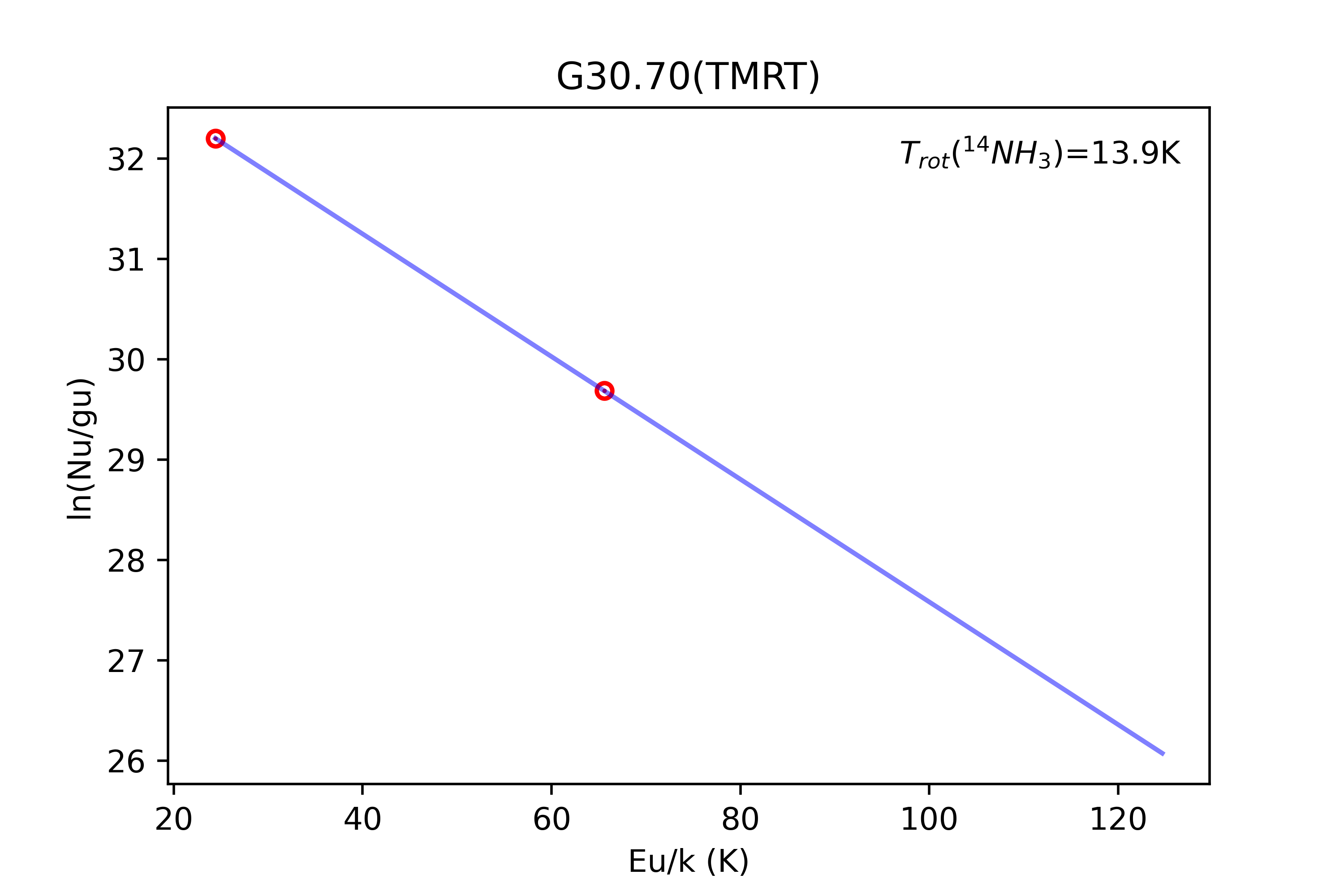}}
	{\includegraphics[width=4.4cm]{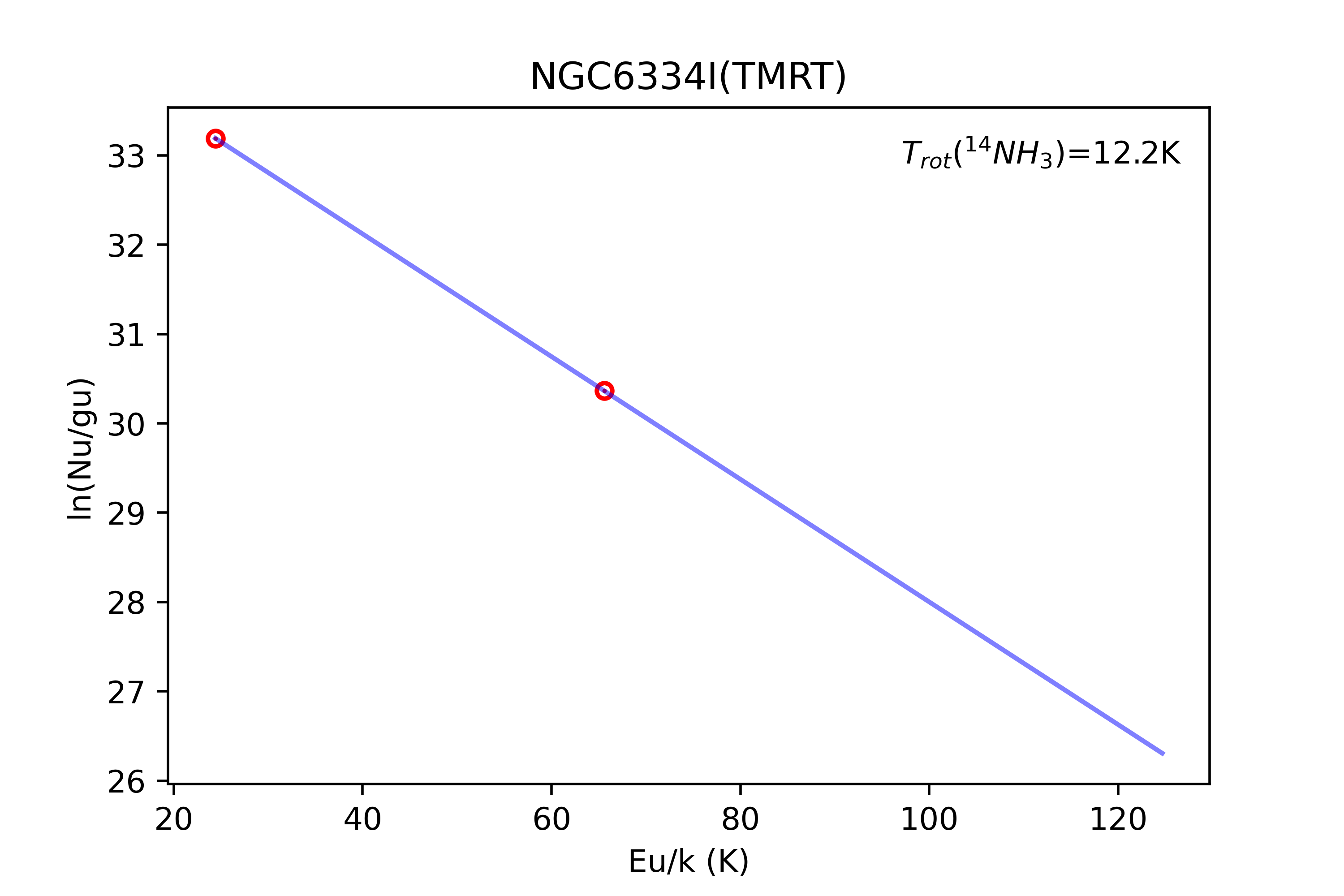}}
	{\includegraphics[width=4.4cm]{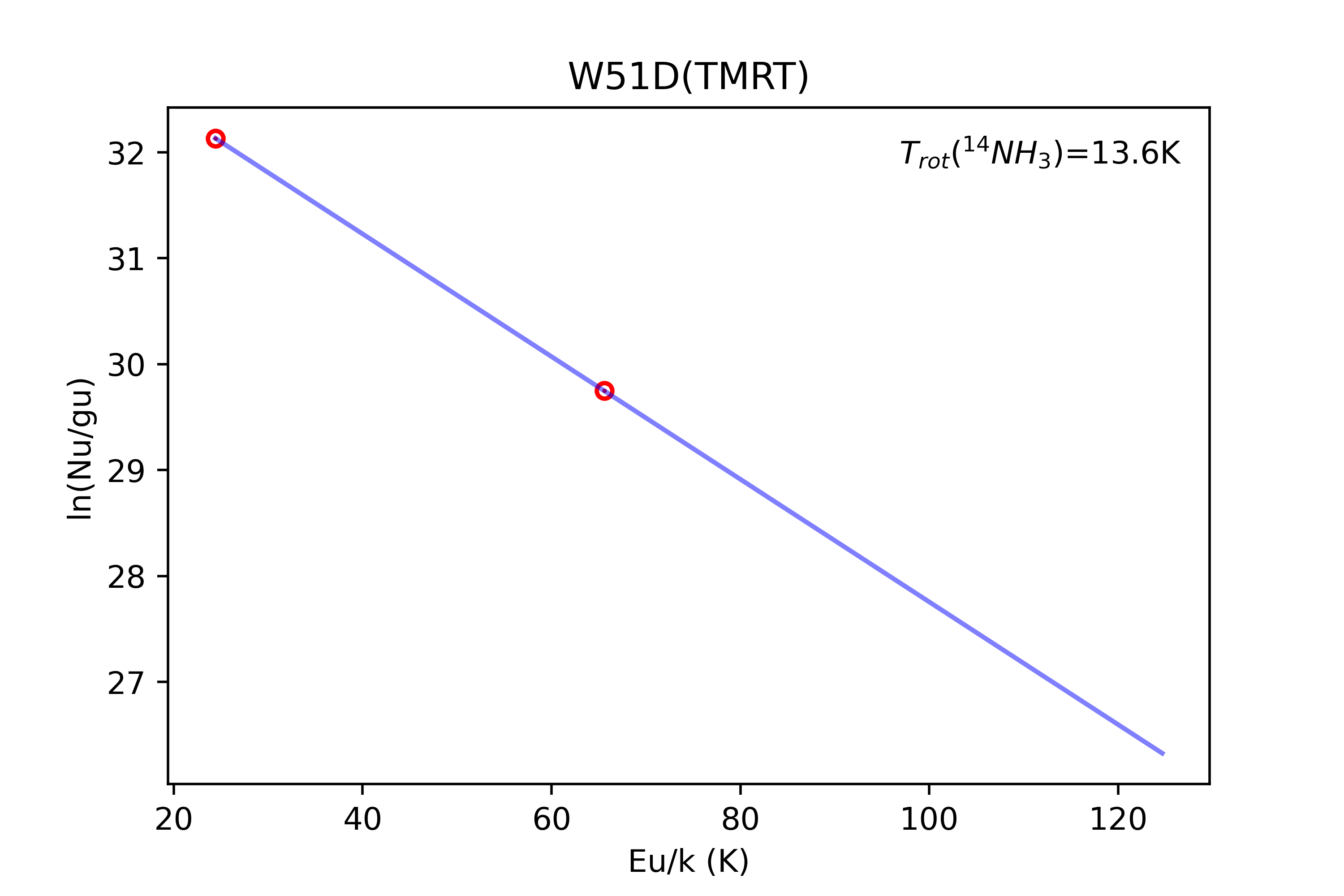}}
	{\includegraphics[width=4.4cm]{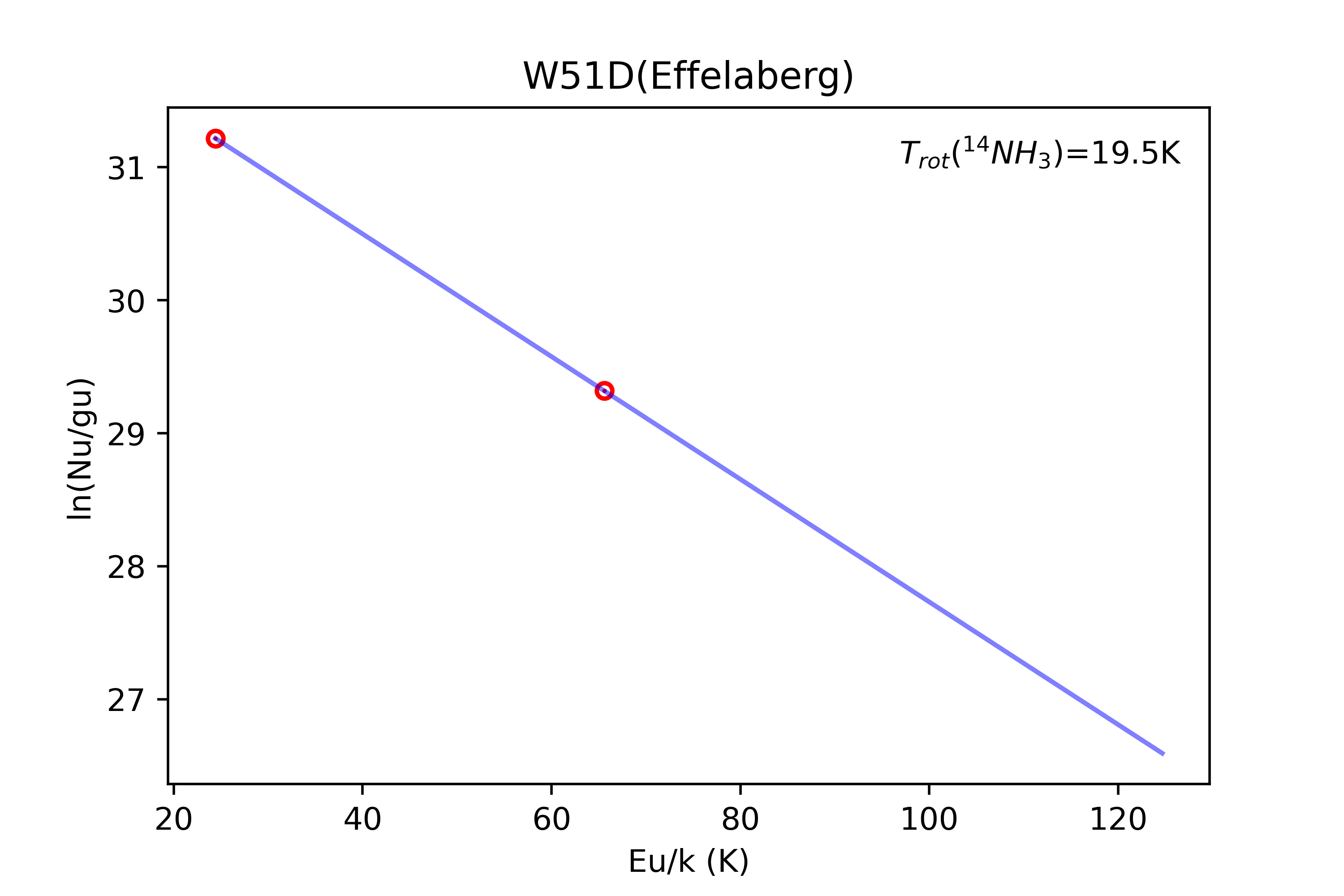}}
	{\includegraphics[width=4.4cm]{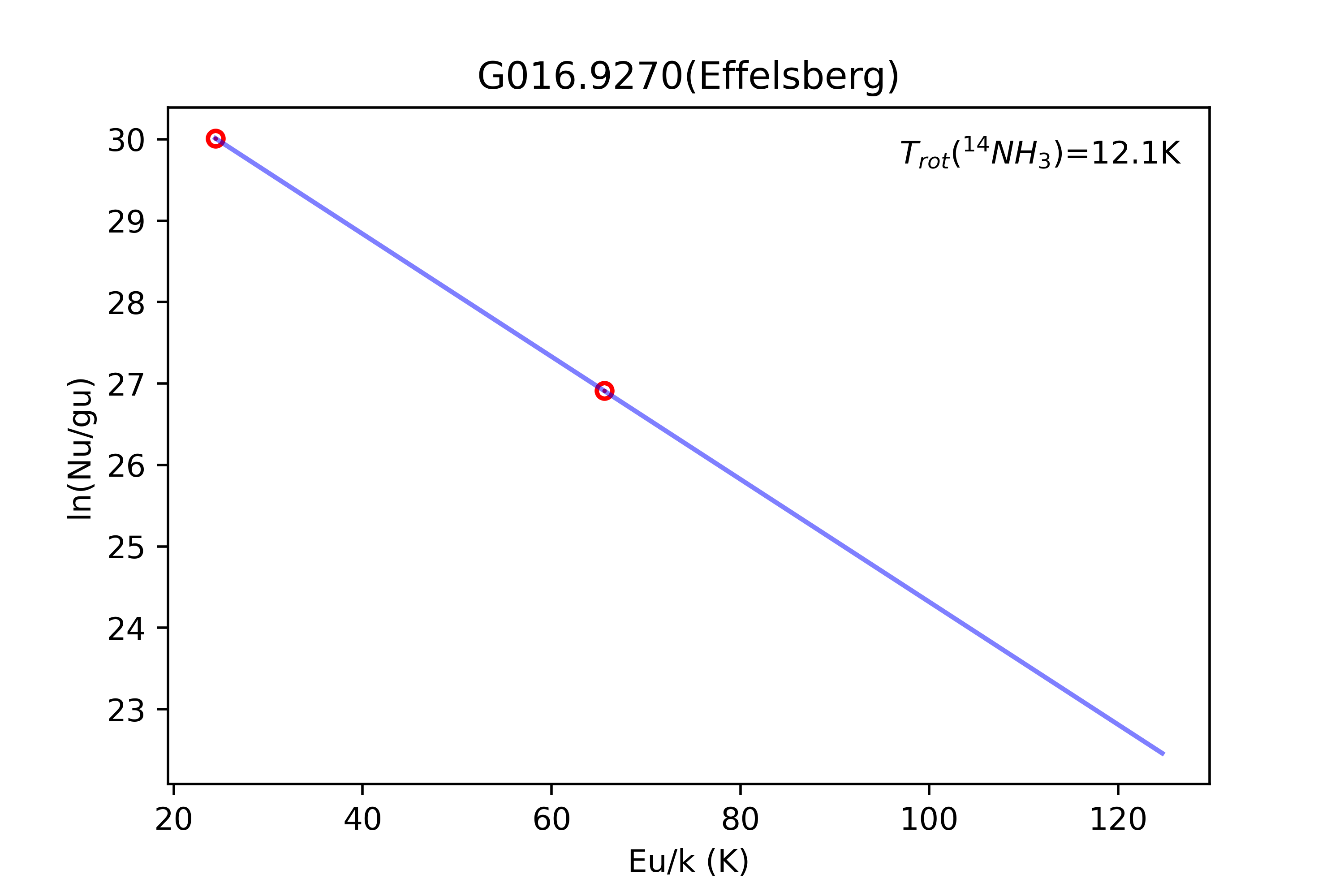}}
	{\includegraphics[width=4.4cm]{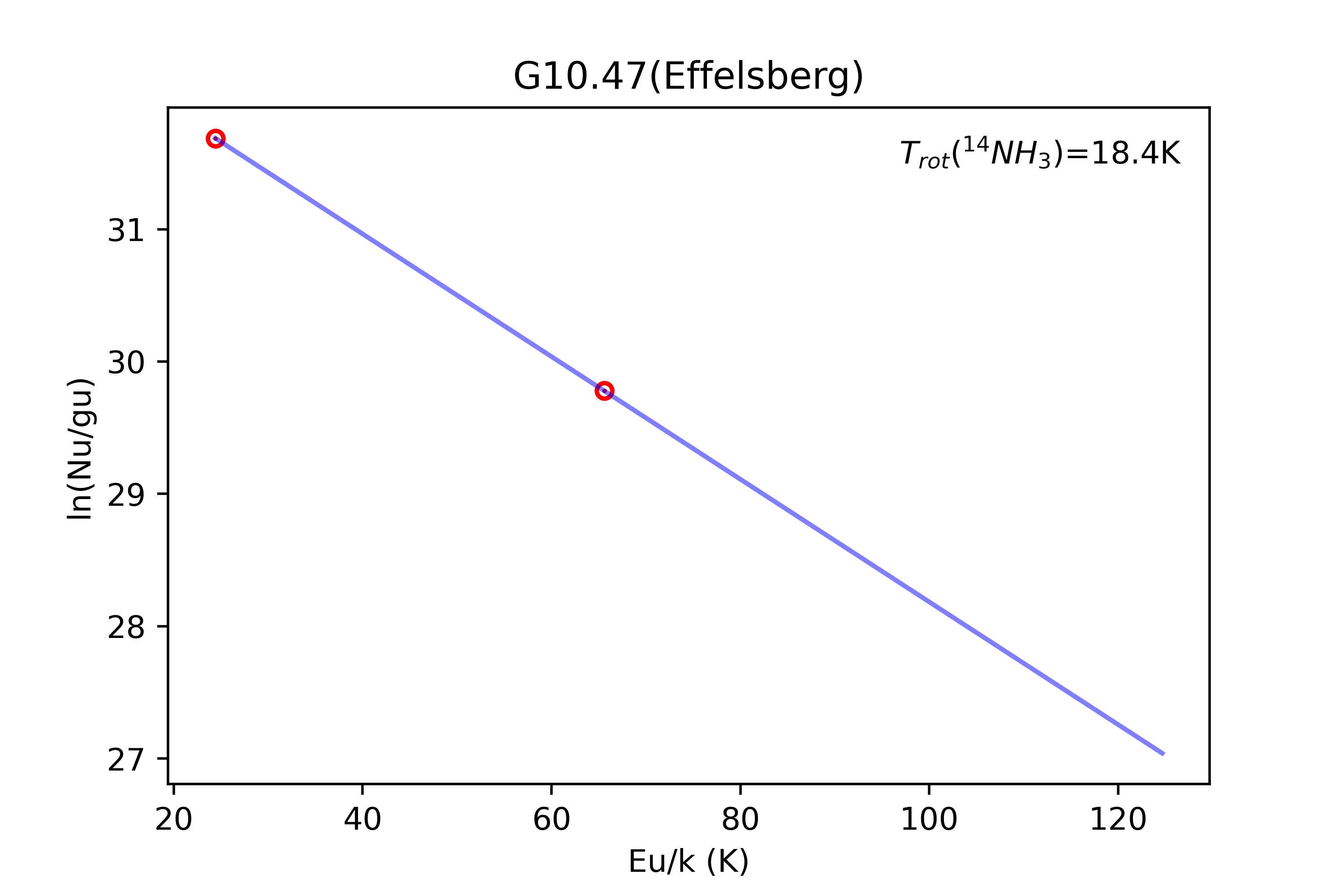}}
	{\includegraphics[width=4.4cm]{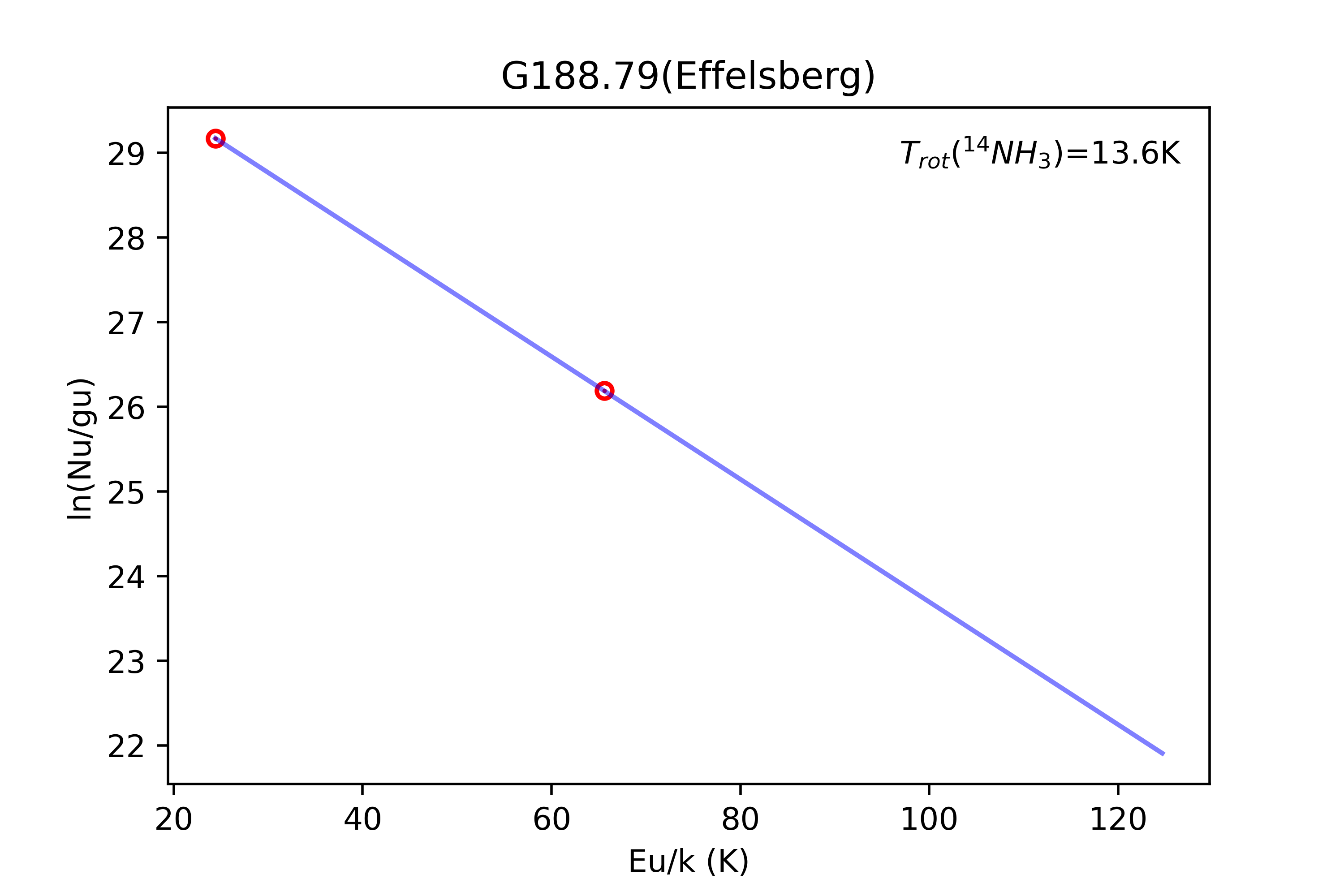}}
	{\includegraphics[width=4.4cm]{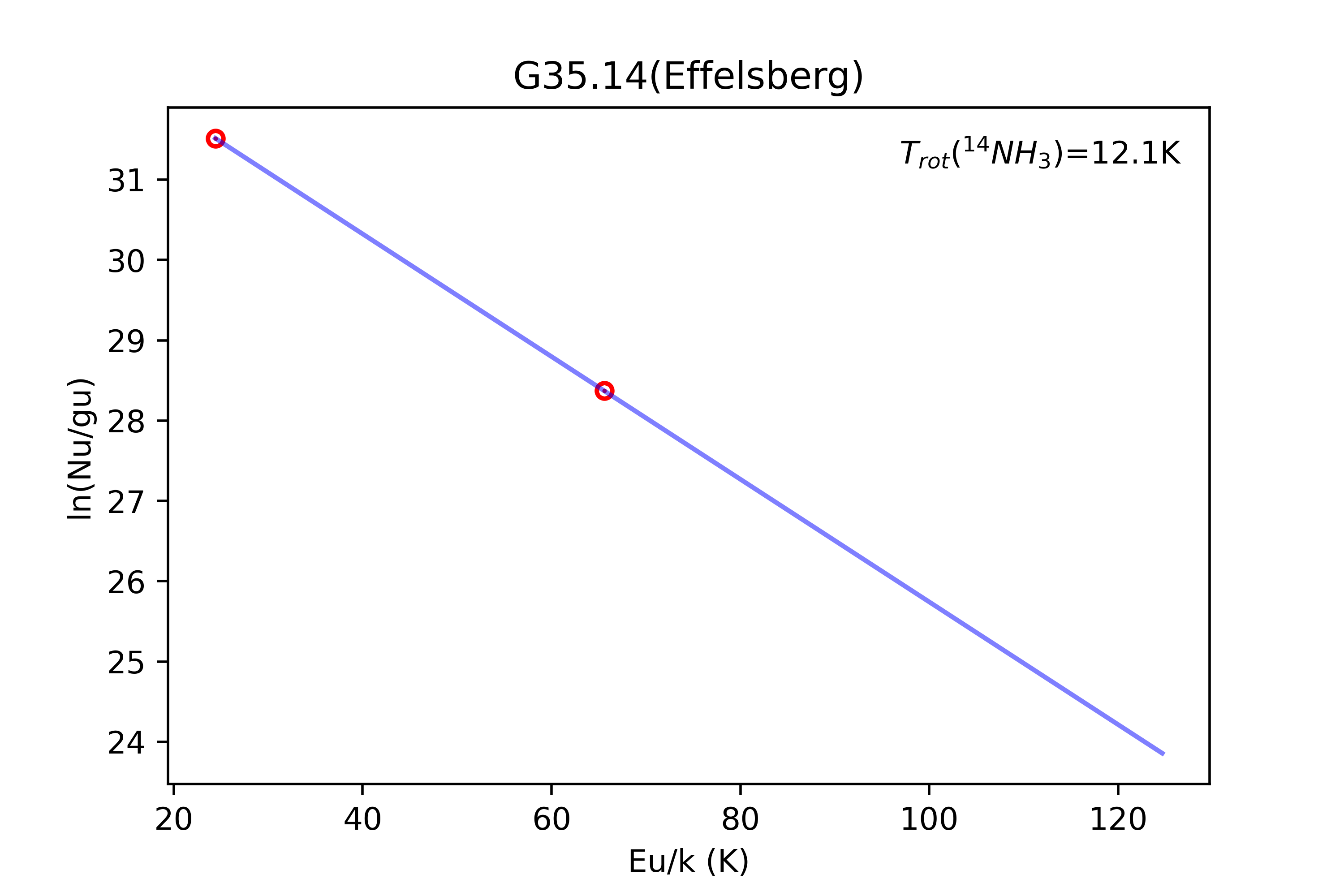}}
	{\includegraphics[width=4.4cm]{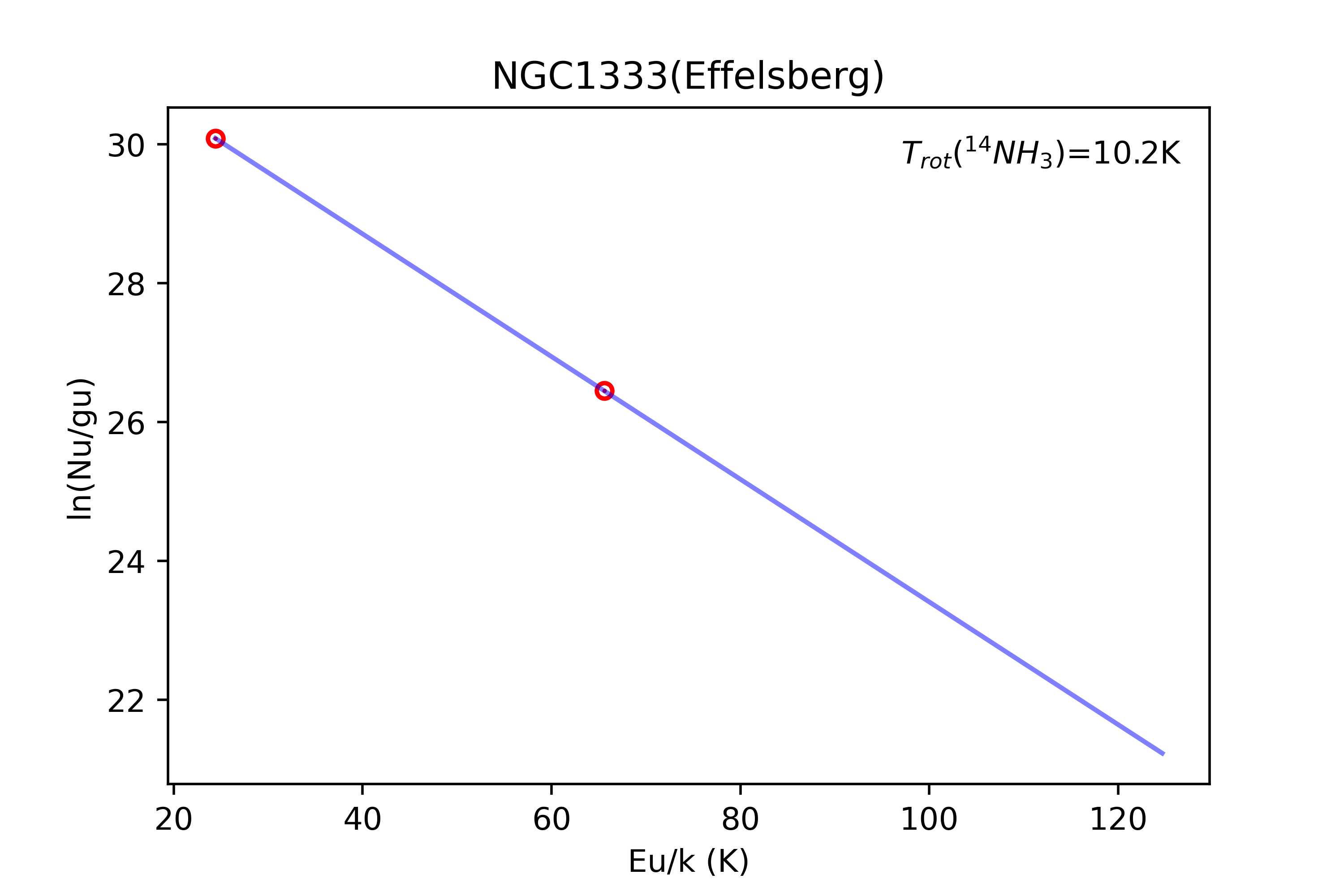}}
	{\includegraphics[width=4.4cm]{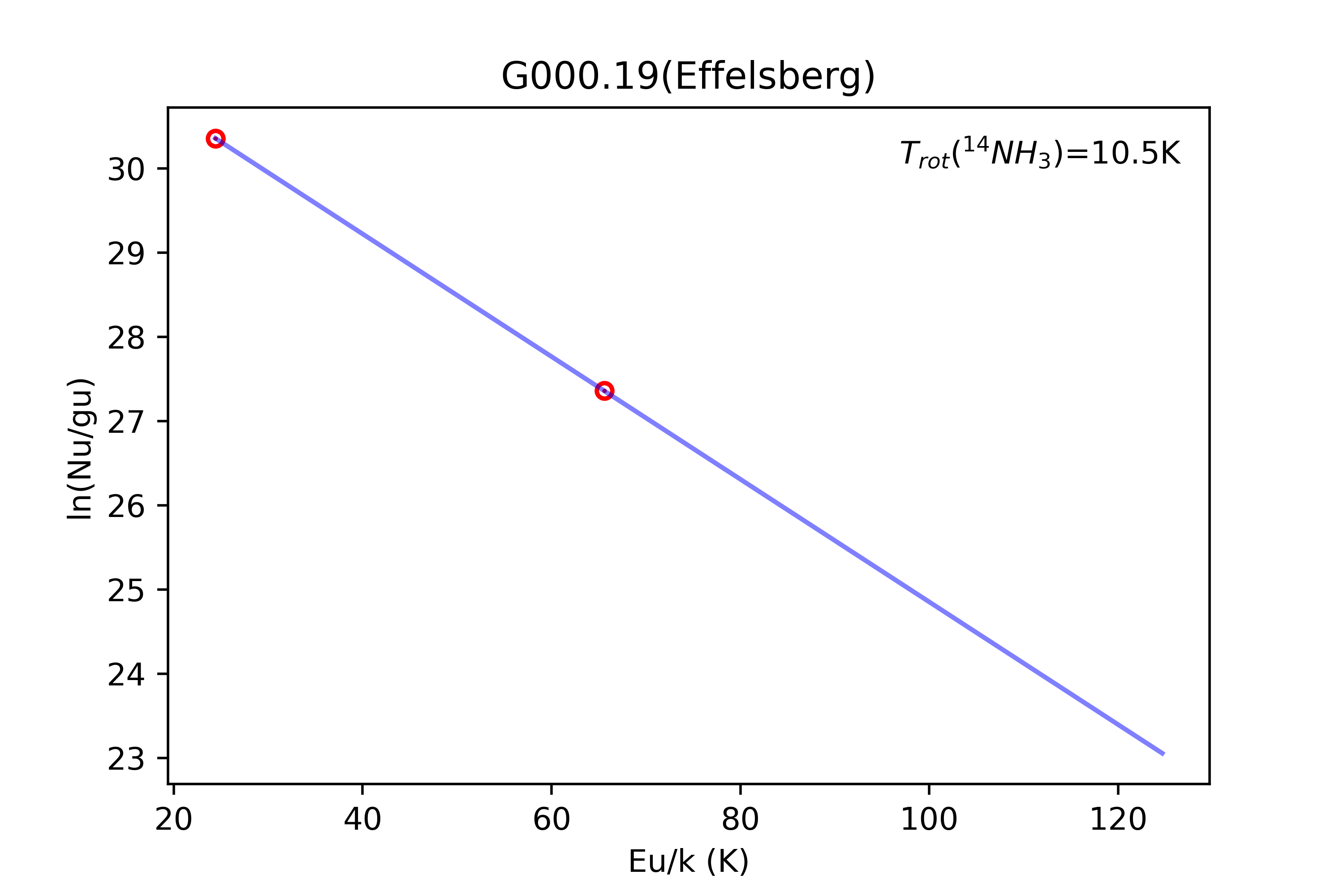}}
	{\includegraphics[width=4.4cm]{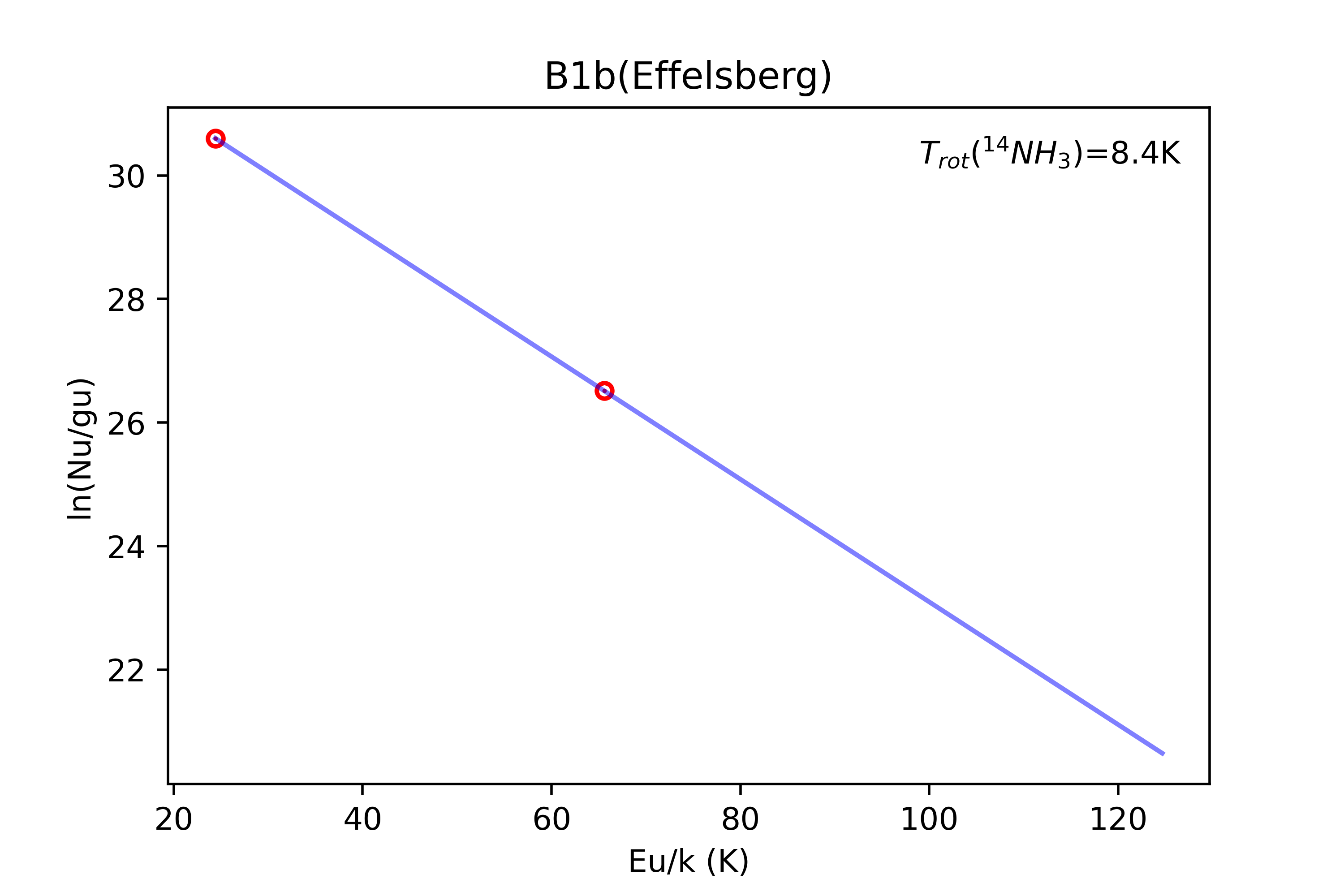}}
	{\includegraphics[width=4.4cm]{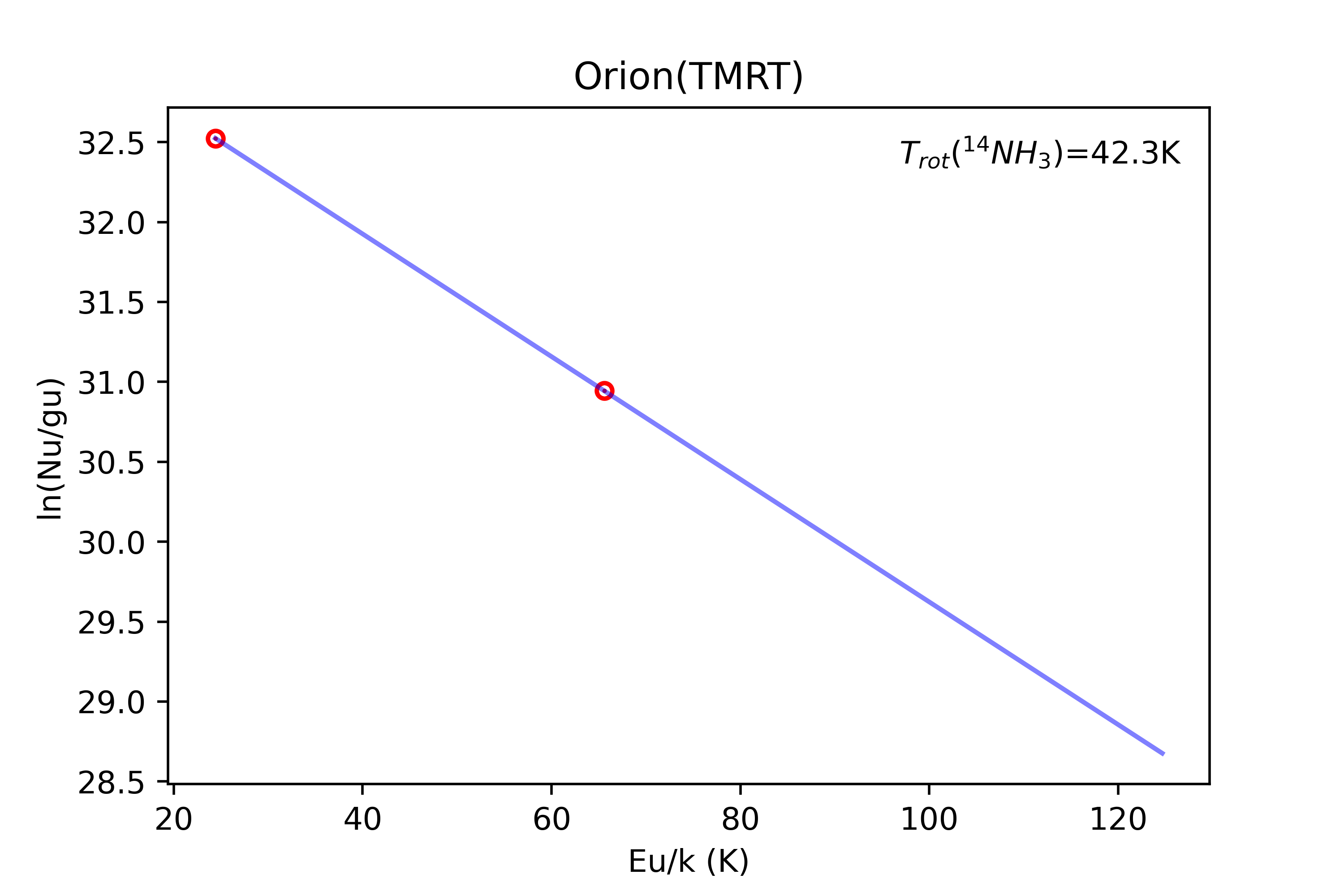}}
	{\includegraphics[width=4.4cm]{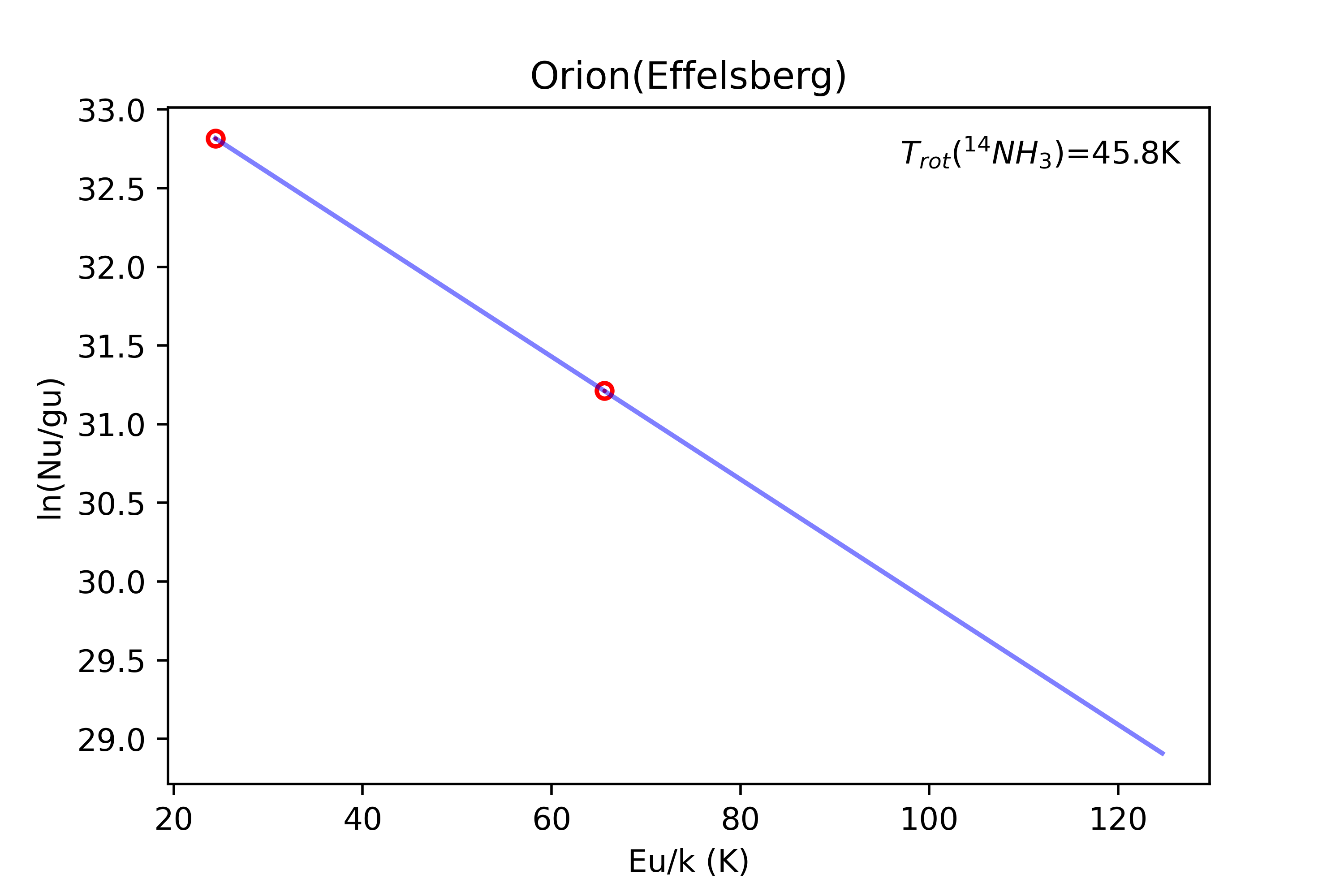}}
	{\includegraphics[width=4.4cm]{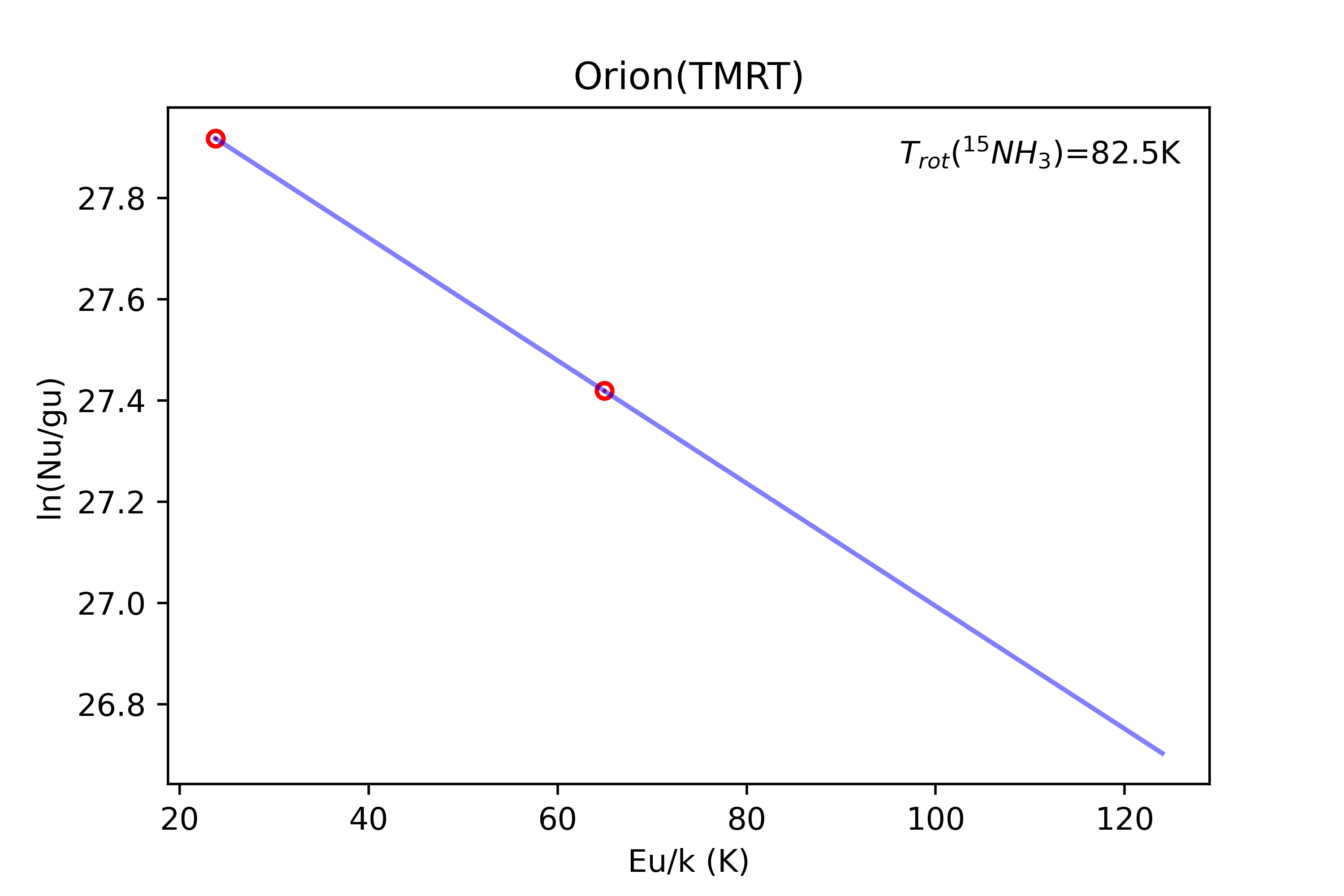}}
	{\includegraphics[width=4.4cm]{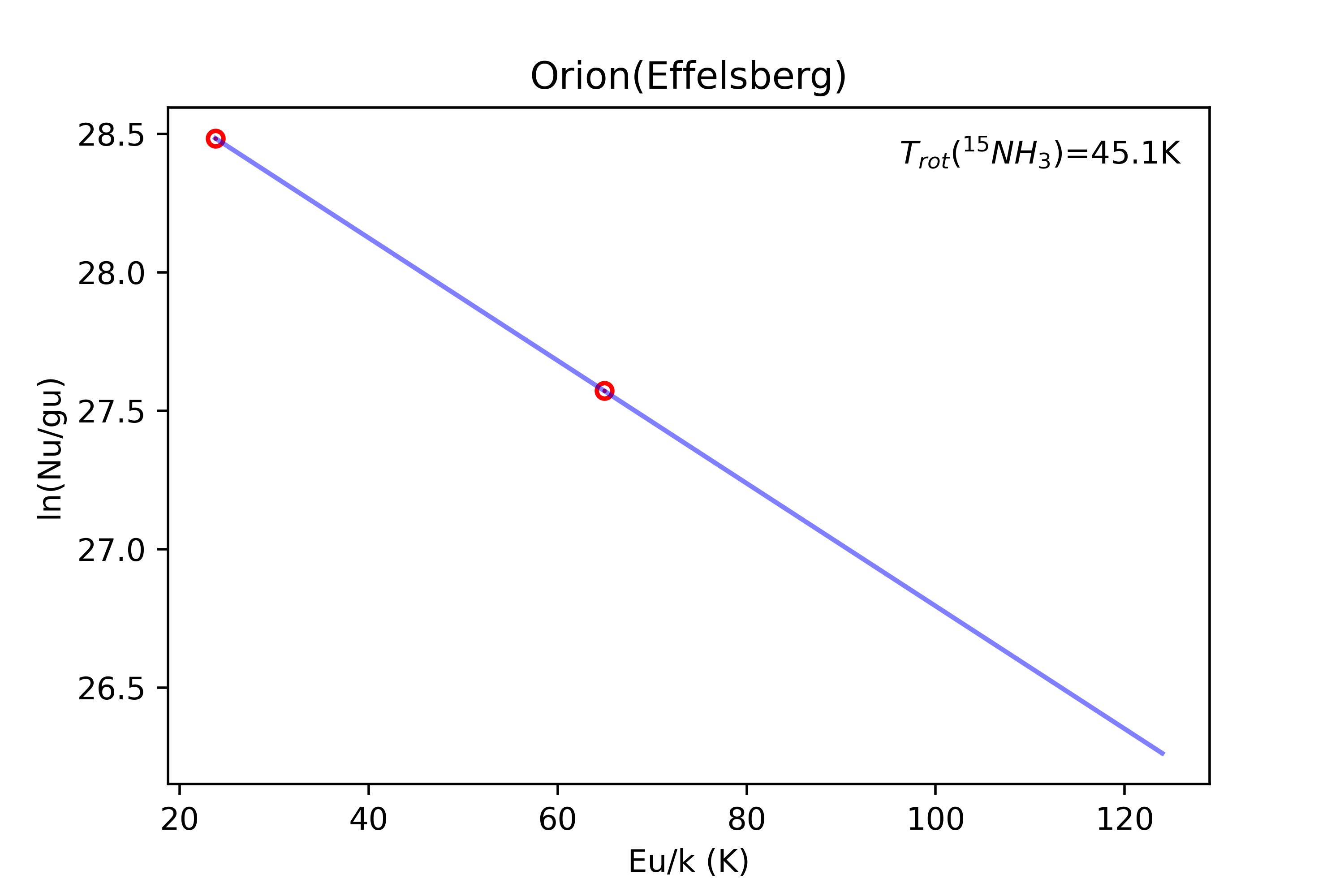}}
	\caption{$^{14}$NH$_{3}$ LTE rotation diagrams for our sample with detected $^{15}$NH$_{3}$ features (see Sect. \ref{sec:Temp}), accounting for opacity effects. This includes all eight sources with TMRT-$^{15}$NH$_{3}$ and the 10 sources with Effelsberg-$^{15}$NH$_{3}$ detections. The two additional panels in the lower right provide $^{15}$NH$_{3}$ rotation diagrams of Orion-KL from Effelsberg-100 m and TMRT - 65 m data. Three sources (G081.7522, W51D and Orion-KL) have $^{14}$NH$_{3}$ detections from both telescopes and are therefore shown twice. $T_{rot}$($^{14}$NH$_{3}$) and in case of the two panels in the lower right also $T_{rot}$($^{15}$NH$_{3}$) can be determined through fits to the data points from the (1, 1) and (2, 2) transitions.}
	\label{14NH3Trotfigure}
\end{figure*}

For the measured (1, 1) and (2, 2) line intensities of our sample, we plotted the rotation diagram, i.e., $ln(N_{u}^{'}/g_{u})$ against $E_{u}/k$. 
The rotational temperature $T_{rot}$ depends on the reciprocal value of the slope (see Figure \ref{14NH3Trotfigure} and Equation (9)). 
The $T_{rot}$ results from this method for our sample are listed in column 10 of Table \ref{tab:tau and T}.
The uncertainties on the rotational temperatures were derived applying error propagation based on equation (9).

\textbf{c) Improved HF fitting method}: This is the updated version of Method in CLASS (Sect. \ref{sec:optical}), which can fit (1, 1) and (2, 2) lines simultaneously. It is included in the Python Package Pyspeckit \citep{Ginsburg11}.
Based on the model spectra produced by the radiative transfer function ($T_{mb}(v)= \eta _{f}[J(T_{ex})-J(T_{bg})][1-e^{-\tau (v)}]$), we can adjust model parameters (the excitation temperature, the line width, etc.) to fit the observed spectra and determine the rotational temperature  \citep{Rodgers08, Camacho20, Keown17}.
$\eta_{f}$ is the beam filling factor assumed to be unity, J$(T_{ex})$ and J$(T_{bg})$ represent the radiation field corresponding to the excitation temperature and the cosmic microwave background temperature of 2.73 K, $\tau (v)$, is the optical depth as a function of frequency.
The following assumptions are adopted for the HF fitting \citep{Ginsburg11}:

(1) Gaussian profiles for the opacity as a function of frequency; 

(2) same excitation temperature for the $^{14}$NH$_{3}$(1, 1) and (2, 2) transitions;

(3) the lines all have the same width;

(4) the multiplet components do not overlap;

(5) LTE is prevailing.

Fixed values for the relative opacities and the frequency shift of each HF component were taken \citep{Mangum15}.
We used the package Pyspeckit to fit the spectra of sources, excluding those three sources with overlapping hyperfine components (G10.47, W51D and Orion-KL, see Sect. \ref{sec:optical}). 
With the exception of NGC\,6334\,I, the $^{14}$NH$_{3}$(1, 1) and (2, 2) groups of the spectra can be fitted well simultaneously (see Figure \ref{HF fitting figure}). 
For NGC\,6334\,I, the two groups of spectra could not be fitted simultaneously with the same value of the excitation temperature. 
To check possible non-LTE effects, we made RADEX\footnote{\url{https://home.strw.leidenuniv.nl/~moldata/radex.html}} calculations for this source and got a $T_{rot}$ value of $\sim$14.3 K with excitation temperatures of 17.1  and 7.3 K for the (1, 1) and (2, 2) lines. The rotation temperature is consistent with the $T_{rot}$ result of 12.2 $\pm$ 6.1 K by the rotation diagram method (LTE), which is used in our later analysis.

For those three sources with blended hyperfine components in the spectra (G10.47, Orion-KL and W51D), an improved method, the Hyperfine Group Ratio (HFGR) method, was used to calculate the rotational temperature. This was developed recently by \citet{Wang20}, which can effectively reduce the uncertainties related to spectral profiles, since only the integrated intensity ratios of groups of hyperfine components are utilized without spectral fitting.
The $T_{rot}$ results for these three sources by the HFGR method, NGC\,6334\,I by RADEX and others by the HF fitting are listed in column 11 of Table \ref{tab:tau and T}.

Comparing the $T_{rot}$ results from the three different methods mentioned above, we find that $T_{rot}$ results from the rotation diagram method are systematically lower than those from the other two methods. 
The $T_{rot}$ results from the intensity ratio method have systematically larger uncertainties, which may be mainly caused by uncertainties of the optical depth, estimated from the ratio of the peak of the main group of hyperfine components and the HF groups giving rise to the inner and outer satellites.
Based on good fitting of the observed spectra in Figure \ref{HF fitting figure}, thus we took the $T_{rot}$ values that arise from the improved HF fitting method for later analysis. 
For those three sources with blended components, their $T_{rot}$ values were taken from the HFGR recipe (see Table \ref{tab:tau and T}).

\begin{figure*}
	\centering
	{\includegraphics[width=8cm]{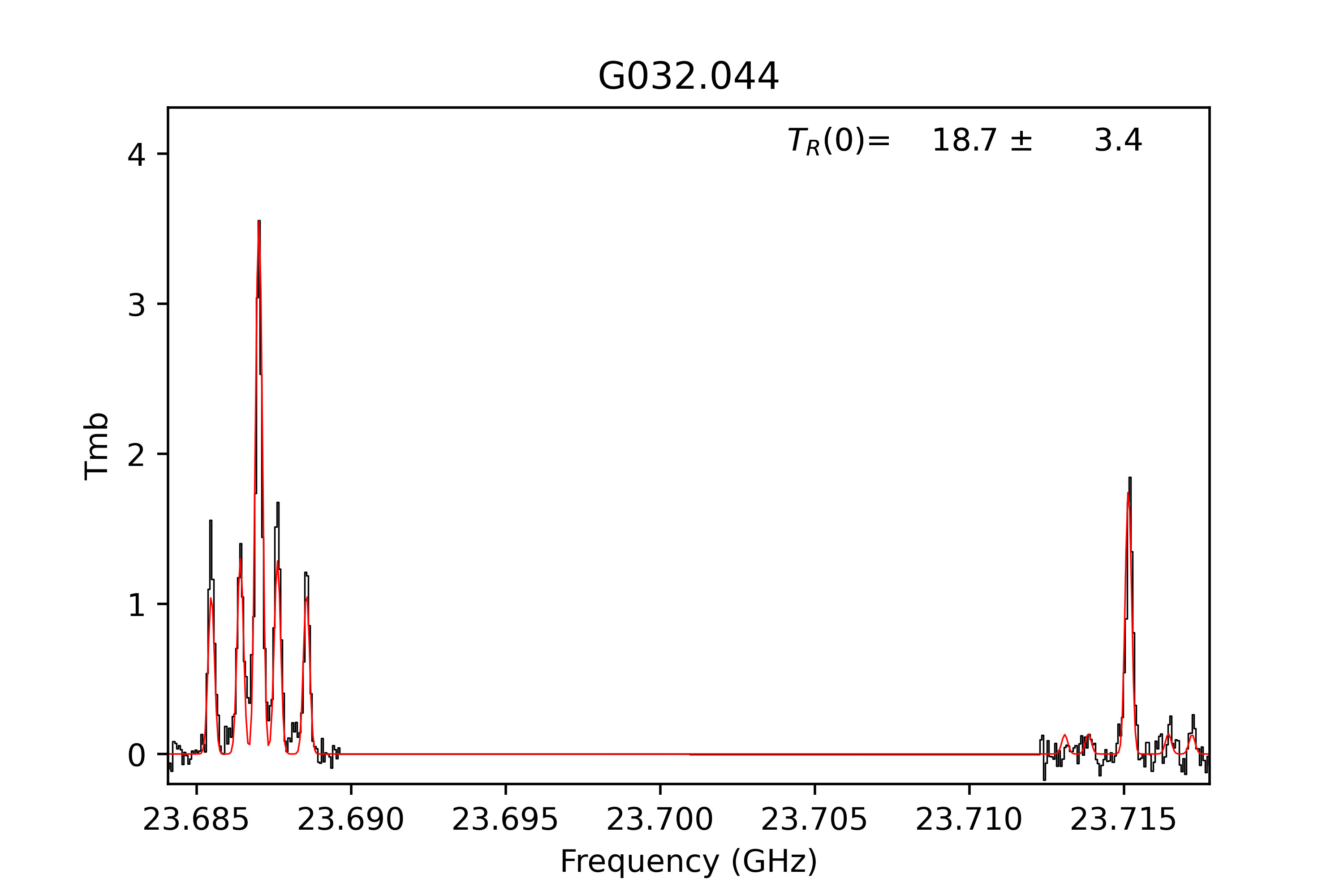}}
	{\includegraphics[width=8cm]{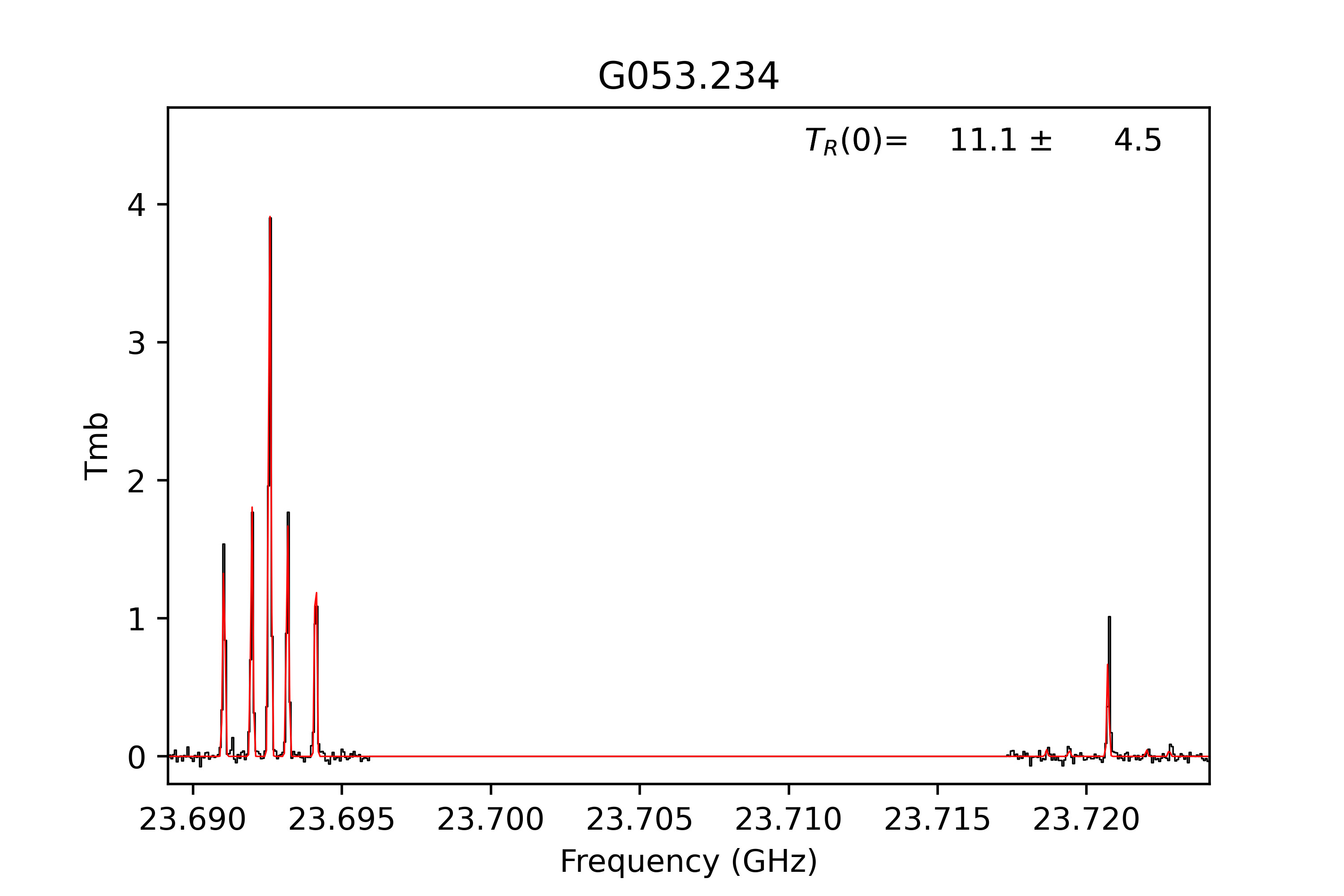}}
	{\includegraphics[width=8cm]{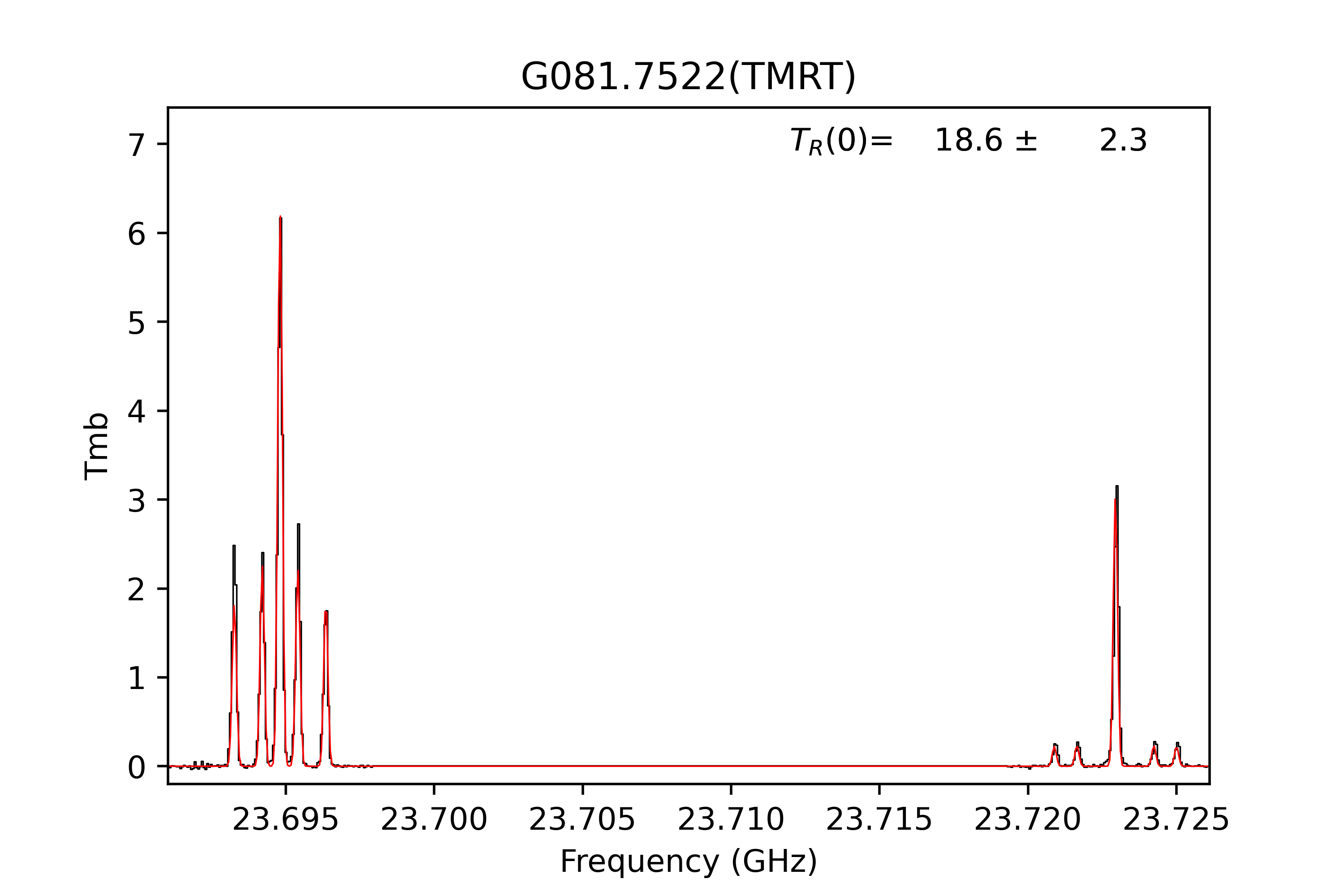}}
	{\includegraphics[width=8cm]{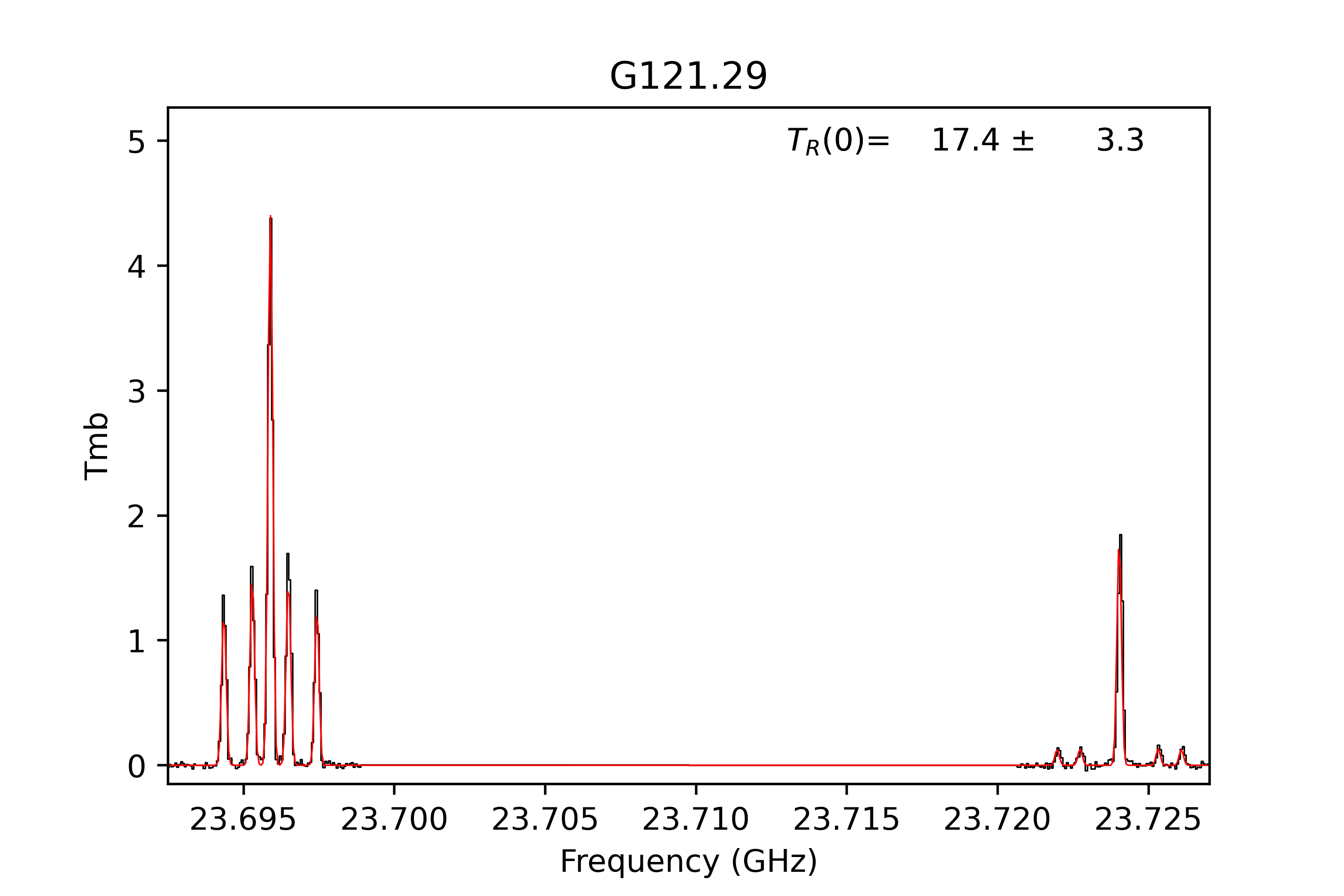}}
	{\includegraphics[width=8cm]{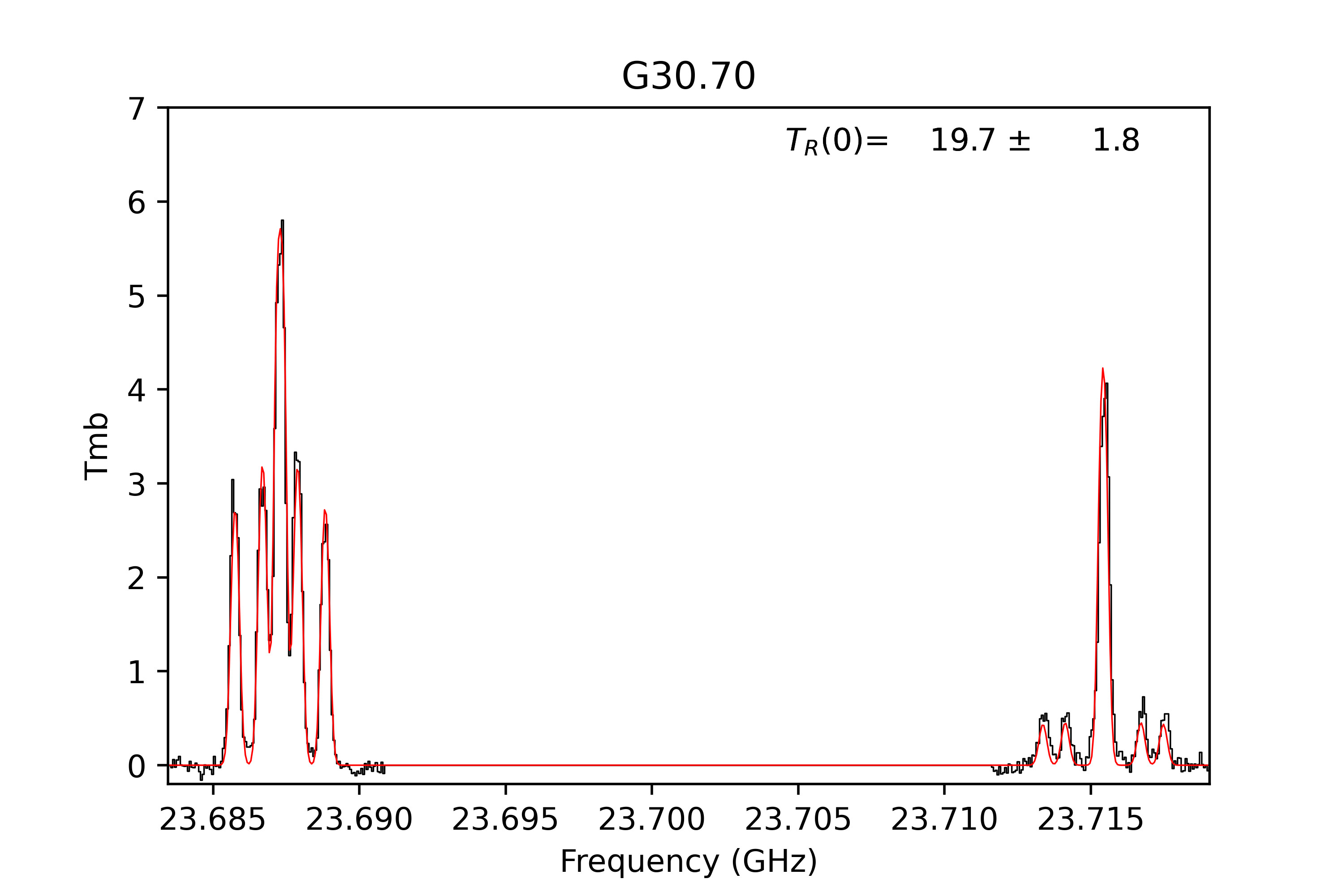}}
	{\includegraphics[width=8cm]{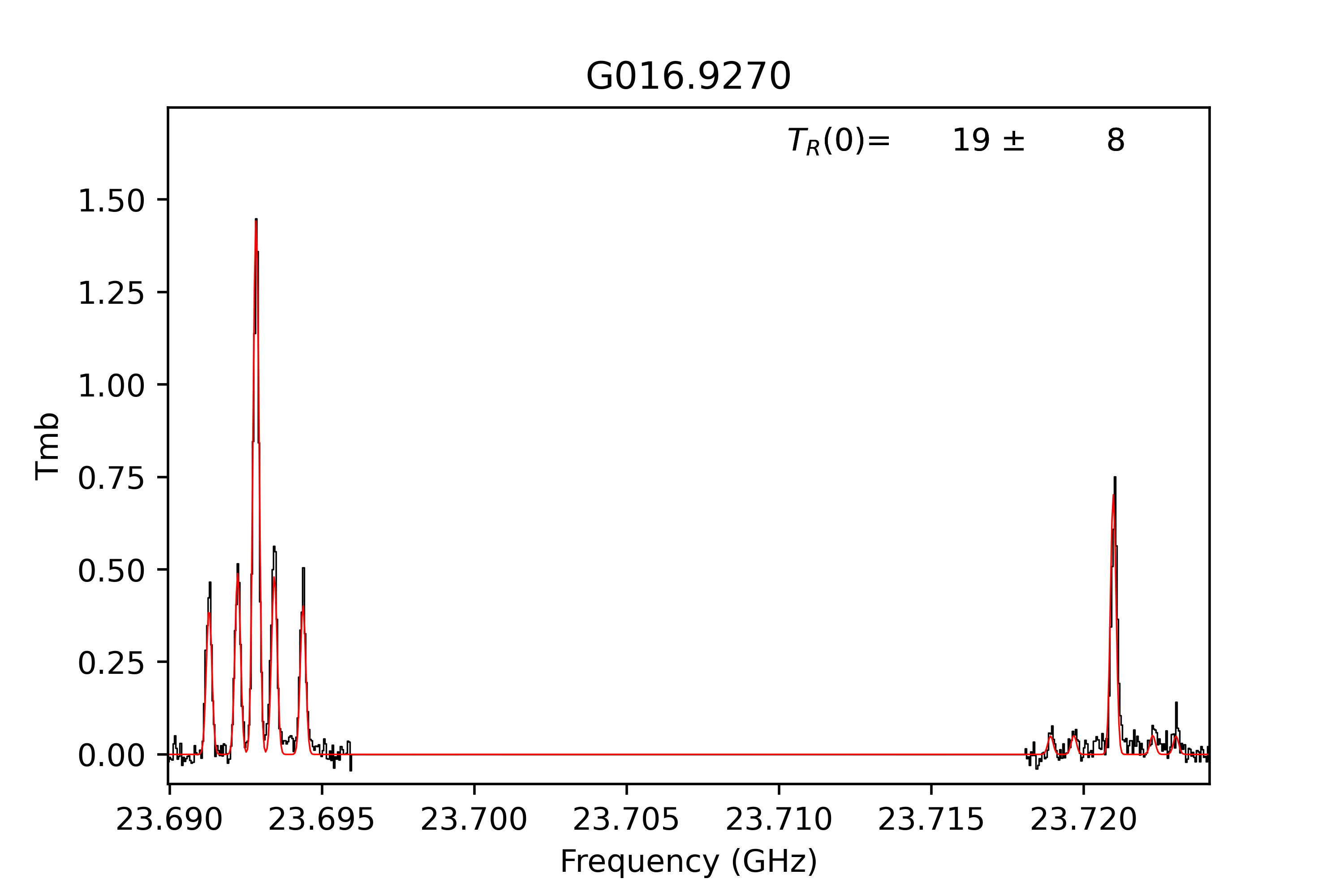}}
	{\includegraphics[width=8cm]{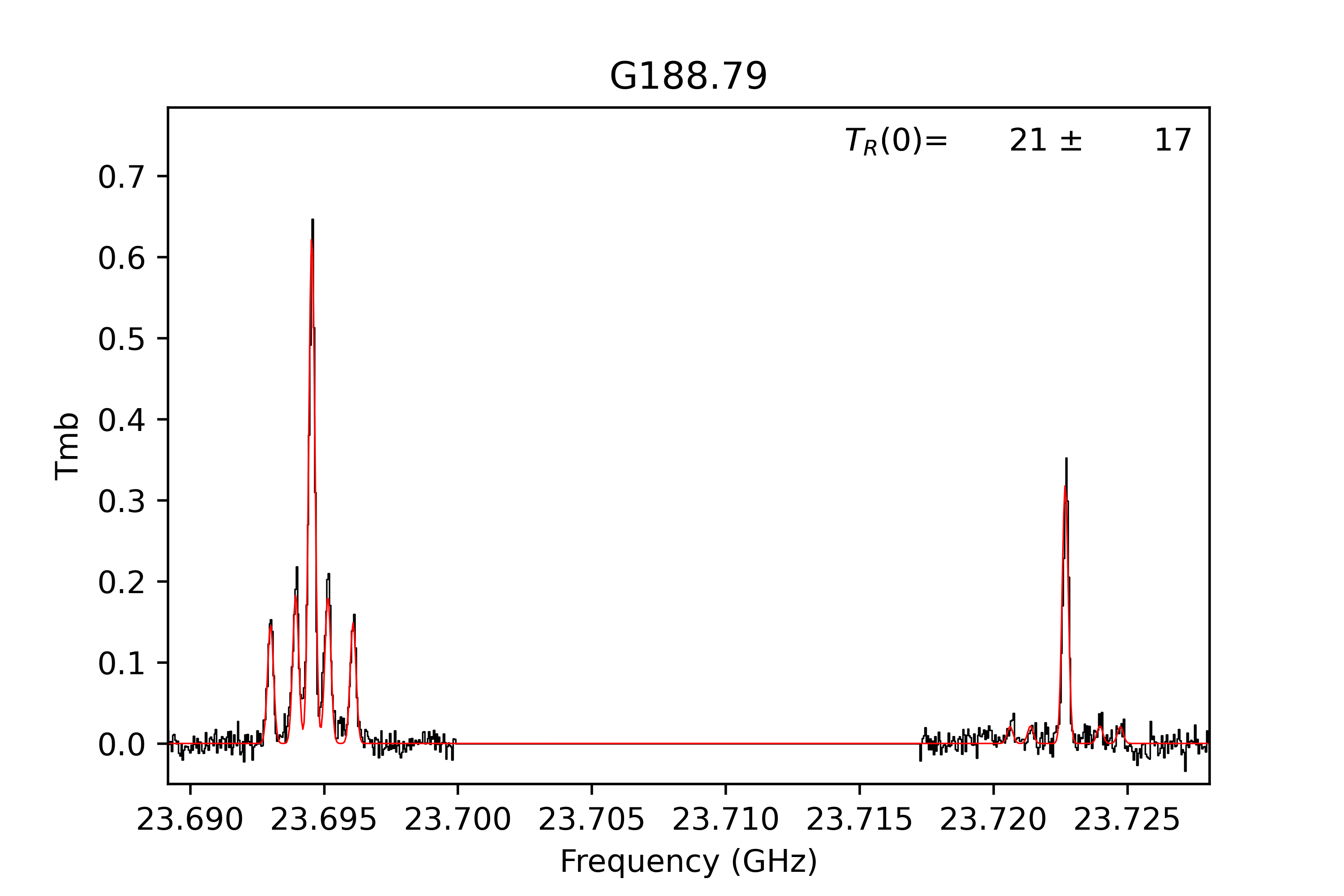}}
	{\includegraphics[width=8cm]{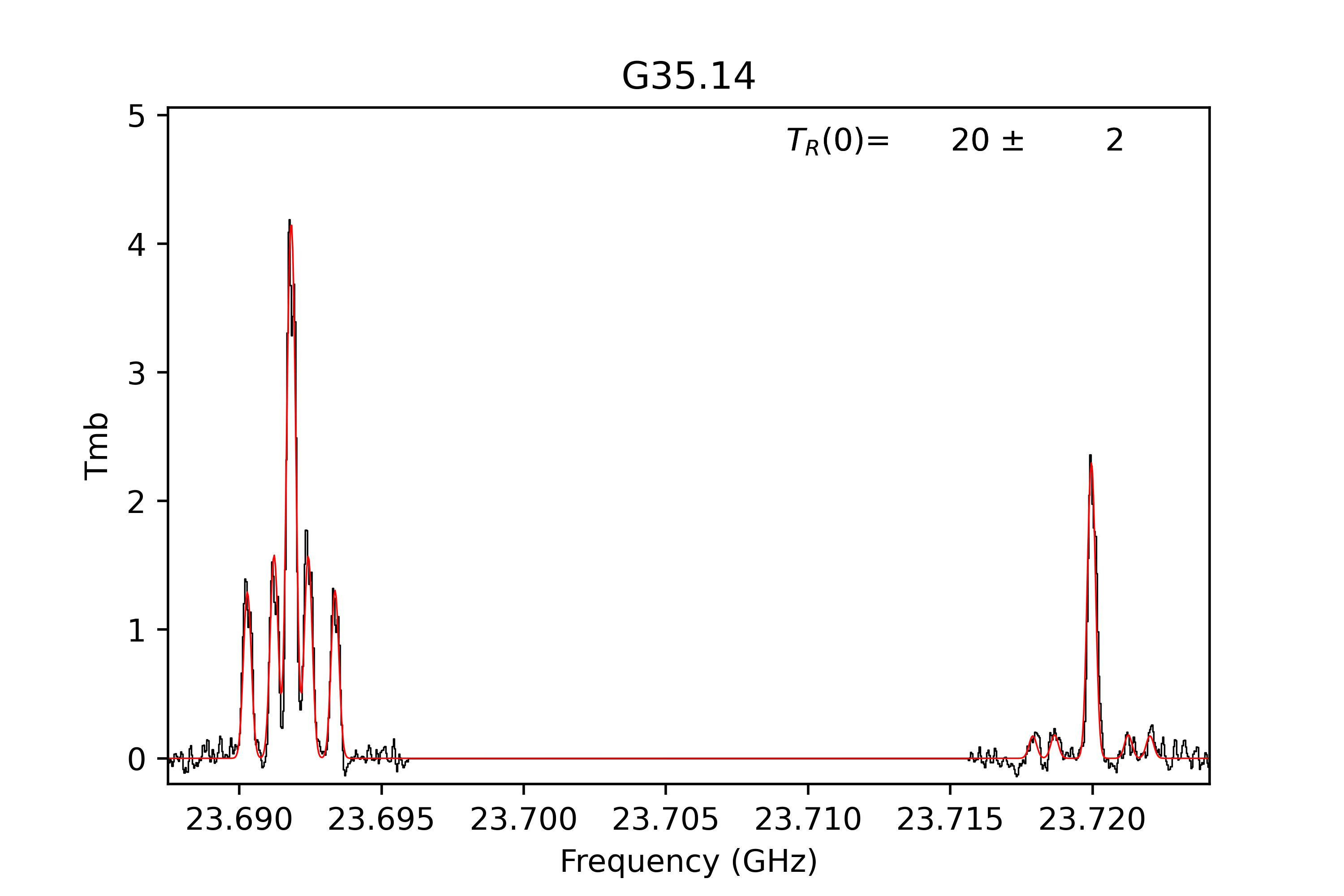}}
\end{figure*}
\begin{figure*}
	\centering	
	{\includegraphics[width=8cm]{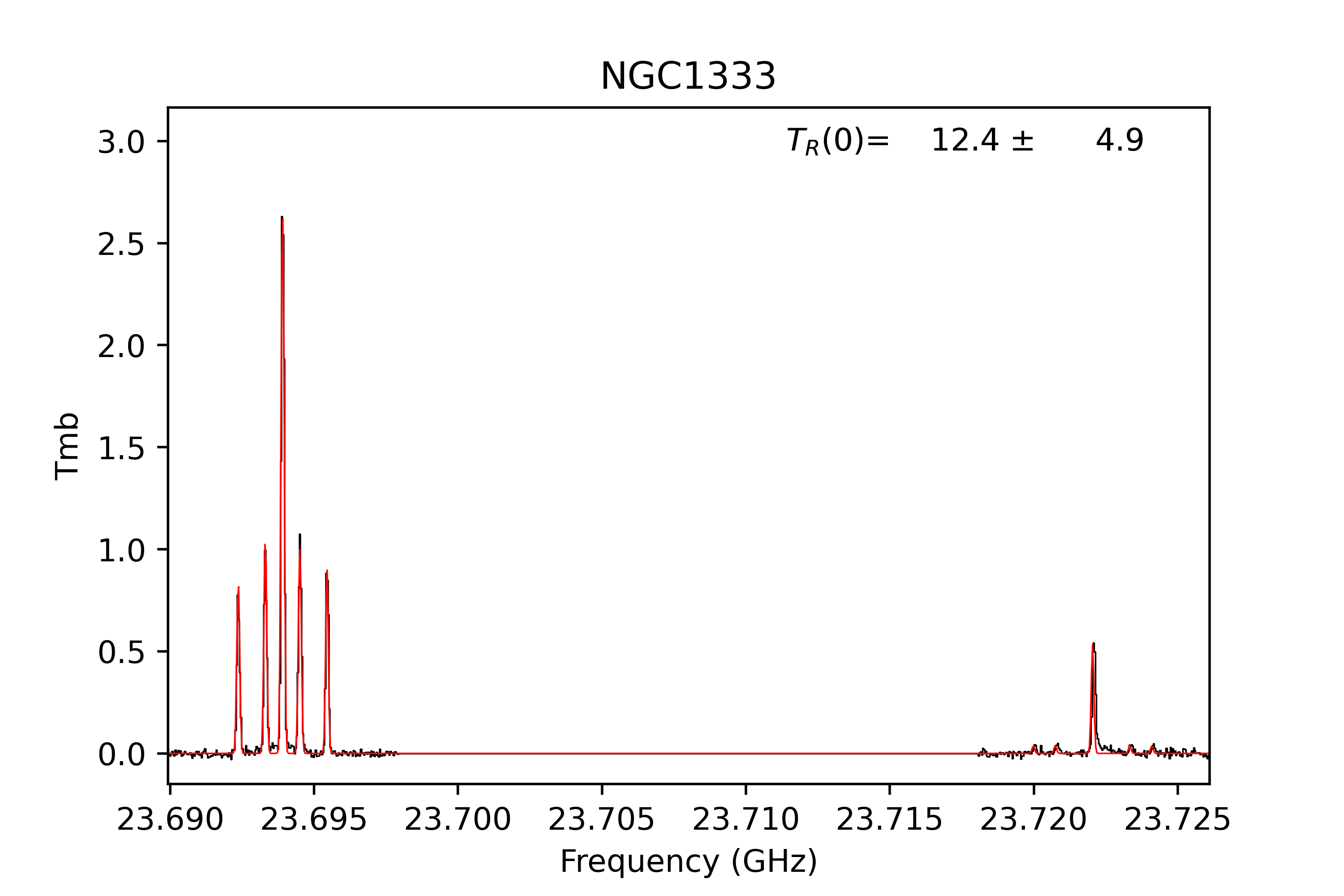}}
	{\includegraphics[width=8cm]{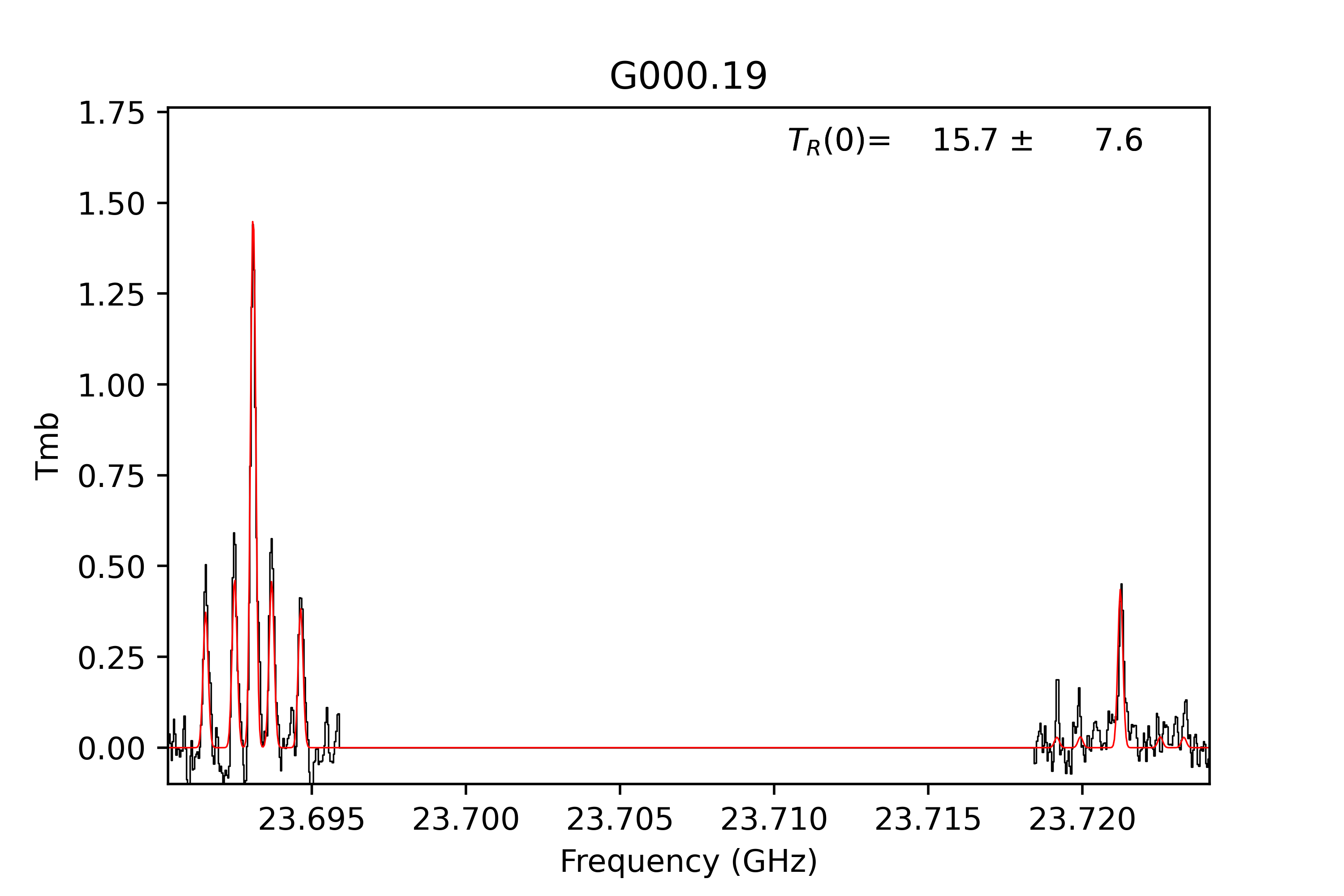}}
	{\includegraphics[width=8cm]{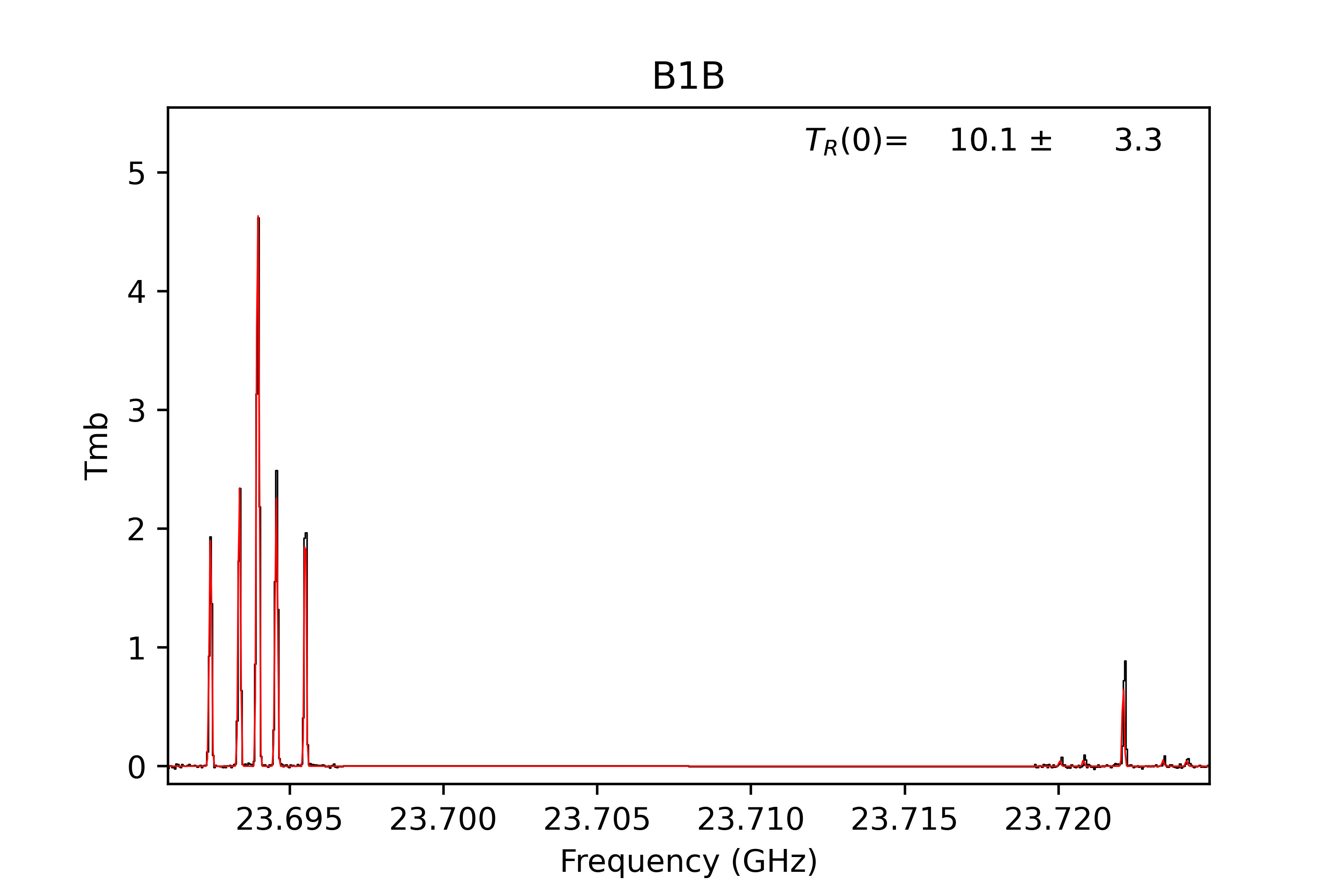}}
	\caption{Profiles on the left hand side of each spectrum represent $^{14}$NH$_{3}$(1, 1) and those on the right $^{14}$NH$_{3}$(2, 2) lines of 11 targets and their fitting lines (red) through the improved HF fitting method. $T_{rot}$ values are indicated at the top right corner of each panel. For three sources with blended hyperfine components in the spectra (G10.47, Orion-KL and W51D), we performed the fit by the HFGR method. }
	\label{HF fitting figure}
\end{figure*}

\textbf{The kinetic temperature}: The conversion of the rotational temperature ($T_{rot}$) into the gas kinetic temperature ($T_{k}$) is a critical part of the NH$_{3}$ inversion line analysis. 
\citet{Tafalla04} provided a detailed $T_{k}$ analysis in their Monte Carlo models using the collision coefficients of \citet{Danby88} and $T_{rot}$ results from an NH$_{3}$ line analysis and derived an expression for accurate gas temperature estimates: 
\begin{equation}
T_{k}=\frac{T_{rot}}{1-\frac{T_{rot}}{\Delta E}ln[1+1.1exp(-\frac{16}{T_{rot}})]}.
\end{equation}
Using this relation between $T_{k}$ - $T_{rot}$ and $T_{rot}$ results derived from the above analysis, we calculated the kinetic temperature values for the sample, which are given in column 12 of Table \ref{tab:tau and T}.


\subsubsection{Column density}\label{sec:column}
As shown in the rotation diagram analysis in Sec. 3.2.2, with the assumption of LTE, the $T_{rot}$ values were obtained from the $^{14}$NH$_{3}$(1, 1) and (2, 2) line intensities and corresponding optical depths for our sample.
According to the derived $T_{rot}$ values and formulae (7) or (8), we can further use the (1, 1) or (2, 2) line intensities and corresponding parameters of $g_{u}$, Q($T_{rot}$) and $E_{u}$ to determine opacity-corrected total column densities of $^{14}$NH$_{3}$.  

For $^{15}$NH$_{3}$, we can carry out a similar analysis to determine the total column density for our sample, but in this case, with $^{14}$N/$^{15}$N commonly in excess of 100, we can realistically assume that all lines are optically thin.
We calculated the total column density of $^{15}$NH$_{3}$ from the formulae (7) or (8), assuming the same $T_{rot}$ value for $^{15}$NH$_{3}$ as for $^{14}$NH$_{3}$ and adopting the spectroscopic parameters of the $^{15}$NH$_{3}$ molecular species (see Table \ref{tab:para}). 
For those 2 sources (NGC\,6334\,I and G10.47) with detections in both $^{15}$NH$_{3}$(1, 1) and $^{15}$NH$_{3}$(2, 2), we took the $^{15}$NH$_{3}$ total column density results from the $^{15}$NH$_{3}$(2, 2) line for later analysis, due to the better quality of the $^{15}$NH$_{3}$(2, 2) spectra with respect to those of the (1, 1) lines. 
For Orion-KL with high quality spectra in both transitions, we also used the rotation diagram method to determine $T_{rot}$ (see Figure \ref{14NH3Trotfigure}) and further determined its total column density. 
The results of the total column density without opacity corrections and the opacity-corrected values of both $^{14}$NH$_{3}$ and $^{15}$NH$_{3}$ of 15 sources are listed in Table \ref{tab:Total_N}.

\clearpage
\startlongtable
\renewcommand\tabcolsep{1.0pt} 
\begin{deluxetable}{lcccccccccccc}
	\tablecaption{Total $^{14}$NH$_{3}$ and $^{15}$NH$_{3}$ column densities without and with opacity corrections and their ratios\label{tab:Total_N}}
	\tablehead{\colhead{Object}	&	\colhead{Telescope}	&\colhead{N$_{t}$($^{14}$NH$_{3}$)}		&\colhead{N$_{t}$($^{15}$NH$_{3}$)}			&\colhead{N$_{t}^{Corr}$($^{14}$NH$_{3}$)}			&\colhead{N$_{t}^{Corr}$($^{15}$NH$_{3}$)}			&\colhead{$\frac{^{14}NH_{3}}{^{15}NH_{3}}$}				&\colhead{$\frac{^{14}NH_{3}}{^{15}NH_{3}}^{Corr}$}			&\colhead{$D_{sun}$}	&\colhead{$D_{\rm GC}$}						&\colhead{Classi-	}		&\colhead{Ref.}&\colhead{Notes}			\\
		&		&\colhead{cm$^{-2}$}			&\colhead{cm$^{-2}$}			&\colhead{cm$^{-2}$}			&\colhead{cm$^{-2}$}			&			&			&\colhead{(kpc)}			&\colhead{(kpc)}			&\colhead{fication}			&	
	}
\colnumbers
	\startdata
	G000.19	&	Effelsberg	&	5.3E+14	&	2.2E+13	&	9.9E+14	&	2.5E+13	&	24 	(10)	&	40 	(13)	&8.4(0.2)	&	0.31(0.15)	&		YSO&Par18&	1	\\
	G10.47	&	Effelsberg	&	2.5E+15	&	3.2E+14	&	6.1E+15	&	4.5E+14	&	8 	(5)	&	13 	(6)	&	8.25(0.11)	&	1.53(0.15)	&	UCH II&Wyr96&	2	\\
	G30.70	&	TMRT	&	4.3E+15	&	4.2E+13	&	1.1E+16	&	4.4E+13	&	102 	(29)	&	253 	(60)	&	5.0(0.7)&	4.6(0.4)	
		&	YSO	&Urq18&	1	\\
	G032.04	&	TMRT	&	2.0E+15	&	4.3E+13	&	4.8E+15	&	5.0E+13	&	46 	(21)	&	97 	(31)	&	5.2(0.5)&	4.8(0.3)		&	YSO	&Coo13&	1	\\
	W51D	&	TMRT	&	3.9E+15	&	3.2E+13	&	1.1E+16	&	3.6E+13	&	121 	(46)	&	295 	(81)	&	5.5(0.4)&	6.2(0.3)		&	YSO	&God15&	1	\\
	&		&		&	9.5E+13	&		&	2.0E+14	&	41 	(25)	&	54 	(27)	&		&		&		&&	2	\\
	&	Effelsberg	&	1.5E+15	&	2.2E+13	&	3.7E+15	&	2.2E+13	&	69 	(42)	&	166 	(62)	&		&		&		&&	1	\\
	&		&		&	4.7E+13	&		&	5.7E+13	&	33 	(56)	&	64 	(62)	&		&		&		&& 2	\\
	G016.92	&	Effelsberg	&	5.4E+14	&	2.5E+13	&	9.8E+14	&	2.7E+13	&	21 	(10)	&	35 	(13)	&	1.81(0.13)&	6.44(0.11)		&	H II	&Urq11&	1	\\
	G35.14	&	Effelsberg	&	2.5E+15	&	2.9E+13	&	4.4E+15	&	3.1E+13	&	86 	(39)	&	143 	(50)	&	2.2(0.2)&	6.5(0.3)		&	IRDC	&Den84&	1	\\
	NGC\,6334\,I	&	TMRT	&	1.2E+16	&	8.4E+13	&	2.5E+16	&	8.2E+13	&	145 	(45)	&	301 	(78)		&	1.34(0.11)&	6.9(0.4)	&	IRDC	&Wil13&	1	\\
	&		&		&	3.1E+14	&		&	6.0E+14	&	39 	(9)	&	48 	(11)	&		&		&		&&	2	\\
	G053.23	&	TMRT	&	1.0E+15	&	6.2E+12	&	2.3E+15	&	7.5E+12	&	167 	(143)	&	314 	(172)	&8.3(0.6)	&	7.3(0.5)	&		YSO	&Urq18&	1	\\
	G081.75	&	TMRT	&	1.8E+15	&	6.9E+12	&	3.5E+15	&	7.5E+12	&	259 	(171)	&	467 	(213)	&	2.35(0.12)&	8.13(0.14)		&	YSO	&Mau15&	1	\\
	&	Effelsberg	&	1.4E+15	&	4.2E+12	&	2.8E+15	&	4.6E+12	&	321 	(157)	&	603 	(213)	&		&		&		&&	1	\\
	NGC\,1333	&	Effelsberg	&	6.9E+14	&	4.8E+12	&	1.4E+15	&	5.7E+12	&	142 	(75)	&	247 	(96)		&	0.21(0.17)&	8.3(0.4)	&	YSO	&Lis10&	1	\\
	Barnard-1b	&	Effelsberg	&	1.3E+15	&	1.7E+13	&	5.8E+15	&	2.5E+13	&	77 	(32)	&	229 	(62)	&0.33(0.15)	&	8.4(0.5)	&		IRDC	&Lis10&	1	\\
	Orion-KL	&	TMRT	&	5.9E+15	&	5.6E+13	&	2.1E+16	&	7.7E+13	&	105 	(39)	&	270 	(72)	&	0.45(0.12)&	8.5(0.6)		&	H II	&Kim08&	1	\\
	&		&		&	1.5E+14	&		&	1.5E+14	&	39 	(10)	&	136 	(45)	&		&		&		&&	$^{15}$NH$_{3}$	\\
	&	Effelsberg	&	7.9E+15	&	9.8E+13	&	3.1E+16	&	1.4E+14	&	80 	(20)	&	212 	(46)	&		&		&		&&	1	\\
	&		&		&	1.4E+14	&		&	1.4E+14	&	55 	(12)	&	215	(47)	&		&		&		&&	$^{15}$NH$_{3}$	\\
	G121.29	&	TMRT	&	1.4E+15	&	1.4E+13	&	2.4E+15	&	1.5E+13	&	100 	(60)	&	159 	(72)	&	1.7(1.3)&	9(2)		&	IRDC	&Ryg10&	1	\\
	G188.79	&	Effelsberg	&	2.3E+14	&	7.6E+12	&	2.5E+14	&	7.7E+12	&	30 	(17)	&	33 	(17)	&	2.14(0.12)&	10.3(1.2)		&	YSO	&Cut03&	1	\\
	\enddata
	\tablecomments{Column(1): source name; Column(2): used Telescope; Columns (3) - (4): $^{14}$NH$_{3}$ and $^{15}$NH$_{3}$ column densities neglecting opacity corrections; Columns (5): column densities N$_{t}^{Corr}$($^{14}$NH$_{3}$) accounting for opacity effects; Columns (6): the corrected column density N$_{t}^{Corr}$($^{15}$NH$_{3}$) was obtained with the assumption of the same $T_{rot}$ as $T_{rot}$($^{14}$NH$_{3}$), which can be derived taking into account the optical depth correction on $^{14}$NH$_{3}$ in the rotation diagram method; Column (7): the ratios of the column densities neglecting opacity corrections. Errors (in parentheses) include standard deviations from the line fitting procedure; Column (8): opacity corrected values of $^{14}$NH$_{3}$/$^{15}$NH$_{3}$; Column (9): heliocentric distance with error, from the Parallax-Based Distance Calculator; Column (10): galactocentric distance with error from the Heliocentric distance;  Column (11): source classification. IRDC: InfraRed Dark Cloud; YSO: Young Stellar Object; H II: associated with an HII region; UCH II: associated with an ultra compact H II region; Column (12): references for the classification from the literature. Par18: \citet{Parsons18}; Wyr96: \citet{Wyrowski96}; Urq18: \citet{Urquhart18}; Coo13: \citet{Cooper13}; God15: \citet{Goddi15}; Urq11: \citet{Urquhart11}; Wil13: \citet{Willis13}; Mau15: \citet{Maud15}; Lis10: \citet{Lis10}; Kim08: \citet{Kim08}; Ryg10: \citet{Rygl10}; Cut03: \citet{Cutri03}; Den84: \citet{Dent84};
	Notes 1: the total column densities of $^{15}$NH$_{3}$ were obtained from the $^{15}$NH$_{3}$(1, 1) line intensity, assuming the same $T_{rot}$ value for $^{15}$NH$_{3}$ as for $^{14}$NH$_{3}$; 
	Notes 2: the total column densities of $^{15}$NH$_{3}$ were obtained from the $^{15}$NH$_{3}$(2, 2) line intensity, assuming the same $T_{rot}$ value for $^{15}$NH$_{3}$ as for $^{14}$NH$_{3}$;  
	Notes $^{15}$NH$_{3}$: the total column densities of $^{15}$NH$_{3}$ were obtained using the $T_{rot}$ of $^{15}$NH$_{3}$ from its $^{15}$NH$_{3}$(1, 1) and (2, 2) line intensities.}
\end{deluxetable}
\subsection{Measured abundance ratios}\label{sec:ratio}

\begin{figure}
	\centering  
	\subfigure[]{
		\includegraphics[width=14cm]{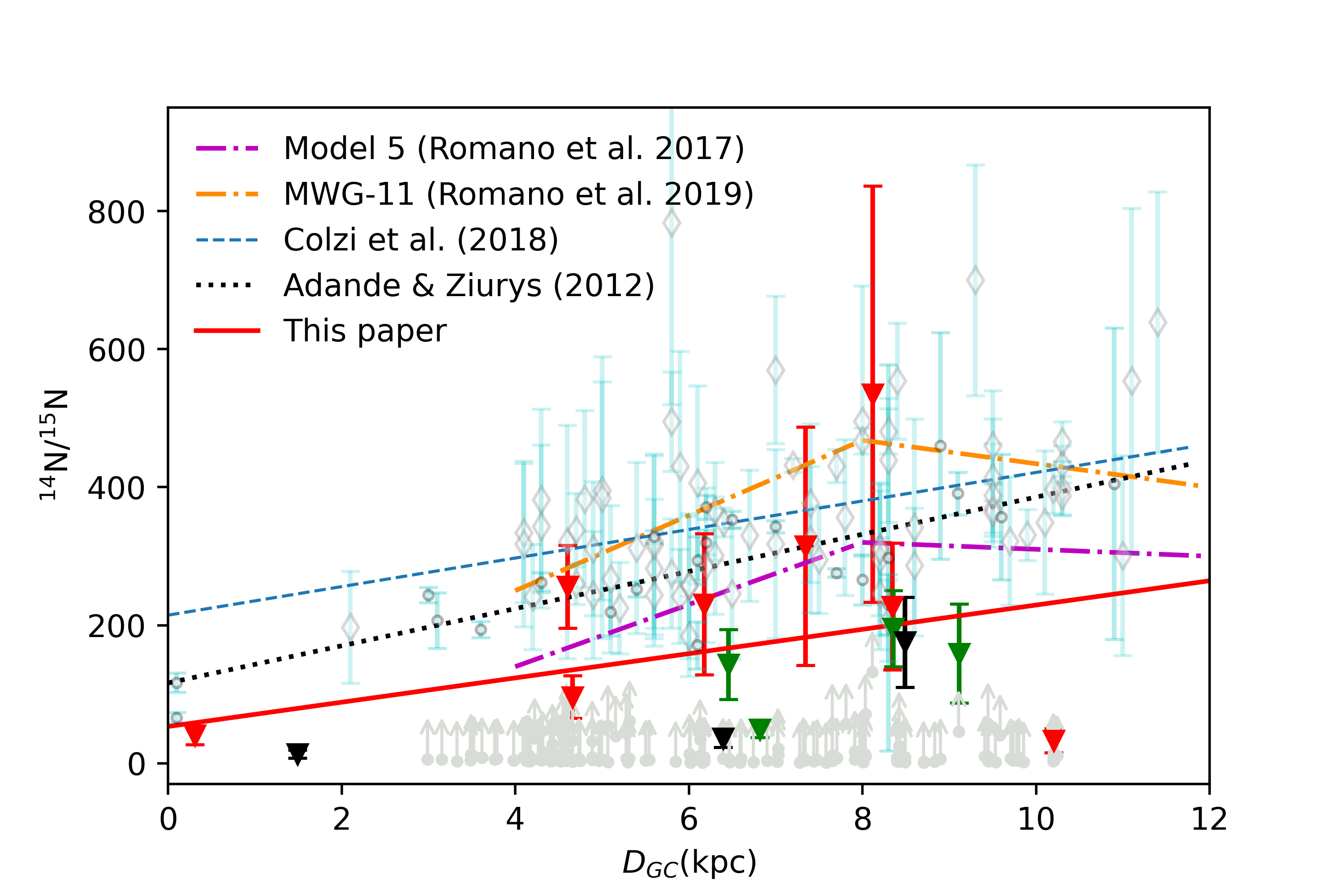}}
	\subfigure[]{
		\includegraphics[width=14cm]{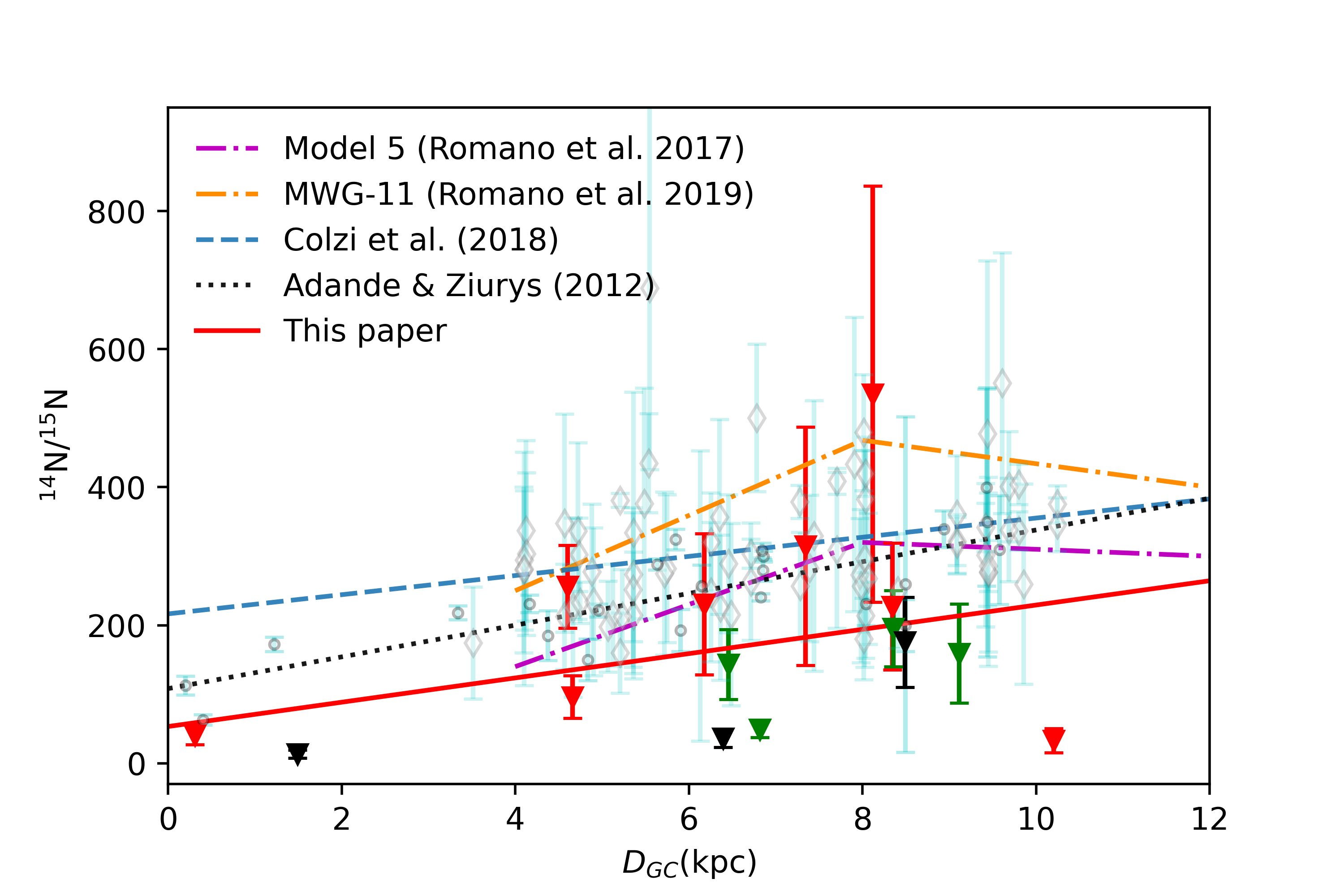}}
	\caption{
		The $^{14}$NH$_{3}$/$^{15}$NH$_{3}$ isotope ratio is plotted as a function of galactocentric distance for our measurements. Our ratios of the opacity corrected total column densities from NH$_{3}$ are reproduced by filled inverted triangles. Green, red and black inverted triangles represent sources at different stages of evolution, i.e.,  IRDCs, YSOs, and regions associated with HII regions, respectively. The red solid line presents the linear fit (no weighting), $\frac{^{14}{\rm NH}_{3}}{^{15}{\rm NH}_{3}} =  (17.50\pm13.14) D_{\rm GC} + (53.91 \pm 91.74)$. Small grey points with arrows denote lower 3$\sigma$ limits of $^{14}$NH$_{3}$/$^{15}$NH$_{3}$ for our sources not detected in $^{15}$NH$_{3}$ (Fig. \ref{14N15N}a). 
		The magenta and yellow dash–dotted line predictions were taken from the most recent Galactic chemical evolution model, that of \citet{Romano17,Romano19}. \textbf{The previous results from HNC \citep{Colzi18a} and CN, and HCN measurements \citep{Dahmen95, Adande12} are shown as grey diamonds and the grey black empty circles, respectively. The linear fit is presented by the dashed line and the dotted line, respectively (Fig. \ref{14N15N}a). Taking the most recent $^{12}$C/$^{13}$C ratios \citep{Yan19} and updating the distances with more recent trigonometric parallax measurements \citep{Reid14,Reid19}, their modified results with the linear fit are plotted together with our results in Fig. \ref{14N15N}b.} Compared with Fig. \ref{14N15N}a, the difference between our results and the modified ratios from other studies become smaller but does not vanish.}
	\label{14N15N}
\end{figure}

As shown in Sect. \ref{sec:column}, we obtained the column densities of $^{14}$NH$_{3}$ and $^{15}$NH$_{3}$ of 15 sources. Based on these results, we estimate the $^{14}$N/$^{15}$N isotope ratios (see Table \ref{tab:Total_N}). 
For those 2 sources (NGC\,6334\,I and G10.47) with $^{15}$NH$_{3}$ (2, 2) detection, the results from their NH$_{3}$(2, 2) lines (with higher quality than their (1, 1) lines, see Sect. \ref{sec:detection}) are used for later analysis because signal-to-noise ratios are higher.
For those sources measured by both the Effelsberg and TMRT telescope (G081.75, W51\,D and Orion-KL), the mean value of their $^{14}$NH$_{3}$/$^{15}$NH$_{3}$ ratios was taken.
For those sources without detection of $^{15}$NH$_{3}$, we estimate the lower limit of $^{14}$NH$_{3}$/$^{15}$NH$_{3}$,
according to the peak temperature of $^{14}$NH$_{3}$ and the 3 {\it rms} value of the $^{15}$NH$_{3}$ line (in grey points with arrows in Figure \ref{14N15N}a). 
The lower limit is mostly around 11, with a mean value of 13, which is smaller than all ratios derived from our 15 detections.
A comparison with previous studies is presented in Sect. \ref{sec:compar} and possible contaminating effects affecting the abundance ratios are discussed in Sect. \ref{sec:obseffect} and \ref{sec:fract}.

\section{Discussion}\label{sec:discussion}

\subsection{Comparisons with previous studies}\label{sec:compar}

Among 15 sources with measured $^{14}$N/$^{15}$N ratio, five sources (Orion-KL, Barnard-1b, NGC\,1333, W51\,D and G000.19 in the Galactic center region) were also measured in previous studies, which therefore provide $^{14}$N/$^{15}$N abundance ratios from a variety of molecular species, namely CN, HNC, HCN, NH$_{2}$D, N$_{2}$H$^{+}$. 
Comparisons show that measured ratios considering opacity effects are basically consistent with previous results within the uncertainties (see details in Table \ref{tab:Compare}).

\textbf{Orion-KL}: This source has been observed in different species to measure the isotope ratio $^{14}$N/$^{15}$N. Assuming LTE conditions, \citet{Hermsen85} derived a $^{14}$NH$_{3}$/$^{15}$NH$_{3}$ ratio of 170$_{-80}^{+140}$. Subsequently, \citet{Adande12} performed observations of HN$^{13}$C, H$^{15}$NC, CN and C$^{15}$N toward this source. They obtained a $^{14}$N/$^{15}$N ratio of 234 $\pm$ 47, from the ratio of the brightness temperatures of the strongest hyperfine components of CN and C$^{15}$N, weighted by the hyperfine relative intensities. From their observations of HN$^{13}$C and H$^{15}$NC using the double isotope method, they got a reasonably consistent ratio of 159 $\pm$ 40. Recently, a relatively lower value of 100 $\pm$ 51 was reported from NH$_{3}$ observations but without considering optical depth corrections \citep{Gong15}. The measurements of this source are, within the uncertainties, consistent with our result of 241 $\pm$ 71.

\textbf{Barnard-1b, NGC\,1333}: The $^{14}$N/$^{15}$N of  Barnard-1b, NGC\,1333 was measured by \citet{Gerin09} and \citet{Lis10}, from observations of NH$_{2}$D and NH$_{3}$, respectively. The results from these two tracers are consistent within uncertainties, with $^{14}$N/$^{15}$N = 334 $\pm$ 50 and 344 $\pm$ 173 \citep[from NH$_{3}$, ][]{Lis10} and 470$^{+170}_{-100}$ and 360$^{+260}_{-110}$ \citep[from NH$_{2}$D, ][]{Gerin09} for Barnard-1b and NGC\,1333. 
For Barnard-1b, the isotope ratio from NH$_{2}$D in \citet{Gerin09} appears to be larger than that from NH$_{3}$ in \citet{Lis10}. \citet{Daniel13} performed observations of multi-tracers (NH$_{3}$, NH$_{2}$D, CN, HCN, N$_{2}$H$^{+}$) to investigate its $^{15}$N-fractionation. Assuming non-LTE conditions, they got similar $^{14}$N/$^{15}$N abundance ratios for all the tracers, independent of the chemical family. And they found a strong dependence of the column density of $^{15}$NH$_{2}$D on the excitation temperature. Using the same observational data of NH$_{2}$D from \citet{Gerin09}, they made a model analysis to obtain the excitation temperature of $^{15}$NH$_{2}$D, instead of assuming the same excitation temperature for $^{15}$NH$_{2}$D and $^{14}$NH$_{2}$D \citep{Gerin09}. Therefore they got a relative accurate $^{14}$N/$^{15}$N value of 230$^{+105}_{-55}$  for this source, with respect to 470$^{+170}_{-100}$ in \citet{Gerin09}. 
This is consistent with our ratio of 229 $\pm$ 62 from the rotation diagram method (LTE), which should reflect non-significant non-LTE effects in our analysis.

\textbf{W51D}: Using Effelsberg data of 13 emission lines of NH$_{3}$ for a rotation diagram analysis (LTE), \citet{Mauersberger87} derived a $^{14}$N/$^{15}$N result of 660 $\pm$ 300 toward W51D, which is larger than our result of 230 $\pm$ 102 from both Effelsberg and TMRT data using the rotation diagram method.
This large difference should be caused by the fact that many NH$_{3}$ lines from levels with high energy above the ground levels were used in their analysis, instead of only the metastable (1, 1) and (2, 2) lines of NH$_{3}$ as in our analysis. Using only the metastable (1, 1) and (2, 2) lines of NH$_{3}$ and $^{15}$NH$_{3}$ in \citet{Mauersberger87}, we performed a consistent analysis and and got smaller $T_{rot}$ and $^{14}$NH$_{3}$/$^{15}$NH$_{3}$ values of 24 $\pm$ 4.1 and 95 $\pm$ 36, which are consistent with our new results. 
In addition, the $^{14}$NH$_{3}$/$^{15}$NH$_{3}$ ratio also depends critically on the population of the NH$_{3}$ non-metastable ammonia levels \citep{Mauersberger87}.
The transition lines from high levels with different excitation conditions should trace denser regions \citep{Goddi15}. 
As our other sources, the measurements from the (1, 1) and (2, 2) lines, presumably representing the bulk of the gas due to their low excitation, are taken for our later analysis.


\textbf{Galactic center region}: As mentioned before (see Sect. \ref{sec:intro}), the only direct measurements towards the Galactic center region obtained a very large value of $^{14}$N/$^{15}$N of $\sim$1000 \citep{Gusten85}, while extrapolations of the trend with galactocentric distances, extrapolated from the disk, indicate much lower values \citep{Adande12,Colzi18a}. Recently, \citet{Mills18} performed VLA mapping on $^{14}$NH$_{3}$ and $^{15}$NH$_{3}$ toward Sgr B2 (N) and measured $^{14}$N/$^{15}$N ratios of $\sim$450 and a lower value of $\sim$200 for the resolved two hot cores N1 and N2, respectively. Our observations toward G000.19, about 30\arcmin away from Sgr A, provide a $^{14}$N/$^{15}$N value of $\sim$40, which is much lower than previous results toward the Galactic center. Actually, non-uniform ratios for other isotopes were reported toward the Galactic center region. \citet{Zhang15} mapped typical molecular clouds (6 sources including Sgr A, Sgr B2, Sgr C and Sgr D) in the $J=1-0$ lines of C$^{18}$O and C$^{17}$O and obtained different ratios of $^{18}$O/$^{17}$O toward those sources, while all their ratios are lower relative to molecular clouds in the Galactic disk. It indicates chemical differentiation of the region, either due to a different origin of the gas, due to different degrees of nuclear processing inside the central molecular zone or due to fractionation effects \citep{Zhang15,Loison19}.
In addition, low isotope ratio values of $^{12}$C/$^{13}$C ($\sim$13) were reported recently toward Orion-KL and other star formation regions from MIR observations data, which is believed to be not biased by chemical effects. As previously mentioned, the Galactic center region is not covered in current Galactic chemical evolution models  \citep[e.g., ][]{Romano17,Romano19}. 
Unlike other isotope ratios all reporting low (though non-uniform) ratios in the Galactic center region, both high and low $^{14}$N/$^{15}$N ratios may be found in this region. This makes nitrogen "special" in this sense: it could imply strong effects due to both nucleosynthesis and chemical fractionation, in spite of the rather large kinetic temperatures in the Galactic center region \citep[e.g. ][]{Ginsburg16}, which needs more measurements and modeling work.

\begin{deluxetable}{llccccc}
	
	\tablecaption{Comparisons with $^{14}$N/$^{15}$N ratios from the literature\label{tab:Compare}}

	\tablehead{	\colhead{Object}	&	\colhead{Species}	&	\colhead{$\alpha$(2000)}	&	\colhead{$\delta$(2000)}	&	\colhead{$\frac{^{14}N}{^{15}N}$}	&	\colhead{Beam size}	&	\colhead{References}		
	}
\colnumbers
	\startdata
	Orion-KL	&	NH$_{3}$	&	05:35:14	&	-05:22:29	&	241 $\pm$ 71	&	40\arcsec	&	2, This paper	\\
	&	NH$_{3}$	&	05:35:14	&	-05:22:29	&	100 $\pm$ 51	&	40\arcsec	&	25, \citet{Gong15}	\\
	&	NH$_{3}$	&	05:35:14	&	-05:22:46	&	170$_{-80}^{+140}$	&	40\arcsec	&	5, \citet{Hermsen85}	\\
	&	HNC	&	05:32:46	&	-05:24:23	&	159 $\pm$ 40	&	63\arcsec	&	 \citet{Adande12}	\\
	&	CN	&	05:32:46	&	-05:24:23	&	234 $\pm$ 47	&	63\arcsec	&	 \citet{Adande12}	\\
	Barnard-1b	&	NH$_{3}$	&	03:33:20	&	+31:07:34	&	229 $\pm$ 62	&	40\arcsec	&	2, This paper	\\
	&	NH$_{3}$	&	03:33:20	&	+31:07:34	&	334 $\pm$ 50	&	33\arcsec	&	2, \citet{Lis10}	\\
	&	NH$_{3}$	&	03:33:20	&	+31:07:34	&	300 $\pm$ 50	&	33\arcsec	&	2, \citet{Daniel13}	\\
	&	NH$_{2}$D	&	03:33:20	&	+31:07:34	&	230$_{-55}^{+105}$	&	29\arcsec	&	\citet{Daniel13}	\\
	&	CN	&	03:33:20	&	+31:07:34	&	290$_{-80}^{+160}$	&	21\arcsec	&	\citet{Daniel13}	\\
	&	HCN	&	03:33:20	&	+31:07:34	&	330$_{-50}^{+60}$	&	29\arcsec	&	\citet{Daniel13}	\\
	&	N$_{2}$H$^{+}$	&	03:33:20	&	+31:07:34	&	400$_{-60}^{+100}$	&	27\arcsec	&	\citet{Daniel13}	\\
	&	HNC	&	03:33:20	&	+31:07:34	&	225$_{-45}^{+75}$	&	28\arcsec	&	\citet{Daniel13}	\\
	&	NH$_{2}$D	&	03:33:20	&	+31:07:34	&	470$_{-100}^{+170}$	&	29\arcsec	&	\citet{Gerin09}	\\
	NGC\,1333	&	NH$_{3}$	&	03:29:11	&	+31:13:26	&	247 $\pm$ 95	&	40\arcsec	&	2, This paper	\\
	&	NH$_{3}$	&	03:29:11	&	+31:13:26	&	344 $\pm$ 173	&	33\arcsec	&	2, \citet{Lis10}	\\
	&	NH$_{2}$D	&	03:29:12	&	+31:13:25	&	360$_{-110}^{+260}$	&	240\arcsec	&	\citet{Gerin09}	\\
	W51\,D	&	NH$_{3}$	&	19:23:39	&	+14:31:07	&	230 $\pm$ 102	&	40\arcsec	&	2, This paper	\\
	&	NH$_{3}$	&	19:23:39	&	+14:31:10	&	660 $\pm$ 300$^{(a)}$	&	40\arcsec	&	6, \citet{Mauersberger87}	\\
	&	NH$_{3}$	&	19:23:39	&	+14:31:10	&	400 $\pm$ 200$^{(b)}$	&	40\arcsec	&	13, \citet{Mauersberger87}	\\
	G000.19	&	NH$_{3}$	&	17:47:52	&	-28:59:59	&	40 $\pm$ 13	&	40\arcsec	&	2, This paper	\\
	Sgr A 	&	NH$_{3}$	&	17:45:52	&	-28:59:59	&	$\sim$1000	&	40\arcsec	&	2, \citet{Gusten85}	\\
	 	&	HCN	&	17:45:52	&	-28:59:59	&	$\sim$510	&	124\arcsec	&	\citet{Wannier81}	\\
	Sgr B2	&	NH$_{3}$	&	17:47:19	&	-28:22.08	&	210 $\pm$ 90	&	33\arcsec	&	12, \citet{Mills18}	\\
	\enddata
	\tablecomments{Column(1): Source name; Column(2): Species; Column(3): Right ascension (J2000); Column(4): Declination (J2000); Column(5): Resulting nitrogen isotope abundance ratio; Column (6): Applied beam size of the telescope; Column(7): References and the number of transition (in case NH$_{3}$ has been used) that were considered for the determination of the column density. (a) and (b): the $^{14}$N/$^{15}$N ratio from the rotation diagram method without or with the populations of non-metastable levels.}
\end{deluxetable}

\subsection{Observational effects}\label{sec:obseffect}

\begin{figure}
	\centering  
	\includegraphics[width=8.5cm]{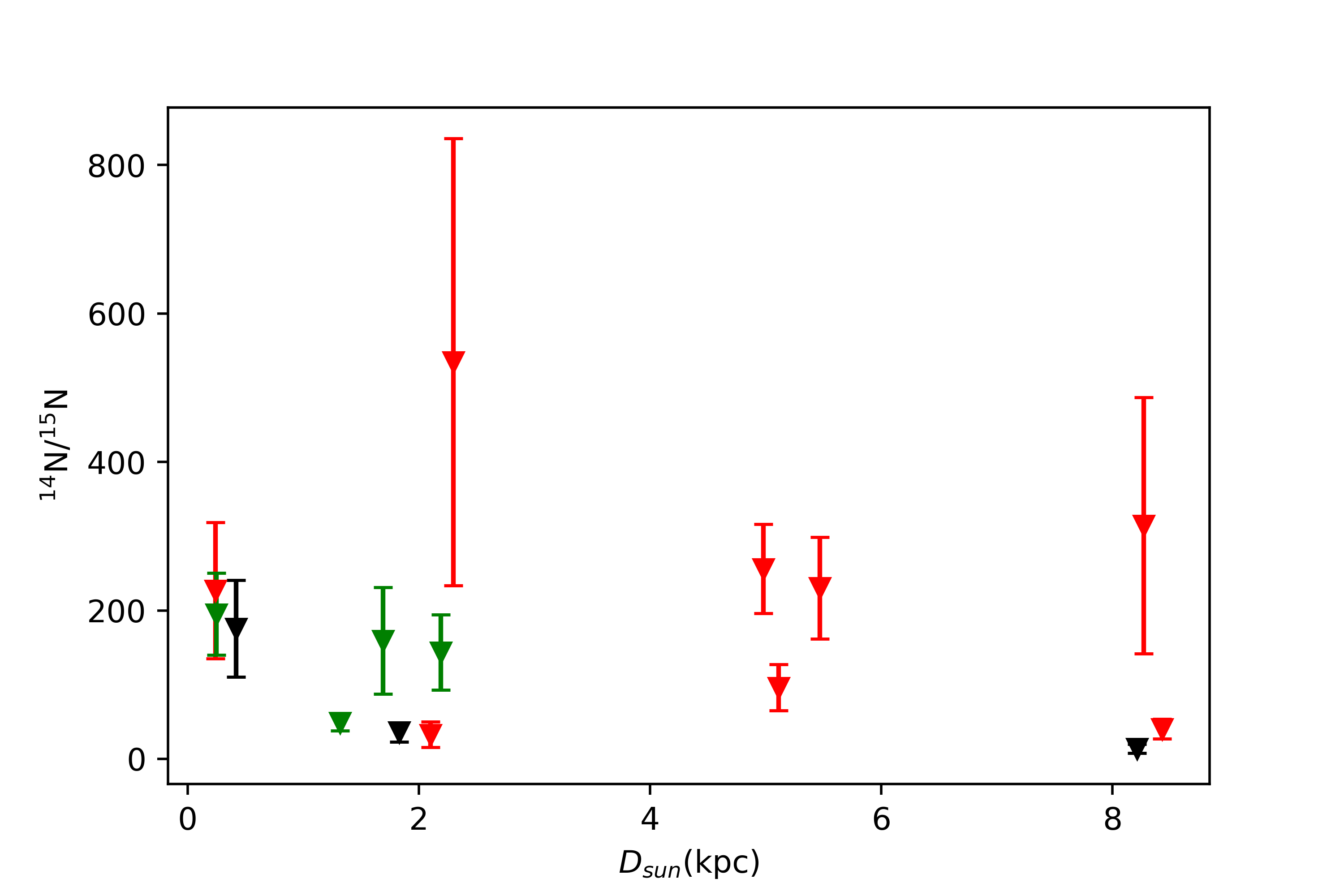}
	\caption{Our $^{14}$N/$^{15}$N isotope ratios from $^{14}$NH$_{3}$ and $^{15}$NH$_{3}$ are plotted against the heliocentric distance (different colours for sources at different evolutionary stages, as in Figures \ref{14N15N}).}
	\label{Dsun}
\end{figure}

Our observations may be biased emphasizing bright sources, with possibly systematically higher $^{14}$NH$_{3}$ opacities, which could lead to uncertain opacity corrections, when trying to determine $^{14}$NH$_{3}$/$^{15}$NH$_{3}$ ratios.
In addition, sources at different distance imply different linear beam sizes covered by the telescope. 
A larger linear size of sources at larger distances may include more relatively diffuse low-density gas of different kinetic temperature, which could affect the isotope ratio results. 
In order to assess possible observational effects on abundance ratios, we plot the abundance ratio against the heliocentric distance in Figure \ref{Dsun}. It shows no systematic dependence between the ratio and the distance, which indicates that observational bias related to beam dilution is not significant. This is supported by the comparison of abundance ratios for the three sources (Orion-KL, W51D and G081.75) detected by both telescopes, which gives consistent ratio values within uncertainties (see Table \ref{tab:Total_N}).

\subsection{Nitrogen fractionation}\label{sec:fract}

To determine accurate isotopic abundance ratios from observed $^{14}$NH$_{3}$/$^{15}$NH$_{3}$ line intensities, the possibility of chemical nitrogen fractionation should also be briefly discussed. 
Although there is a number of dedicated papers \citep{Rodgers08, Lis10, Roueff15, Colzi18b,Wirstrom18, Viti19, Loison19}, N-fractionation is still a matter of debate. 

The main mechanism assumed to cause nitrogen fractionation are isotope-exchange reactions. N isotope exchange reactions normally occur at low temperatures, with $^{15}$N enhancement in CO-depleted dense gas at low temperatures of \textless 10 K \citep{Adams81,Terzieva00,Charnley02,Rodgers08,Fontani15,Colzi18b,Loison19}. However, such reactions at low temperature should be inhibited by an entrance barrier and thus the $^{14}$N/$^{15}$N ratios do not change with time \citep{Roueff15,Wirstrom18}, which has also been demonstrated observationally  by \citet{Fontani15} and \citet{Colzi18b}.

Another possible mechanism for the N-fractionation, isotope selective photo-dissociation, was proposed by \citet{Heays14,Visser18,Furuya18}. $^{14}$N/$^{15}$N fractionation is believed to be predominantly caused by isotope-selective photodissociation of N$_{2}$ rather than isotope exchange reactions \citep{Furuya18}. 

\begin{figure}
	\centering  
	\includegraphics[width=8.5cm]{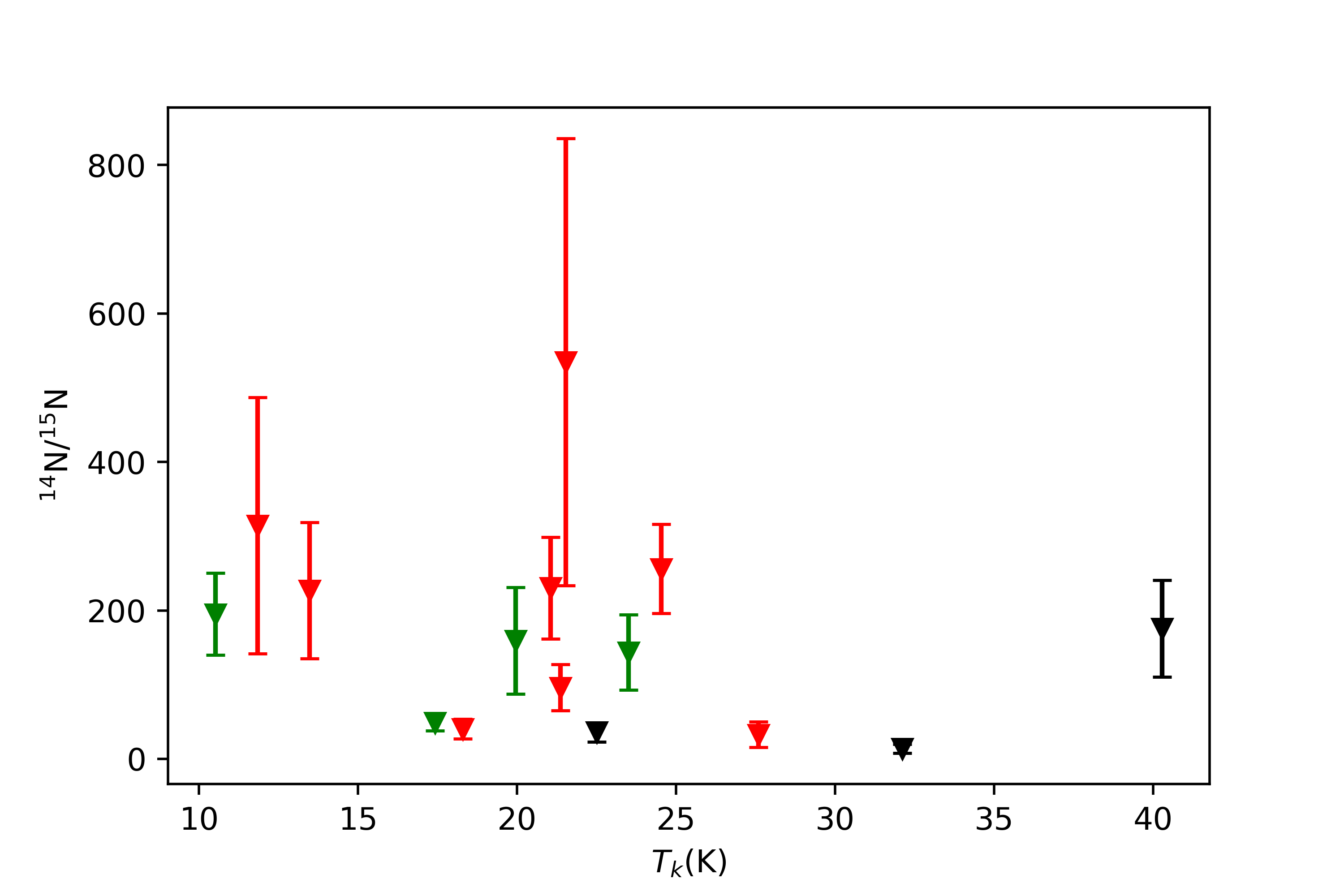}
	\caption{The $^{14}$NH$_{3}$/$^{15}$NH$_{3}$ ratio is plotted against gas kinetic temperature of our sample (different colours for sources at different evolutionary stages, as in Figures \ref{14N15N} and \ref{Dsun})}.
	\label{Tkin}
\end{figure}

All our sources have known kinetic temperatures larger than 10 K, which may imply that $^{14}$NH$_{3}$/$^{15}$NH$_{3}$ ratios are not seriously affected by the N-fractionation effect. The plot of $^{14}$NH$_{3}$/$^{15}$NH$_{3}$ against the kinetic temperature of sources (Figure \ref{Tkin}) shows no significant correlation, which may indicate that fractionation effects are not a decisive factor affecting measurements. 
However, nitrogen fractionation may be scale-dependent, possibly representing a local effect and observations with highly different beam sizes might provide different $^{14}$N/$^{15}$N values \citep{Colzi19}. Our measurements from single dish telescopes with a relatively larger beam size may include more relatively diffuse low-density gas, which could be affected by the interstellar radiation field. Observations with high resolution should be helpful to probe the N-fractionation effect in both the molecular cores and outskirts and to determine accurate ratio values of $^{14}$N/$^{15}$N.

\subsection{A Galactic interstellar $^{14}$N/$^{15}$N gradient?}\label{sec:gradient}

Figure \ref{14N15N} plots measured $^{14}$N/$^{15}$N isotope ratios from $^{14}$NH$_{3}$/$^{15}$NH$_{3}$ (inverted triangles) against galactocentric distance. Our measurement suggests that the isotope ratio increases with galactocentric distance. Our sources belong to different stages of massive star formation including 4 sources in IRDCs (green inverted triangles), 8 associated with YSOs (red inverted triangles) and 3 next to HII regions (black inverted triangles). 
Comparisons show that both measured ratios and the $^{14}$N/$^{15}$N gradient with galactocentric radius are independent of the evolutionary stage.  
An unweighted linear fit was used to fit data (red solid line), in order not to bias those results toward low values with small error bars. Our data provide a weak radial gradient of $\frac{^{14}{\rm NH}_{3}}{^{15}{\rm NH}_{3}} =  (17.50 \pm 13.14) D_{\rm GC} + (53.91 \pm 91.74)$, with a Pearson's rank correlation coefficient of R = 0.35\footnote{Pearson's rank correlation coefficient is defined in statistics as the measurement of the strength of the relationship between two variables and their association with each other. It indicates a weak correlation between 0.3 and 0.8.}. 

For comparison, previous measurements from HCN and/or HNC in \citet{Adande12} and \citet{Colzi18a} are added as grey empty circles and diamonds, respectively. 
Fits to both data sets are shown as dashed and dotted lines in Fig. \ref{14N15N}, respectively . 
We find that our ratios tend to be slightly smaller than those previous results, though a similar trend can be found. 
However, those results may suffer from uncertainties, using relatively early measurements of $^{12}$C/$^{13}$C \citep{Milam05}. 
The most recent results related to carbon isotope ratios, the only ones leading to a self-consistent interpretation of sulfur isotope ratios through the use of double isotope ratios involving $^{12}$C/$^{13}$C \citep{Yu20}, is reported by \citet{Yan19}. 
They presented observations of the K-doublet lines of H$_{2}$CO and H$_{2}^{13}$CO at the C ($\sim$5 GHz) and Ku ($\sim$15 GHz) bands toward a large sample of Galactic molecular clouds. 
Thus we modify those previous results \citep{Adande12,Colzi18a}, taking the most recent $^{12}$C/$^{13}$C ratios \citep{Yan19} and distance values from trigonometric parallax measurements \citep{Reid14,Reid19}. 
The modified previous results are plotted together with our results shown in Figure \ref{14N15N}b. 
We find that their ratios decrease by about 12 - 15\% and the difference between their and our results becomes smaller. This is visualized by the smaller gap between the two fitted (red solid and blue-dashed) lines. However, the difference does not vanish. This may suggest that (1) our approach using exclusively the lowest metastable inversion lines of ammonia leads to too small ratios, (2) that fractionation plays a role and/or (3) that the use of double isotope ratios adds uncertainties in the determination of nitrogen isotope ratios.

Theoretical models for Galactic chemical evolution (GCE) are important tools in understanding the isotopic ratio evolution in the Galaxy. 
Recently, new GCE models were developed to track the cosmic evolution of the CNO isotopes in the interstellar medium (ISM) of galaxies, yielding powerful constraints on their stellar initial mass function (IMF) \citep{Romano17,Zhang18,Romano19}. The theoretical $^{14}$N/$^{15}$N gradient across the Milky Way disk is shown in the magenta dash-dotted curves \citep[Model-5 in Figure 5 in ][]{Romano17} and the yellow dash-dotted curves \citep[Model-11 in Figure 6 in ][]{Romano19} in Figure \ref{14N15N}. Nucleosynthesis prescriptions in Model 5 \citep{Romano17} adopted the yields for low- and intermediate-mass stars, massive stars, super-AGB stars and nova, while different initial rotation velocities for low metallicity massive stars were also considered in Model-11 \citep{Romano19}. The trend of measured $^{14}$N/$^{15}$N isotope ratios increasing with galactocentric distance is consistent with predictions of both models. And it is interesting that measurements show a ”tentative indication” of the trend to decrease from 8 up to 10 kpc (but based on only 3 sources), which is similar to predictions from both models.
More data from the Galactic center and the sources with large distance (\textgreater 8kpc) as well as more Galactic disk values with smaller uncertainty would still be highly desirable to better constrain this gradient.

\section{Summary}\label{sec:summary}
To investigate the nitrogen abundance ratio varying across the Galaxy, we performed systematic observations of the ($J$, $K$) = (1, 1), (2, 2) and (3, 3) transitions of $^{14}$NH$_{3}$ and $^{15}$NH$_{3}$ toward a large sample of 210 sources in the Galactic disk with the TMRT - 65 m and Effelsberg - 100 m telescopes. 
Through TMRT observations, a total of 141 objects were detected in the $^{14}$NH$_{3}$ lines. 
8 out of them were detected in $^{15}$NH$_{3}$.
In order to detect $^{15}$NH$_{3}$ lines in more sources, 36 sources with strong $^{14}$NH$_{3}$ signals were selected to be observed by the Effelsberg-100 m telescope. The $^{15}$NH$_{3}$(1, 1) line was detected in 10 sources, including 3 sources (NGC\,6334\,I, Orion-KL and W51D) also detected by the TMRT. Thus, 15 sources were detected in the $^{15}$NH$_{3}$(1, 1) line and 4 among them were also detected in $^{15}$NH$_{3}$(2, 2) and $^{15}$NH$_{3}$(3, 3). Our results include:

1) Physical parameters of the gas emitting ammonia lines for these 15 sources with detections of NH$_{3}$ and $^{15}$NH$_{3}$ are determined from their spectral data, including optical depths, rotation and kinetic temperatures and total column densities. The opacity-corrected total column densities of $^{14}$NH$_{3}$ and $^{15}$NH$_{3}$ are used to estimate their $^{14}$N/$^{15}$N ratio.

2) An observational bias due to bright sources and/or effects related to different linear beam sizes is not found for our measured ratios of $^{14}$N/$^{15}$N. This is supported by the fact that no systematic variations appear between the isotopic ratios and heliocentric distances and consistent ratios of the three sources detected by both the TMRT - 65 m and the Effelsberg-100 m telescopes. 
Fractionation remains insignificant for isotope ratios, as indicated by the correlation between abundance ratios and the kinetic temperature $T_{k}$.
This indicates that fractionation as a temperature dependent effect does not play a dominant role for our results. Other chemical processes could include the presence of a notable UV-field and related isotope selective fractionation \citep[e.g. ][]{Weiss01}.

3) Our measured $^{14}$N/$^{15}$N isotope ratios increase with galactocentric distance, which confirms the Galactic radial gradient proposed by previous studies. An unweighted linear fit gives $\frac{^{14}{\rm NH}_{3}}{^{15}{\rm NH}_{3}} =  (17.50 \pm 13.14) D_{\rm GC} + (53.91 \pm 91.74)$, with a Pearson's  rank correlation coefficient of R = 0.35, which matches the trend predicted by Galactic chemical evolution models. More data from the Galactic center and the sources with large galactocentric distance (\textgreater 8kpc) as well as more Galactic disk values with smaller uncertainty would still be desirable to better confirm and quantify this gradient.

\begin{acknowledgements}
	
We wish to thank the anonymous referee for careful reading and detailed comments, which helped to improve the manuscript. We also thank the operators and staff at both the TMRT and Effelsberg stations for their assistance during our observations, and Dr. X. Chen for his TMRT data of NGC\,6334\,I and nice comments. This work is supported by the Natural Science Foundation of China (No. 12041302, 11590782).  C.H. acknowledges support by Chinese Academy of Sciences President's International Fellowship Initiative under Grant No.2021VMA0009. Y. T. Y. is a member of the International Max Planck Resraech School (IMPRS) for Astronomy and Astrophysics at the Universities of Bonn and Cologne. Y. T. Y. would like to thank the China Scholarship Council (CSC) for support. J. J. Q. thanks for support from the NSFC (No. 12003080), the China Postdoctoral Science Foundation funded project (No. 2019M653144), the Guangdong Basic and Applied Basic Research Foundation (No. 2019A1515110588), and the Fundamental Research Funds for the Central Universities, Sun Yat-sen University (No. 19lgpy284). X.D.T. acknowledges support by the Heaven Lake Hundred-Talent Program of Xinjiang Uygur Autonomous 432 Region of China and the National Natural Science Foundation of China under grant 11903070 and 11433008. 
\end{acknowledgements}


\appendix 
\startlongtable

\begin{deluxetable}{lccccclccccc}
	
	\tablecaption{Our source list for $^{14}$NH$_{3}$ and $^{15}$NH$_{3}$ observations \label{tab:sample}}
	
	\tablehead{\colhead{Object}	&	\colhead{Telescope}	&	\colhead{$\alpha$(2000) $\delta$(2000)}		&	\colhead{time}	&	\colhead{Molecule}	&	\colhead{r.m.s.}	&	\colhead{Object}	&	\colhead{Telescope}	&	\colhead{$\alpha$(2000) $\delta$(2000)}		&	\colhead{time}	&	\colhead{Molecule}	&	\colhead{r.m.s.}	\\
		\colhead{}	&		&	($^h \; ^m \; ^s$)   ($^{\circ} \; ^{\prime} \; ^{\prime\prime}$)		&	\colhead{(min)}	&	\colhead{}	&	\colhead{(mK)}	&	\colhead{}	&		&	($^h \; ^m \; ^s$)   ($^{\circ} \; ^{\prime} \; ^{\prime\prime}$)		&	\colhead{(min)}	&	\colhead{}	&	\colhead{(mK)}	
	}
	\startdata
	G032.04	&	TMRT	&	18:49:36	-00:45:37.1	&	11	&	\textbf{$^{14}$NH$_{3}$(1,1)}	&	47	&	G192.60	&	TMRT	&	06:12:54.02	+17:59:23.3	&	30	&	\textbf{$^{14}$NH$_{3}$(1,1)}	&	85 	\\
	&		&			&		&	\textbf{$^{14}$NH$_{3}$(2,2)}	&	66	&		&		&			&		&	\textbf{$^{14}$NH$_{3}$(2,2)}	&	68 	\\
	&		&			&		&	\textbf{$^{14}$NH$_{3}$(3,3)}	&	62	&		&		&			&		&	\textbf{$^{14}$NH$_{3}$(3,3)}	&	75 	\\
	&		&			&		&	\textbf{$^{15}$NH$_{3}$(1,1)}	&	78	&		&		&			&		&	$^{15}$NH$_{3}$(1,1)	&	95 	\\
	&		&			&		&	$^{15}$NH$_{3}$(2,2)	&	77	&		&		&			&		&	$^{15}$NH$_{3}$(2,2)	&	85 	\\
	&		&			&		&	$^{15}$NH$_{3}$(3,3)	&	78	&		&		&			&		&	$^{15}$NH$_{3}$(3,3)	&	71 	\\
	G053.23	&	TMRT	&	19:29:33	+18:01:00.6	&	65	&	\textbf{$^{14}$NH$_{3}$(1,1)}	&	26	&	G192.60	&	TMRT	&	06:12:53.99	17:59:23.70	&	9	&	\textbf{$^{14}$NH$_{3}$(1,1)}	&	71 	\\
	&		&			&		&	\textbf{$^{14}$NH$_{3}$(2,2)}	&	29	&		&		&			&		&	\textbf{$^{14}$NH$_{3}$(2,2)}	&	74 	\\
\enddata
\end{deluxetable}
\tablecomments{Column (1): source name; Column(2): used Telescope; Column(3): Right ascension (J2000) and Declination (J2000); Column(4): total integration time; Column(5): molecular line detections in boldface; Column(6): the Root-Mean-Square ({\it rms}) noise value.}


\begin{thebibliography}{}

\bibitem[Adande \& Ziurys(2012)]{Adande12} Adande G. R., Ziurys L. M. 2012, \apj, 744, 194
\bibitem[Adams(1981)]{Adams81} Adams S. 1981, \apj, 247, L123
\bibitem[Audouze et al.(1975)]{Audouze75} Audouze J., Lequeux J., \& Vigroux L. 1975, \aap, 43, 71
\bibitem[Camacho et al.(2020)]{Camacho20} Camacho V., Vázquez-Semadeni E., Palau A., Busquet, G., Zamora-Avilés, M. 2020, \apj, 903, 46
\bibitem[Charnley \& Rodgers et al.(2002)]{Charnley02} Charnley S. B. \& Rodgers S. D. 2002, \apj, 569, 137
\bibitem[Clayton(2003)]{Clayton03} Clayton D. D. 2003, Handbook of Isotopes in the Cosmos: Hydrogen to Gallium. Cambridge Univ. Press, Cambridge
\bibitem[Colzi et al.(2018a)]{Colzi18a} Colzi L., Fontani F., Rivilla V. M., et al. 2018a, \mnras, 478, 3693
\bibitem[Colzi et al.(2018b)]{Colzi18b} Colzi, L., Fontani, F., Caselli, P., et al.\ 2018b, \aap, 609, A129
\bibitem[Colzi et al.(2019)]{Colzi19} Colzi L., Fontani F., Caselli P. et al. 2019, \mnras, 485,5543
\bibitem[Cooper et al.(2013)]{Cooper13} Cooper H. D. B., Lumsden S. L., Oudmaijer R. D. et al. 2013, \mnras, 1125, 1157
\bibitem[Cutri et al.(2003)]{Cutri03} Cutri R. M., Skrutskie M. F., van Dyk S. et al. VizieR, 2003, 246, 2246
\bibitem[Cyganowski et al.(2013)]{Cyganowski13} Cyganowski C. J., Koda J., Rosolowsky E., et al. 2013, \apj, 764, 61
\bibitem[Dahmen et al.(1995)]{Dahmen95} Dahmen G., Wilson T. L., \& Matteucci F. 1995, \aap, 295, 194
\bibitem[Danby et al.(1988)]{Danby88} Danby G., Flower D. R., Valiron P., Schilke P., \& Walmsley C. M. 1988, \mnras, 235, 229
\bibitem[Daniel et al.(2013)]{Daniel13}  Daniel1 F., Gérin M., Roueff E. et al. 2013, \aap, 560, A3
\bibitem[Dent et al.(1984)]{Dent84} Dent W. R. F., Little L. T., White G. J. \mnras, 210, 173
\bibitem[Fontani et al.(2015)]{Fontani15} Fontani F., Caselli P., Palau A. et al. 2015, \apjl, 808, L46
\bibitem[Furuya \& Aikawa(2018)]{Furuya18} Furuya, K. \& Aikawa, Y.\ 2018, \apj, 857, 105
\bibitem[Gerin et al.(2009)]{Gerin09} Gerin M., Marcelino N., Biver N. et al. 2009, \aap, 498, L9
\bibitem[Ginsburg et al.(2011)]{Ginsburg11} Ginsburg A., \& Mirocha J. 2011, PySpecKit: Python Spectroscopic Toolkit, Astrophysics Source Code Library, ascl:1109.001
\bibitem[Ginsburg et al.(2016)]{Ginsburg16} Ginsburg, A., Henkel, C., Ao, Y., et al.\ 2016, \aap, 586, A50
\bibitem[Goddi et al.(2015)]{Goddi15} Goddi C., Henkel C., Zhang Q., et al. 2015, \aap, 573, 109
\bibitem[Goldsmith \& Langer(1999)]{Goldsmith99} Goldsmith P. F., \& Langer W. D. 1999, \apj, 517, 209
\bibitem[Gong et al.(2015)]{Gong15} Gong Y., Henkel C., Thorwirth S. et al. 2015, \aap, 581, A48
\bibitem[Gravity Collaboration et al.(2018)]{Gravity18} Gravity Collaboration, Abuter R., Amorim A. et al. 2018, \aap, 615, L15 
\bibitem[G{\"u}sten \& Ungerechts(1985)]{Gusten85} G{\"u}sten R., \& Ungerechts H. 1985, \aap, 145, 241
\bibitem[Heays et al.(2014)]{Heays14} Heays A. N., Visser R., Gredel R. et al 2014, \aap, 562, 61
\bibitem[Hermsen et al.(1985)]{Hermsen85} Hermsen W., Wilson T. L., Walmsley C. M., and Batrla W. 1985, \apj, 146, 134
\bibitem[Ho \& Townes(1983)]{Ho83} Ho P. T. P., \& Townes C. H. 1983a, ARA\&A, 21, 239
\bibitem[Izzard et al.(2004)]{Izzard04} Izzard R. G., Tout C. A., Karakas A. I., Pols O. R. 2004, \mnras, 350, 407
\bibitem[Keown et al.(2017)]{Keown17} Keown J., Di Francesco J., Kirk, H. et al. 2017, \apj, 850, 3
\bibitem[Kim et al.(2008)]{Kim08} Kim M. K., Hirota T., Honma M. et al. 2008, PASJ, 60, 991
\bibitem[Li et al.(2003)]{Li03} Li D., Goldsmith P. F., \& Menten K. 2003, \apj, 587, 262
\bibitem[Li et al.(2016)]{Li16} Li J., Shen Z.-Q., Wang J. Z. et al. 2016, \apj, 824, 136
\bibitem[Lis et al.(2010)]{Lis10} Lis D. C., Wootten A., Gerin M., Roueff E. 2010, \aap, 52, L26
\bibitem[Loison et al.(2019)]{Loison19} Loison J. C., Wakelam V., Gratier P., et al. 2019, \mnras, 484, 2747
\bibitem[Loison et al.(2020)]{Loison20} Loison J. C., Wakelam V., Gratier P. et al. 2020, \mnras, 498, 4663
\bibitem[Longmore et al.(2007)]{Longmore07} Longmore S. N., Burton M. G., Barnes P. J. et al. 2007, \mnras, 379, 535
\bibitem[Mangum et al.(1992)]{Mangum92} Mangum J. G., Wootten A., \& Mundy L. G. 1992, \apj, 388, 467
\bibitem[Mangum et al.(2015)]{Mangum15} Mangum J. G., Shirley Y. L. 2015, \pasp, 127,266
\bibitem[Maud et al.(2015)]{Maud15} Maud L. T., Moore T. J. T., Lumsden S. L. et al. 2015, \mnras, 645, 665
\bibitem[Mauersberger et al.(1987)]{Mauersberger87} Mauersberger R., Henkel C., Wilson T. L., 1987, \aap, 173, 352
\bibitem[Mei et al.(2020)]{Mei20} Mei Y., Chen X., Shen Z. Q., Li B. 2020, \apj, 898, 157 
\bibitem[Milam et al.(2005)]{Milam05} Milam S. N., Savage C., Brewster M. A., Ziurys L. M., \& Wyckoff S. 2005, \apj, 634, 1126
\bibitem[Mills et al.(2018)]{Mills18} Mills E. A. C., Corby J., Clements A.R. 2018, \apj, 869, 121
\bibitem[Ott et al.(1994)]{Ott94} Ott M., Witzel A., Quirrenbach A., et al. 1994, \aap, 284, 331 
\bibitem[Parsons et al.(2018)]{Parsons18} Parsons H., Dempsey J. T., Thomas H. S. et al. 2018, \apjs, 234, 22
\bibitem[Pickett et al.(1998)]{Pickett98} Pickett H. M., Poynter R. L., Cohen E. A. et al. 1998,  JQSRT, 60, 883
\bibitem[Ragan et al.(2011)]{Ragan11} Ragan S. E., Bergin E. A., \& Wilner D. 2011, \apj, 736, 163
\bibitem[Reid et al.(2014)]{Reid14} Reid M. J., Menten K. M., Brunthaler A., et al. 2014, \apj, 783, 130
\bibitem[Reid et al.(2019)]{Reid19} Reid M. J., Menten K. M., Brunthaler A. et al. 2019, \apj, 885, 131
\bibitem[Rodgers \& Charnley et al.(2008)]{Rodgers08} Rodgers S. D., \& Charnley S. B. 2008, \mnras, 385, L48
\bibitem[Roman-Duval et al.(2009)]{Roman09} Roman-Duval J., Jackson J. M., Heyer M. et al. 2009, \apj, 699, 1153
\bibitem[Romano \& Matteucci(2003)]{Romano03} Romano D., Matteucci F. 2003, \mnras, 342, 185
\bibitem[Romano et al.(2017)]{Romano17} Romano D., Matteucci F., Zhang Z. -Y., Papadopoulos P. P., Ivison R. J. 2017, \mnras, 470, 401
\bibitem[Romano et al.(2019)]{Romano19} Romano D., Matteucci F., Zhang Z. -Y. et al. 2019, \mnras, 490, 2838
\bibitem[Rosolowsky et al.(2009)]{Rosolowsky09} Rosolowsky E., Dunham M. K. et al. 2009, \apjs, 188: 123
\bibitem[Roueff et al.(2015)]{Roueff15} Roueff E., Loison J. C., Hickson K. M. 2015, \aap, 576, A99
\bibitem[Rygl et al.(2010)]{Rygl10} Rygl K. L. J., Brunthaler A., Reid M. J., et al. 2010, \aap, 511, 2
\bibitem[Tafalla et al.(2004)]{Tafalla04} Tafalla M., Myers P. C., Caselli P., \& Walmsley C. M. 2004, \aap, 416, 191
\bibitem[Terzieva \& Herbst(2000)]{Terzieva00} Terzieva R., \& Herbst. E. 2000, \mnras, 317, 563
\bibitem[Urquhart et al.(2018)]{Urquhart18} Urquhart J. S., K{\"o}nig C., Giannetti A. et al. 2018, \mnras, 1059, 1102
\bibitem[Urquhart et al.(2011)]{Urquhart11} Urquhart J. S., Morgan L. K., Figura C. C. et al. 2011, \mnras, 418, 1689
\bibitem[Visser et al.(2018)]{Visser18} Visser R., Bruderer S., Cazzoletti P. et al 2018, \aap, 615, 75
\bibitem[Viti et al.(2019)]{Viti19} Viti S., Fontani F., Jiménez-Serra I., Holdship J. 2019, \mnras, 486, 4805
\bibitem[Wang et al.(2020)]{Wang20} Wang S., Ren Z. Y., Li D. et al. 2020 \mnras, 1, 19
\bibitem[Wannier et al.(1981)]{Wannier81} Wannier P. G., Linke R. A., \& Penzias A. A. 1981, \apj, 247, 522
\bibitem[Wei{\ss} et al.(2001)]{Weiss01} Wei{\ss}, A., Neininger, N., Henkel, C., et al.\ 2001, \apjl, 554, L143. doi:10.1086/321711
\bibitem[Wiescher et al.(2010)]{Wiescher10} Wiescher M., G{\"o}rres J., Uberseder E., Imbriani G., Pignatari M. 2010, ARNPS, 60, 381
\bibitem[Willis et al.(2013)]{Willis13} Willis S., Marengo M., Allen L. et al. 2013, \apj, 96, 20
\bibitem[Wilson \& Rood(1994)]{Wilson94} Wilson T. L., \& Rood R. T. 1994, ARA\&A, 32, 191
\bibitem[Wilson(1999)]{Wilson99} Wilson T. L. 1999, RPPh, 62, 143	
\bibitem[Wirstr{\"o}m \& Charnley(2018)]{Wirstrom18} Wirstr{\"o}m E. S., \& Charnley S. B., 2018, \mnras, 474,3720
\bibitem[Wyrowski and Walmsley(1996)]{Wyrowski96} Wyrowski F. and Walmsley C. M. 1996, \apj, 314,265
\bibitem[Yan et al.(2019)]{Yan19} Yan Y. T., Zhang J. S., Henkel C., Mufakharov T., 2019, \apj, 877, 154
\bibitem[Yu et al.(2020)]{Yu20} Yu H. Z., Zhang J. S., Henke C., et al. 2020, \apj, 899, 145
\bibitem[Zhang et al.(2015)]{Zhang15} Zhang J. S., Sun L. L., Riquelme D. et al. 2015, \apjs, 219, 28
\bibitem[Zhang et al.(2018)]{Zhang18} Zhang Z. Y., Romano D., Ivison R. J., Papadopoulos Padelis P., Matteucci F. 2018, Natur, 558, 260
\bibitem[Zhou et al.(2020)]{Zhou20} Zhou D. D., Wu G., Esimbek J., Henkel C., Zhou J.J. 2020, \aap, 640, 114
\end{thebibliography}
\end{document}